\documentclass[onecolumn,draftclsnofoot,12pt]{IEEEtran}
\usepackage[pdftex]{graphicx}
\usepackage{amsfonts}
\usepackage{amsmath}
\usepackage{amssymb,amsthm}
\usepackage{bbm}
\usepackage{float}
\usepackage{graphicx}
\usepackage{tikz}
\usepackage{xcolor}
\usepackage{verbatim}
\usepackage[footnotesize]{caption}
\usepackage{subcaption}
\usepackage[colorlinks=true,linkcolor=black,anchorcolor=black,citecolor=black,filecolor=black,menucolor=black,runcolor=black,urlcolor=black]{hyperref}
\usepackage{multirow}
\usepackage{framed} 
\usepackage{booktabs}
\usepackage{subcaption}

\newcommand{\expU}[1]{e^{#1}}
\newcommand{\SINR}{\mathsf{SINR}}

\newcommand{\SThres}{\tau}
\newcommand{\Ball}{\mathcal{B}}
\newcommand{\expects}[2]{\mathbb{E}_{#1}\left[#2\right] }

\newcommand{\expS}[1]{\exp\left(#1\right)}
\newcommand{\prim}{\mathrm{p}}
\newcommand{\seco}{\mathrm{s}}
\newcommand{\noise}{\sigma^2}
\newcommand{\gst}[1]{g_\mathrm{st}\left(#1\right)}
\newcommand{\gsr}[1]{g_\mathrm{sr}\left(#1\right)}
\newcommand{\gpt}[1]{g_\mathrm{pt}\left(#1\right)}
\newcommand{\gpr}[1]{g_\mathrm{pr}\left(#1\right)}
\newcommand{\dd}{\mathrm{d}}
\newcommand{\X}{\mathbf{X}}
\newcommand{\x}{\mathbf{x}}
\newcommand{\Y}{\mathbf{Y}}

\renewcommand{\ast}{a_\mathrm{st}}
\newcommand{\asr}{a_\mathrm{sr}}
\newcommand{\apt}{a_\mathrm{pt}}
\newcommand{\apr}{a_\mathrm{pr}}
\newcommand{\bst}{b_\mathrm{st}}
\newcommand{\bsr}{b_\mathrm{sr}}

\newcommand{\bpr}{b_\mathrm{pr}}
\newcommand{\qst}{q_\mathrm{st}}

\newcommand{\qpr}{q_\mathrm{pr}}
\newcommand{\rcvdp}[2]{P_{#2\rightarrow#1}}
\newcommand{\tbr}[1]{}
\newcommand{\tbrb}[1]{}

\newtheorem{theorem}{Theorem}
\newtheorem{corr}{Corollary}
\newtheorem{lemma}{Lemma}
\newtheorem{remark}{Remark}
\newtheorem{example}{Example}

\IEEEoverridecommandlockouts

\begin{document}
%----------------------------------------------------------------------------------------------------- 
\pagenumbering{gobble}  
\title{On the Coverage of Cognitive mmWave Networks with Directional Sensing and Communication}
\author{Shuchi Tripathi, Abhishek K. Gupta, SaiDhiraj Amuru \vspace{-.1in}
\thanks{S. Tripathi and A. Gupta are with the Department of Electrical Engineering, Indian Institute of Technology Kanpur,  India. S. Amuru is with the Department of Electrical Engineering, Indian Institute of Technology Hyderabad, India. Email: \{shuchi, gkrabhi\}@iitk.ac.in, asaidhiraj@ee.iith.ac.in. A part of the paper was presented in ASILOMAR 2021 \cite{tripathi2021coverage}}  
	}
	
%----------------------------------------------------------------------------------------------------- 
\maketitle
%----------------------------------------------------------------------------------------------------- 

\begin{abstract}
Millimeter-waves' propagation characteristics create prospects for spatial and temporal spectrum sharing in a variety of contexts, including cognitive spectrum sharing (CSS). However, CSS along with omnidirectional sensing, is not efficient at mmWave frequencies due to their directional nature of transmission, as this limits secondary networks' ability to access the spectrum. This inspired us to create an analytical approach using stochastic geometry to examine the implications of directional cognitive sensing in mmWave networks. We explore a scenario where multiple secondary transmitter-receiver pairs coexist with a primary transmitter-receiver pair, forming a cognitive network. The positions of the secondary transmitters are modelled using a homogeneous Poisson point process (PPP) with corresponding secondary receivers located around them. A threshold on directional transmission is imposed on each secondary transmitter in order to limit its interference at the primary receiver. We derive the medium-access-probability of a secondary user along with the fraction of the secondary transmitters active at a time-instant. To understand cognition's feasibility, we derive the coverage probabilities of primary and secondary links. We provide various design insights via numerical results. For example, we investigate the interference-threshold's optimal value while ensuring coverage for both links and its dependence on various parameters. We find that directionality improves both links' performance as a key factor. Further, allowing location-aware secondary directionality can help achieve similar coverage for all secondary links.
\end{abstract}

%----------------------------------------------------------------------------------------------------- 
\section{Introduction}  
%-------------------------------------------------------------------------------------------------

The licensed operators traditionally have exclusive control over a certain band of the spectrum which led to underutilization of a significant portion of the spectrum outside of a few peak hours \cite{durantini2013spectrum}. The need for more flexible spectrum management gives birth to a variety of spectrum sharing techniques, including CSS, also known as secondary license sharing \cite{AG2020}. Here, a primary operator lends the secondary operators a portion of its licenced spectrum under specific constraints to ensure no or minimal effect on its quality of service (QoS). These limitations may include priority to the primary transmission and provisioning a threshold limit on the secondary interference \cite{gupta2016restricted}. CSS allows secondary devices to perform opportunistic transmission depending on the band's availability and the geographical positions of the devices under these restrictions. One example of this is the scenario where each secondary device continuously senses the licensed channel and uses the channel only when all restrictions are satisfied. This sensing, hereby known as cognitive channel sensing (CCS), enables a secondary device to efficiently utilise the sparsely used frequency bands allocated to primary operators. As the modern generation of wireless networks is moving towards mmWave \cite{tripathi2021millimeter}, incumbent services at these frequencies along with various types of proposed communication services require us to efficiently share this spectrum \cite{fccnoi14}. The good news is that the increased blockage sensitivity and use of highly directed antenna at the mmWave frequencies encourages the ideal spectrum sharing conditions resulting in a notable reduction in the degree of interference and higher spatial reuse of the spectrum. Although it was shown that spectrum sharing is feasible without any coordination \cite{AG2020,GuptaAndHea16}, it will be important to understand how much CCS can help at mmWave frequencies.

%-----------------------------------------------------------------------------------------
\subsection{Related work}
%--------------------------------------------------------------------------------------

Authors in \cite{nguyen12} studied the system level analysis of a CCS based wireless network consisting of primary and multiple secondary devices at the conventional frequency with omnidirectional antennas. They considered a cognitive model where a secondary device can only transmit if it does not sense any primary signal above some threshold and developed a stochastic geometry based analytical framework to derive the secondary transmission probability and SINR distribution. Note that stochastic geometry is a tractable tool to study the performance of wireless networks \cite{AndGupDhi2016}. The omnidirectional sensing at traditional frequencies was used in \cite{zhang2014spectrum} to quantify the relation between proximity to the primary-transmission range and diverse spectrum access opportunity of secondary devices with stochastic geometry. However, omnidirectional sensing gives rise to distance-dependent CCS, which provides channel access opportunities to secondary devices based on their distance from the primary receiver. This results in a circular region around the primary receiver where a secondary transmitter is not allowed to transmit, known as the primary receiver's protection zone.

Now, coming to communication at mmWave frequencies, it requires different analytical frameworks \cite{bai2014coverage,AndrewsmmWaveTut2016} due to many differences from its traditional counterparts. It was shown that the use of many antennas in mmWave systems leads to directional communication which provides higher SINR gain and enables spectrum sharing naturally without any explicit coordination \cite{GuptaAndHea16}. Since applying complex coordination schemes may not be desirable to operators, \cite{gupta16} studied the impact of simple transmit restrictions on secondary coverage performance for cognitive mmWave networks. The work assumed omnidirectional CCS at both primary and secondary ends and showed that under directional communication, the transmit-based restriction can improve performance. However, protection zones were radially symmetric around the primary receivers due to omnidirectional CCS. Authors in \cite{yamashita2018exclusive} considered a directional communication based cognitive system with a single primary receiver and many secondary receivers spread over a region. The region is divided into small grids with different channel usage probabilities for the secondary devices. Simulation results verified that the primary receiver's protection zone is not symmetric and a larger protection zone is needed in high antenna gain directions. To understand it better, consider a scenario in which a secondary transmitter is placed very close to the primary receiver but has a very low antenna gain toward the primary receiver. Undoubtedly, the secondary transmitter's orientation allows it to utilise the channel without interfering with the primary transmission. Similarly, additional freedom can be given to all secondary devices located in the primary receiver's low gain direction. On the other hand, a symmetrical circular protection zone at mmWave frequencies will unnecessarily prevent many secondary transmitters from using the channel, which will result in the under-use of the spectrum. Thus, the locations of the secondary transmitters as well as the \textit{orientations of the associated secondary links} with respect to the primary link affect channel access options under directional CCS. As stated above, the use of directional CCS in mmWave raises fundamental differences in the overall system and it is important to evaluate such systems. Some recent works \cite{lagen2018lbt}-\cite{li18} 
%%%%%%%%%%%% Don't Delete %%%%%%%%%%%%%%%%
%\cite{lagen2018lbt, lagen18, daraseliya2021coexistence, ramisetti2020methods, li18} 
%%%%%%%%%%%%%%%%%%%%%%%%%%%%%%%%%%%%%%%%%
have taken the simulation-based approach to discuss the importance of directional sensing in the cognitive mmWave network. The work in \cite{lagen2018lbt} addressed the role of directional CCS in improving the coverage performance from the perspective of secondary transmitters only. Authors in \cite{lagen18, daraseliya2021coexistence} have studied circumstances when potential interfering secondary links can be shut down based on the exact location information of the associated secondary receivers. Authors in \cite{ramisetti2020methods},\cite{ li18} have shown that directional CCS performed at both ends of the secondary link has the advantage of eliminating hidden sources of interference to the primary link. However, a systematic approach to an analytical evaluation of the gain that directional CCS can provide in mmWave cognitive networks in comparison to that of omnidirectional CCS is not available in the literature. Also, the analytical investigation of directional CCS in a mmWave cognitive network - where both primary and secondary links are operating at the same mmWave band - has not been done in the past. Both of these are the focus of our work.

%-----------------------------------------------------------------------------------------
\subsection{Contributions} \label{subsect:contribution}
%--------------------------------------------------------------------------------------

In this paper, we analyze a mmWave cognitive network consisting of a single primary and multiple secondary links with directional channel sensing and communication, all operating at the same frequency band. In particular, we have the following contributions:

\begin{itemize}
\item We consider a mmWave cognitive network consisting of a single primary and multiple secondary links distributed as a PPP. We develop an analytical framework to study the impact of directional CCS in this network. In particular, we consider that the primary operator restricts the secondary links to transmit only when its signal power (which includes antenna directional gains and instantaneous fading) received at the primary receiver is below a certain threshold. We show that this results in an asymmetric protection zone around the primary where the secondary transmitter can not exist.

\item We compute the transmission opportunities for the typical secondary transmitter. We also compute coverage probability for both, the primary and secondary links. We would like to highlight that the analysis is significantly different than that done with omnidirectional CCS due to the presence of directional gains associated with channel sensing opportunities of secondary transmitters, which requires many non-trivial changes.

\item We also describe the effects of different parameters on the coverage of both types of links - primary and secondary - to aid in our quest for design insights that will help us to improve the performance of the network. For example, we show the independence of secondary transmitters' channel access probabilities and activity ratios from the secondary network density, the existence of the trade-off between antenna gain and beam-width of the directional antenna with dominating effect of the antenna's beamwidth over the gain, the role of primary and secondary antennas on activity and links' performance, and dependence of transmission threshold $\rho$ on system parameters together with its selection criteria for better network performance.
\end{itemize}

%--------------------------------------------------------------------------------------
\subsection{Notation} 
%--------------------------------------------------------------------------------------
For a location $\x$,  $x=||\x||$. $\Ball(\x,r)$ denotes a ball with radius $r$ and centre $\x$. The $x\angle \theta$ denotes the location with radial distance $x$ and angle $\theta$. For a location $\x$, the $\angle \x$ denotes the angle of vector $\x$ with respect to the x-axis. Here, $ \mathrm{p}, \ \mathrm{s}, \ \mathrm{t}$ and $ \mathrm{r} $ stand for primary, secondary, transmitter, and receiver, respectively. $\Gamma(a, b) = \int_{0}^{b} t^{a - 1} e^{-t}\, \mathrm{d}t $ denotes the lower incomplete Gamma function.

%--------------------------------------------------------------------------------------
\section{System model} 
%--------------------------------------------------------------------------------------

In this paper, we consider a cognitive mmWave communication consisting of a single primary link coexisting with multiple secondary links with directional CCS. 
%--------------------------------------------------------------------------------------
\subsection{Network model}
%--------------------------------------------------------------------------------------

We consider a single primary link with $\X_{\mathrm{p}}$ and $\Y_{\mathrm{p}}$ denoting the locations of the primary transmitter and receiver, respectively. Without loss of generality, we can assume that the primary receiver location is fixed at the origin \textit{i.e.} $\Y_{\mathrm{p}} = \mathbf{o}$ and the primary link, $\X_{\mathrm{p}} - \Y_{\mathrm{p}}$ is aligned with the X-axis as shown in 
%-------------------------------------------------
Fig. 1(a). 
%\ref{fig:PrimarySystemModel}.
%------------------------------------------------
Let $r_\prim$ represent the link distance between the primary transmitter and receiver. 
\vspace*{-0.5cm}
\begin{figure}[ht!]
\centering
{\includegraphics[scale=0.15]{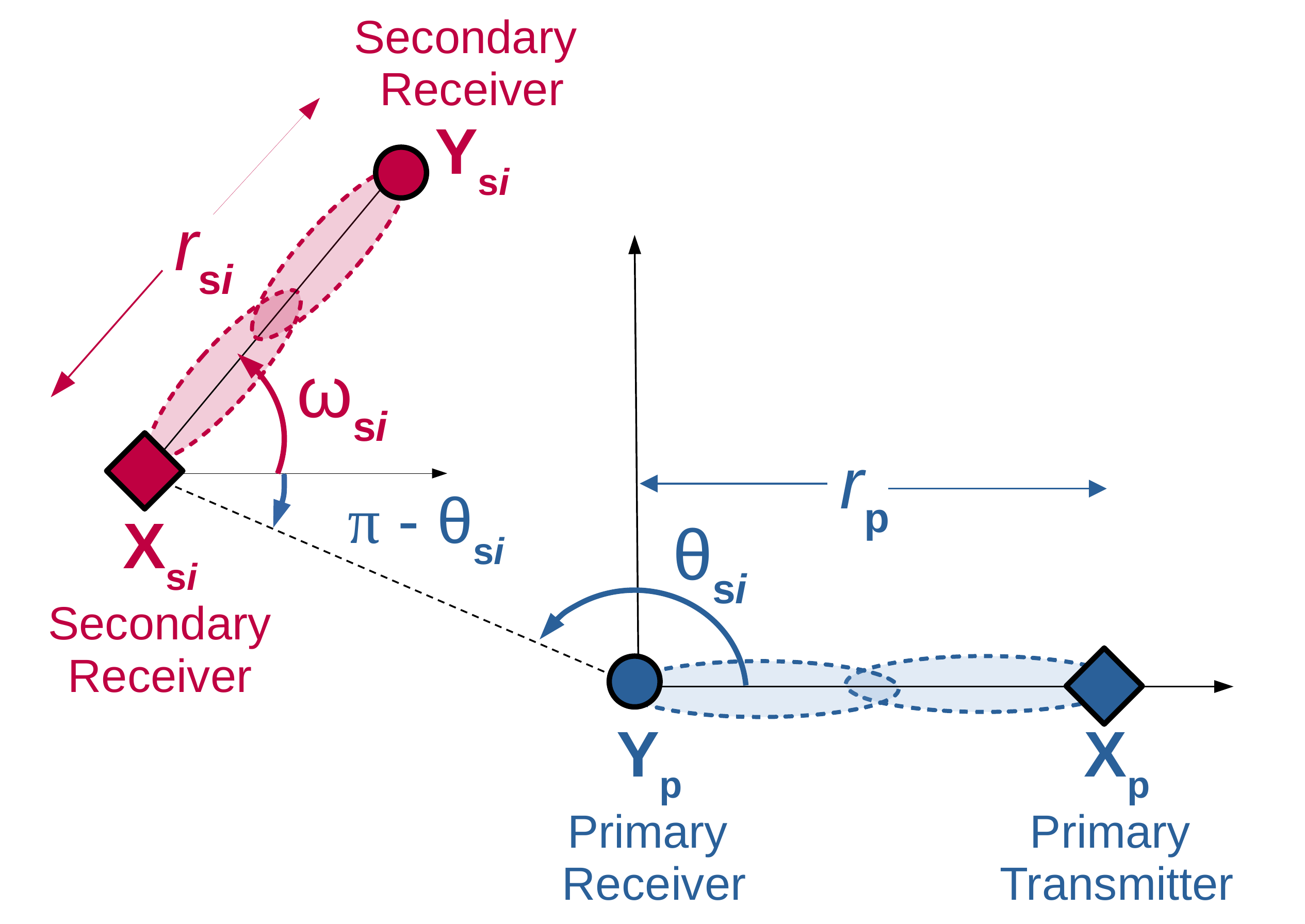}\label{fig:PrimarySystemModel}}
{\includegraphics[scale=0.15]{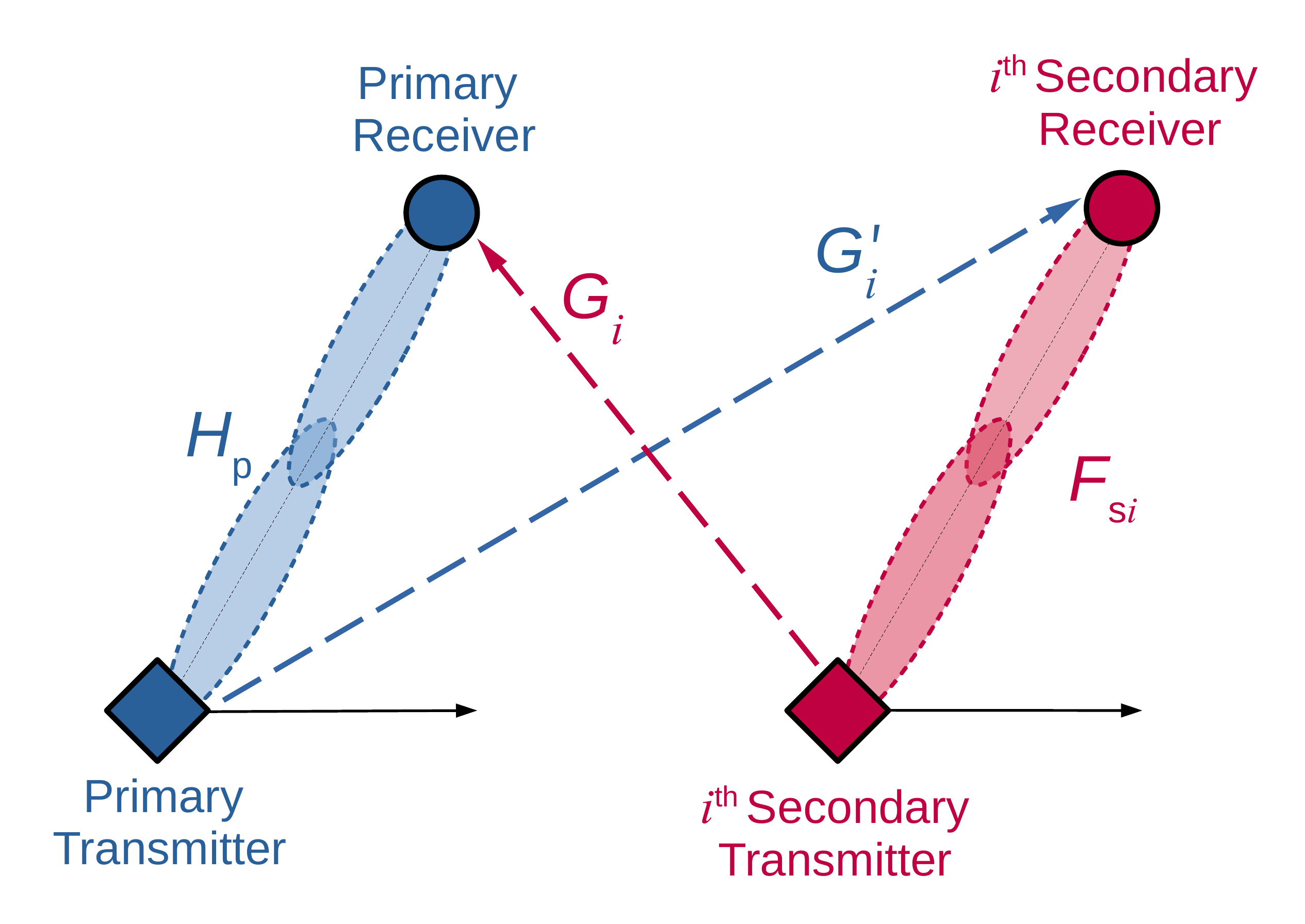}\label{fig:fading}}\\
\vspace*{-0.3cm}
{\small (a) \hspace{3.5cm} (b)}
\vspace*{-0.3cm}
\caption{An illustration showing the system model (a) the location of the single primary link and $i^{\mathrm{th}}$ secondary transmitter-receiver link, (b) the fading coefficients between various devices.}
\end{figure}
\vspace*{-0.5cm}

The locations of secondary transmitters are distributed as a 2D-homogeneous Poisson point process $\Phi_{\mathrm{s}}$ with density $\lambda_{\mathrm{s}}$  \cite{AndrewsmmWaveTut2016} \textit{i.e.} $\Phi_{\mathrm{s}} = \{ \X_{\mathrm{s}i} \}$, where $\X_{\mathrm{s}i}$ is the location of $i^\mathrm{th}$ transmitter. Let $\theta_{\mathrm{s}i}$ represents the angular direction of the $i^\mathrm{th}$ secondary transmitter with respect to the primary link, $\Y_{\mathrm{p}} - \X_{\mathrm{p}}$. Its corresponding receiver is located at $\Y_{\seco i}$.  The length of $i^\mathrm{th}$ secondary link $\X_{\mathrm{s}i}-\Y_{\mathrm{s}i}$ is $r_{\mathrm{s}i}$ and it makes an angle $\omega_{\mathrm{s}i}$ in the anti-clockwise direction from the X-axis (See 
%-------------------------------------------------
Fig. 1(a))
%\ref{fig:PrimarySystemModel}). 
%-------------------------------------------------
Based on the specific application, the values of $\omega_{\mathrm{s}i}$ and $r_{\mathrm{s}i}$ can be fixed or they can be taken as uniform random variables with some distribution \textit{e.g.} uniform. We can see $\left( \omega_{\mathrm{s}i}, \ r_{\mathrm{s}i} \right) $ as \textit{marks} of $\X_{\mathrm{s}i}$, hence, $\left\{\X_{\mathrm{s}i}, \ \left(\omega_{\mathrm{s}i}, \ r_{\mathrm{s}i} \right) \right\}$ is a marked PPP which gives the complete information about the locations of secondary transmitter and receiver pairs. For simplicity, we assume that all secondary receivers are oriented differently with the fixed distance from their assigned transmitters. In other words, the values of $r_{\mathrm{s}i}$ are fixed {\em i.e} $r_{\mathrm{s}i}=r_\seco$, while $\omega_{\mathrm{s}i}$ is a random variable uniformly distributed between 0 and $2\pi$. However, note that the presented analysis can be trivially extended to the general distribution of $r_{\mathrm{s}i}$ and $\omega_{\mathrm{s}i}$. Let $p_\prim$ and $p_\seco$ denote the transmit power of the primary and secondary transmitter, respectively.

\vspace*{-0.2cm}
\begin{figure}[ht!]
\centering
{\includegraphics[scale=0.1]{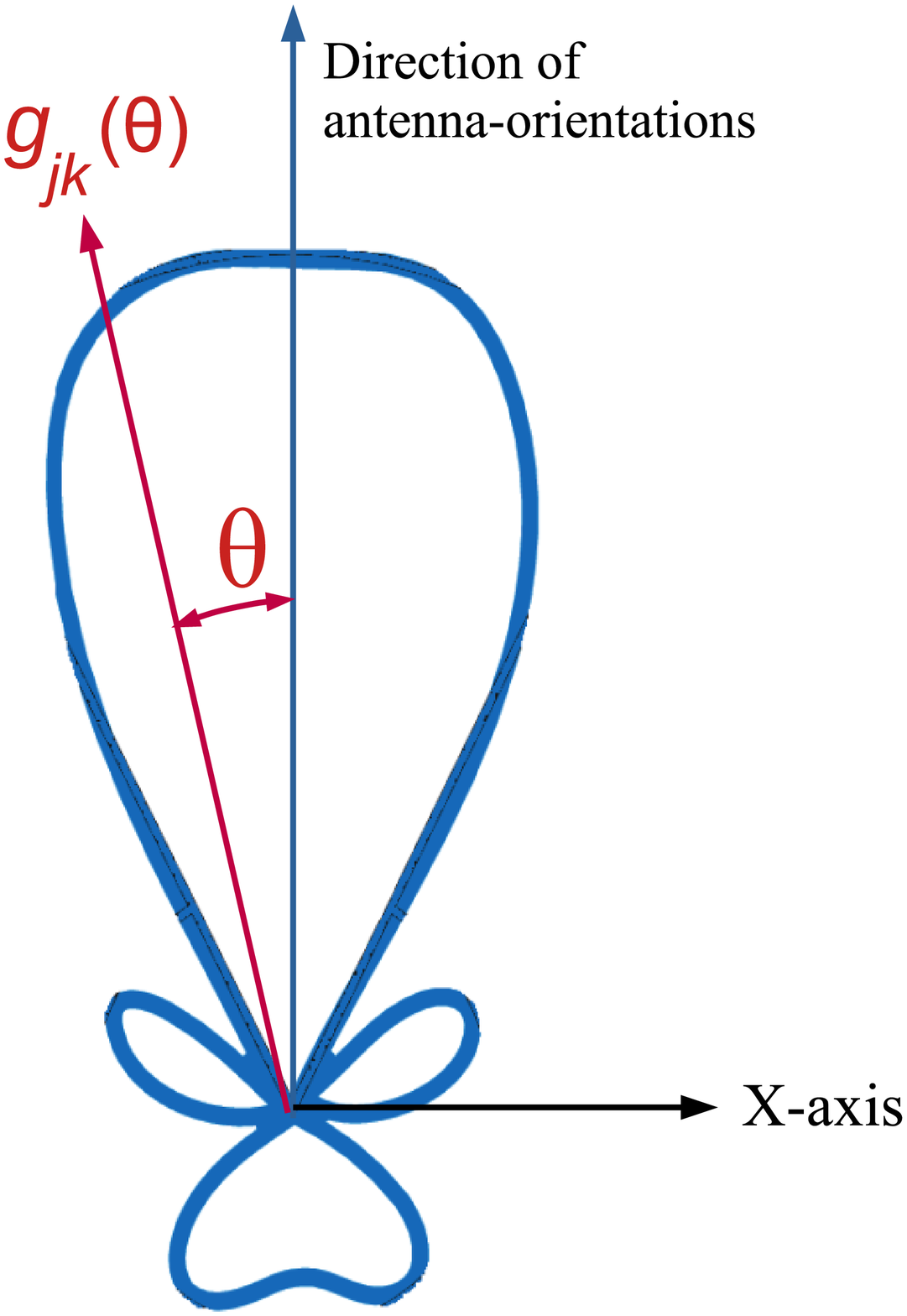}
\label{fig:antennadirection}}
{\includegraphics[scale=0.1]{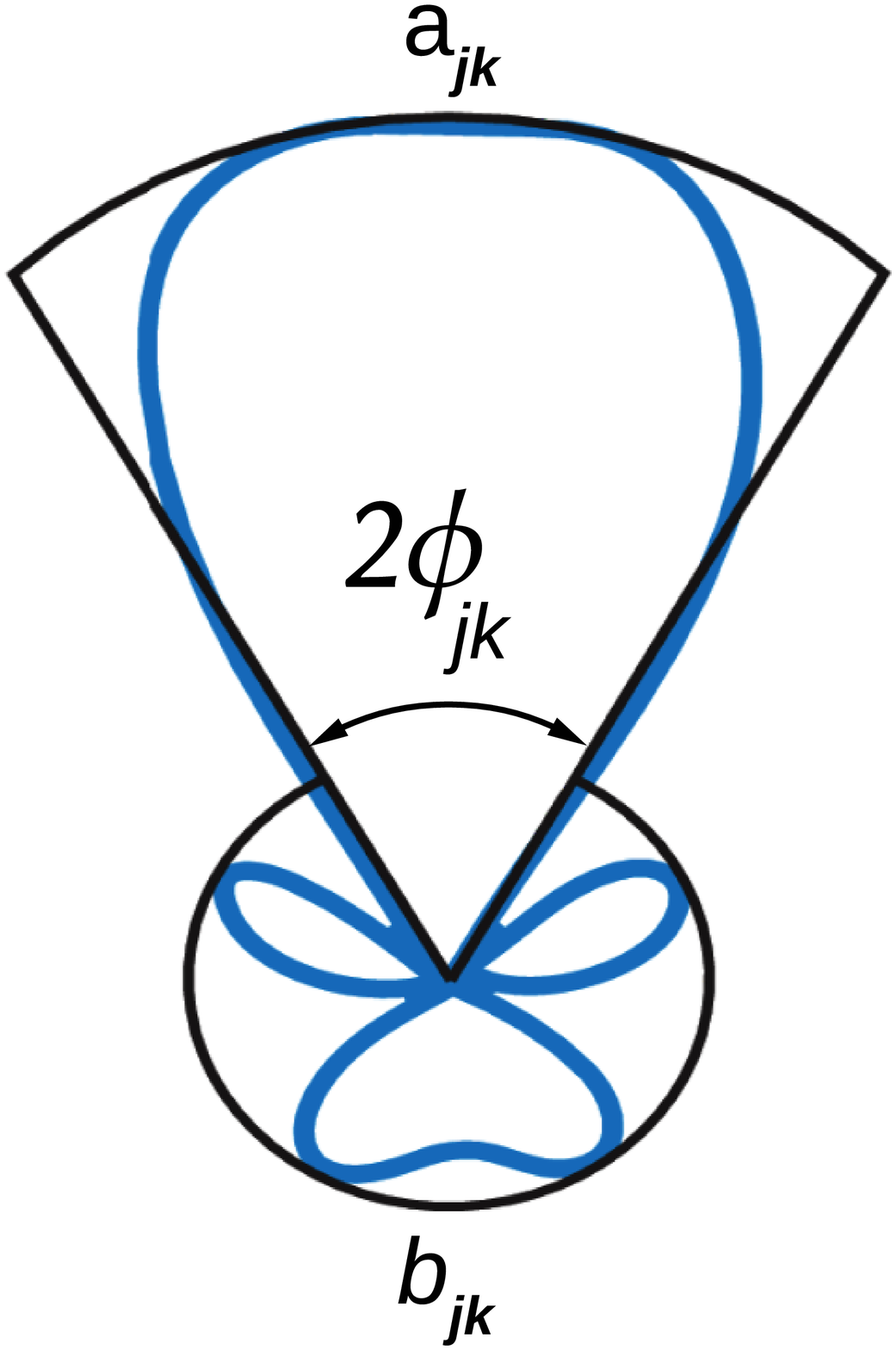} 	\label{fig:sectorbeam}} \\
\vspace*{-0.3cm}
{\small (a) \hspace{1.5cm} (b)}
\vspace*{-0.3cm}
\caption{Antenna beam patterns, illustrating (a) the $g_{jk}(\theta)$ with respect to the direction of antenna-orientation, (b) the sectorized beam pattern approximation.}
\end{figure}
\vspace*{-1cm}

%--------------------------------------------------------------------------------------
\subsection{Directional communication} 
%--------------------------------------------------------------------------------------

We consider that the device $k$ of type $j$ is equipped with  $ M_{jk} $ antenna elements, where $j \in \{ \mathrm{p}, \ \mathrm{s} \} $ with $\prim$ and $\seco$ denoting primary and secondary types, and $ k \in \{ \mathrm{t}, \ \mathrm{r} \} $ denoting transmitter and receiver device. The beam pattern of this device is denoted by $g_{jk}(\theta)$ where $\theta \in [-\pi, \pi)$ is the angle with respect to the beam orientation (see 
%-------------------------------------------------
Fig. 2(a)).
%\ref{fig:antennadirection}). 
%-------------------------------------------------
Note that for omni-directional CCS $g_{jk}(\theta) = 1,\,\forall \,\theta$. To simplify the results, we will also consider the following two special beam patterns occasionally.

%---------------------------
\subsubsection{Sectorized beam pattern} The beam pattern of a  multi-antenna system can be well approximated by a sectorized beam pattern, as given in \cite{AndrewsmmWaveTut2016,li2017design}. 
\begin{align}
g_{jk}(\theta) = \begin{cases}
a_{jk} \qquad &\text{if}\ \lvert \theta \rvert \leq \phi_{jk}/2 \\
b_{jk} \qquad &\text{if}\ \lvert \theta \rvert > \phi_{jk}/2,
\end{cases} \label{eq:gainApproximation}
\end{align} 
where $\phi_{jk} $ is the beamwidth, $ a_{jk} $ is the main lobe gain and $ b_{jk} $ is the side lobe gain (see 
%-------------------------------------------------
Fig. 2(b)).
%\ref{fig:sectorbeam}).
%-------------------------------------------------
Beamwidth is a reciprocal function of $M_{jk}$. The probability $ q_{jk} $ of having $a_{jk}$ gain at a random direction uniformly picked around the antenna is given by 
\begin{align}
q_{jk} = \frac{\phi_{jk} }{ 2 \pi }. \label{eq:WsiPrabability}
\end{align}

Similar to \cite{AndGupDhi2016, AndrewsmmWaveTut2016}, the antenna gains of primary and secondary devices should follow $a_{jk}q_{jk} + b_{jk}(1 - q_{jk}) = 1$ such that transmit powers $p_\prim$ and $p_\seco$ remain independent of beamwidth. For a uniform linear array (ULA), the beamwidth $\phi_{jk}$ is given as $\kappa/M_{jk}$ for some constant $\kappa$ and $a_{jk} = M_{jk}$. Hence, $q_{jk} = \kappa/(2\pi M_{jk}) = \kappa'/M_{jk}$ where $\kappa'= \kappa/2\pi$. Further, the side lobe gain is $b_{jk} = (1 - q_{jk} M_{jk})/(1 - q_{jk}) = (1 - \kappa')/(1 - \kappa'/M_{jk})$.

%---------------------------
\subsubsection{Ideal beam pattern} It is a special case of the sectorized beam pattern with zero side lobe gain \textit{i.e.} $ b_{jk} = 0 $.

%--------------------------------------------------------------------------------------
\subsection{Fading} 
%--------------------------------------------------------------------------------------

Let $H_{\mathrm{p}}$ and $F_{\mathrm{s}i}$ represent the  fading coefficient for the primary link and the $i^{\mathrm{th}}$ secondary link, respectively. Similarly, $G_i$ and $G^{'}_i$ represent the cross-link fading coefficients for the $i^{\mathrm{th}}$-secondary-transmitter-to-primary-receiver and the primary-transmitter-to-$i^{\mathrm{th}}$-secondary-receiver, respectively (see 
%-------------------------------------------------
Fig. 1(b)).
%\ref{fig:fading}). 
%-------------------------------------------------
The channel between the primary and secondary systems is assumed to have Rayleigh co-link and cross-link fading characteristics for the sake of analysis' simplicity. However, the analytical results presented in this paper are easily adaptable to different fading distributions as shown in \cite{gupta2020does, ghatak2021deploy}.

%--------------------------------------------------------------------------------------
\subsection{Cognitive communication and protection zones} 
%--------------------------------------------------------------------------------------

This work considers directional sensing in addition to directional communication. The received power at the primary receiver due to the $i^{\mathrm{th}}$ secondary transmitter is
\begin{align}
\rcvdp{\mathrm{p}}{\mathrm{s}i} = p_\seco G_i g_{\mathrm{pr}} \left( \theta_{\mathrm{s}i} \right) g_{\mathrm{st}} \left( \theta_{\mathrm{s}i} - \pi - \omega_{\mathrm{s}i} \right)x_{\mathrm{s}i}^{- \alpha},
\end{align}
where $\alpha$ is the path-loss exponent. Here, $g_{\mathrm{pr}}$ and $ g_{\mathrm{st}}$ are the antenna beam pattern of the primary receiver and secondary transmitter, respectively. Note that the directional gain of the secondary transmitter depends upon $\omega_{\mathrm{s}i}$ which means that $\rcvdp{\mathrm{p}}{\mathrm{s}i}$ takes account of its associated secondary receiver's relative location. To protect the primary communication, the network restricts each secondary transmitter-receiver pair from causing interference at the primary receiver above a threshold $\rho$, termed as \textit{transmit-restriction-threshold}. Specifically, the $i^{\mathrm{th}}$ secondary transmitter at location $\X_{\mathrm{s}i}$ can transmit only if the received power $\rcvdp{\mathrm{p}}{\mathrm{s}i}$ at the primary receiver is less than $\rho$. Let us denote this event by $E_{i}$ \textit{i.e.} $E_{i} = \left \{ \rcvdp{\mathrm{p}}{\mathrm{s}i} < \rho \right \}$. Let $U_{i}$ be an indicator for the occurrence of event $E_{i}$ \textit{i.e.}
\begin{align}
U_{i} &= \mathbbm{1} \left ( p_\seco G_i g_{\mathrm{pr}} \left( \theta_{\mathrm{s}i} \right) g_{\mathrm{st}} \left( \theta_{\mathrm{s}i} - \pi - \omega_{\mathrm{s}i} \right) x_{\mathrm{s}i}^{- \alpha} < \rho \right ). \label{eq:Indicator}
\end{align}

Thus, $U_{i}$ represents the transmission activity of the $i^{\mathrm{th}}$ secondary transmitter-receiver pair with directional CCS. This restriction creates a dynamic protection zone around the primary receiver where secondary links with a particular orientation cannot remain active depending on their distance and angular location relative to the primary receiver. 

\begin{remark}
Note that the indicator function under omnidirectional CCS is $
U_{i, \mathrm{omni}} = \mathbbm{1} ( p_\seco G_i x_{\mathrm{s}i}^{- \alpha}  < \rho ).$
\end{remark} 

To get a better understanding of the primary receiver's protection zone for directional CCS, in 
%-------------------------------------------------
Fig. 3,
%\ref{fig:SectionE}, 
%-------------------------------------------------
we illustrate two examples of considered model with unit fading ($G_{i} = 1$), constant secondary orientation ($\omega_{\mathrm{s}i} = 0$ and $\pi$ respectively) and sectorized antenna-beam pattern \eqref{eq:gainApproximation}. Since $\rcvdp{\mathrm{p}}{\mathrm{s}i}$ depends on $\theta_{\mathrm{s}i}$, $\omega_{\mathrm{s}i}$ and $x_{\mathrm{s}i}$, we can divide the entire 2D-space into four regions with directional gain being $\apr\ast$ (shown by $\mathsf{S_1}\cup \mathsf{S_2}$), $\apr\bst$ ($\mathsf{S_3}\cup \mathsf{S_4}$), $\bpr\bst$ ($\mathsf{S_5}\cup \mathsf{S_6}$) and $\bpr\ast$ ($\mathsf{S_7}\cup \mathsf{S_8}$), where $\mathsf{S}_j$'s are shown in 
%-------------------------------------------------
Fig. 3.
%\ref{fig:SectionE}.
%------------------------------------------------- 
We can further divide each region into two segments where the inner segment has $\rcvdp{\mathrm{p}}{\mathrm{s}i} \geq \rho$ and the outer segment has $\rcvdp{\mathrm{p}}{\mathrm{s}i} < \rho$. Hence, the union of all inner segments, $\mathsf{S_1}$, $\mathsf{S_3}$, $\mathsf{S_5}$ and $\mathsf{S_7}$, form the protection zone as the secondary transmitters lying in this zone are not allowed to transmit.
\begin{figure}[ht!]
\centering
{\includegraphics[scale=0.15]{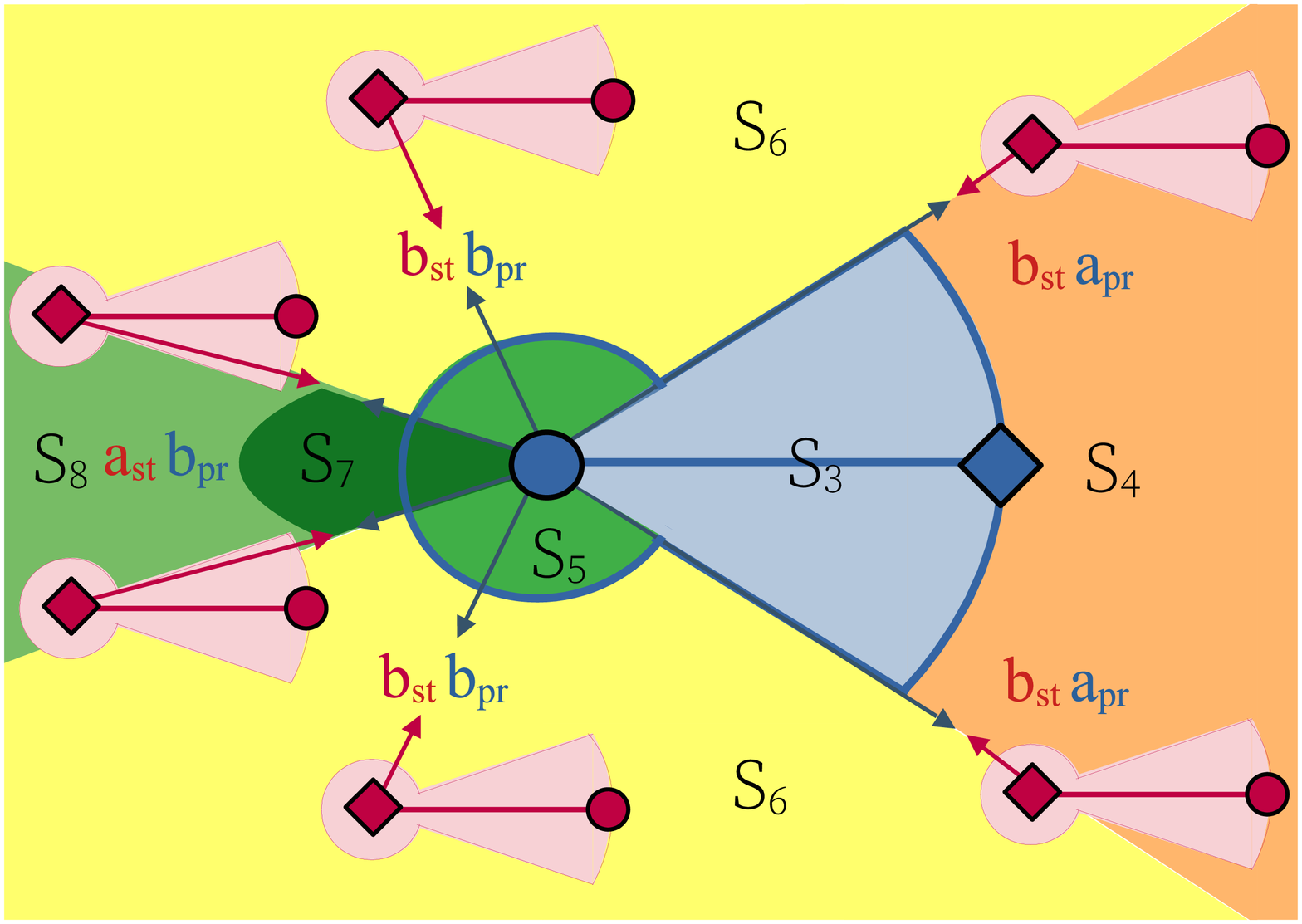}
\label{fig:SectionE--0}}
{\includegraphics[scale=0.15]{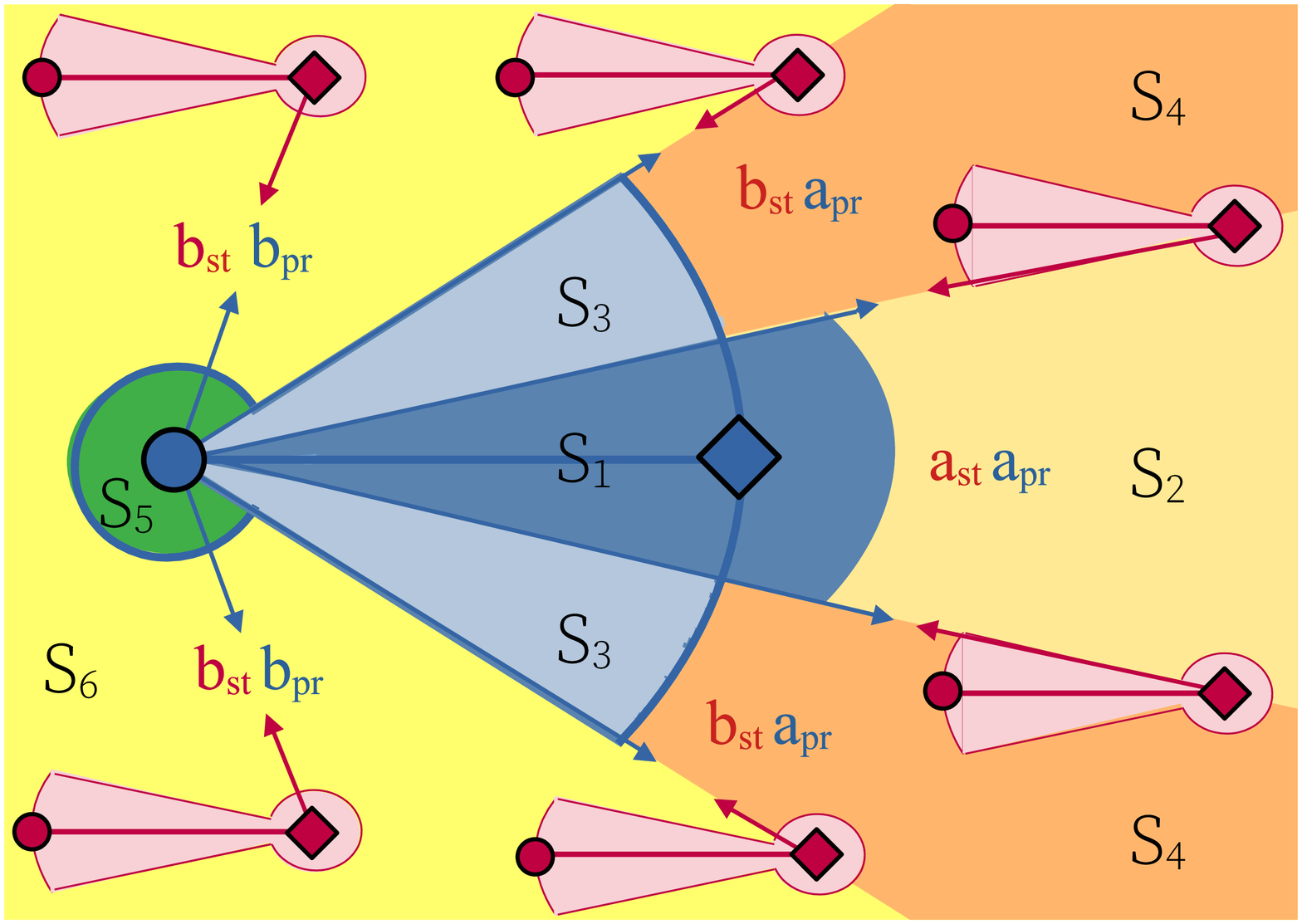} 	\label{fig:SectionE--pi}} \\
\vspace*{-0.3cm}
{\small (a) $\omega_{\mathrm{s}i} = 0$ \hspace{2cm} (b) $\omega_{\mathrm{s}i} = \pi$}
\vspace*{-0.3cm}
\caption{An illustration of secondary-transmitter-to-primary-receiver-antenna-link-gains for different segments of the primary receiver's protection zone.}	
\label{fig:SectionE}
\end{figure}
\vspace*{-0.5cm}

We can determine the boundary between the inner and outer segments in each region. For example, the boundary between $\mathsf{S}_1$ and $\mathsf{S}_2$ lies where $\rcvdp{\mathrm{p}}{\mathrm{s}i} \lvert \mathsf{S}_1 = \rcvdp{\mathrm{p}}{\mathrm{s}i} \lvert \mathsf{S}_2 = \rho$ which is at $x_\seco = (p_\seco\apr\ast/\rho  )^{1/\alpha}$. Similarly, other boundaries can be derived.  We can observe that due to directional gains, these regions are radially asymmetrical and depends on both, primary and secondary link orientations. The presence of fading and secondary link orientations adds further randomization to these regions, making the protection zone more asymmetrical and dynamic.  

%-------------------------------------------------------------------------------------- 
\section{Analysis}
%--------------------------------------------------------------------------------------

In this section, we analyze the considered cognitive mmWave network from the perspective of transmission opportunities for secondary devices and its impact on the coverage performances of primary and secondary devices.

%--------------------------------------------------------------------------------------
\subsection{Medium access probability (MAP)} \label{subsect:MAP}
%--------------------------------------------------------------------------------------

As it was explained earlier, the channel access ability of a secondary transmitter depends on its location and orientation of the associated secondary link. To quantitatively understand this dependence, we define MAP of the $i^{\mathrm{th}}$ secondary transmitter as the probability that it can transmit \cite{nguyen12} {\em i.e.}  $p_{\mathrm{m}i} = \mathbb{P} \left(U_{i} = 1\right)$, which is given in the following lemma.

%---------------------------------- THEOREM 1 -----------------------------------

\begin{lemma} \label{lemma:MAP}
The MAP of the $i^{\mathrm{th}}$ secondary transmitter located at $\X_{\mathrm{s}i}= x_{\mathrm{s}i} \angle{ \theta_{\mathrm{s}i} }$ is given as
\begin{align}
p_{\mathrm{m}i}  = 1 - \exp \left( - { \rho x_{\mathrm{s}i}^{ \alpha } / p_\seco g_{\mathrm{pr}} \left( \theta_{\mathrm{s}i} \right) g_{\mathrm{st}} \left( \theta_{\mathrm{s}i} - \pi - \omega_{\mathrm{s}i} \right) } \right). \label{eq:MAP}
\end{align} 
\end{lemma}

%---------------------------- PROOF -----------------------------
\begin{IEEEproof} 
Using \eqref{eq:Indicator}, we can write
\begin{align}
p_{\mathrm{m}i} &= \mathbb{P} \left(p_\seco G_i x_{\mathrm{s}i}^{- \alpha} g_{\mathrm{pr}} \left( \theta_{\mathrm{s}i} \right) g_{\mathrm{s}i} \left( \theta_{\mathrm{s}i} - \pi - \omega_{\mathrm{s}i} \right) < \rho \right) %\nonumber \\
%&
= \mathbb{P} \left( G_i < { \rho x_{\mathrm{s}i}^{ \alpha } }/{p_\seco g_{\mathrm{pr}} \left( \theta_{\mathrm{s}i} \right) g_{\mathrm{st}} \left( \theta_{\mathrm{s}i} - \pi - \omega_{\mathrm{s}i} \right) } \right). \nonumber 
\end{align} 
Using the distribution of $G_i$, we get the desired result.
\end{IEEEproof}

From \eqref{eq:MAP}, we can observe that the MAP depends upon the ratio $\rho/p_{\seco}$. Varying $\rho$ and $p_{\seco}$ while \textit{maintaining their ratio will not affect} the MAP of a secondary transmitter. In other words, the doubling of $\rho$ or halving of $p_\seco$ cause similar effects on the MAP. 

\begin{remark}
Note that MAP of the $i^{\mathrm{th}}$ secondary transmitter under omni-directional CCS is $p_{{\mathrm{m}i}, \mathrm{omni}} =  1 - \exp \left(- \rho x_{\mathrm{s}i}^{ \alpha } / p_\seco \right)$.
\end{remark} 

Clearly $p_{{\mathrm{m}i}, \mathrm{omni}}$, as function of $x_{\mathrm{s}i}$ is a distance-dependent but orientation-independent variable. On the other hand, the same can not be true for the MAP with directional CCS. From \eqref{eq:MAP}, we can observe that the MAP of the $i^{\mathrm{th}}$ secondary transmitter depends upon its location $\X_{\mathrm{s}i} = x_{\mathrm{s}i} \angle{ \theta_{\mathrm{s}i} }$ (which includes distance and angle both) with respect to the primary receiver. Thus, depending on the location, the MAP of some secondary transmitters are higher in comparison to other secondary transmitters, even when they are at the same distance. Similarly, due to the presence of secondary-link orientation $\omega_{\mathrm{s}i}$ in $ g_{\mathrm{st}}$, two secondary transmitters at the same location observe different MAP if one is pointing towards while other is pointing outwards from the primary receiver. 

This asymmetry depends on the directionality also. As the number of antennas increases, gain in the direction of the main and side lobes increases and decreases respectively. Therefore, all secondary transmitters falling inside the main lobe region observe a significant decrease in their MAPs and an increase in MAPs in the side lobe regions. Further, as the beamwidth narrows with the number of antennas, the main lobe region reduces. There is a trade-off between the beamwidth and gain of the directional antenna, making it difficult to understand how directional CCS impacts MAP. The MAP given in Lemma \ref{lemma:MAP} can be simplified for the ideal beam pattern as given in the following corollary.

%-------------------- COROLLARY 1 ----------------
\begin{corr}\label{corollary:MAPidealBeam}
Under ideal beam pattern approximation, MAP of the $i^{\mathrm{th}}$ secondary transmitter under ideal beam pattern assumption is  
\begin{align}
p_{\mathrm{m}i, \mathrm{ideal}} = \begin{cases}
		1 - \exp \left( - \frac{ \rho x_{\mathrm{s}i}^{ \alpha } } {p_{\seco} a_{\mathrm{pr}} \ a_{\mathrm{st}} } \right) \ &\mathrm{if }\ \    
	%\begin{aligned} 
	%&
	\lvert \theta_{\mathrm{s}i} \rvert \leq \phi_{\mathrm{pr}}, \, %\\ 
	%&
	\lvert \theta_{\mathrm{s}i} - \pi - \omega_{\mathrm{s}i} \rvert \leq \phi_{\mathrm{st}}. 
	%\end{aligned} 
	\\
		1 \ &\mathrm{otherwise}.
		\end{cases} \label{eq:MAPspecialCase}
\end{align}
\end{corr}

%---------------------------- PROOF -----------------------------
\begin{IEEEproof} 
Let $ D_{\mathrm{s}i} = \gpr{\theta_{\mathrm{s}i}} \gst{\theta_{\mathrm{s}i} - \pi - \omega_{\mathrm{s}i}}$. Using \eqref{eq:gainApproximation} with $ b_{\mathrm{pr}} = b_{\mathrm{st}} = 0 $, we get  
\begin{align*}
D_{\mathrm{s}i} = \begin{cases}
a_{\mathrm{pr}} \cdot a_{\mathrm{st}} 
&\quad \mathrm{when} \quad 
%\begin{aligned} 
%&
\lvert \theta_{\mathrm{s}i} \rvert \leq \phi_{\mathrm{pr}}, 
\,
%\\ 
%&
\lvert \theta_{\mathrm{s}i} - \pi - \omega_{\mathrm{s}i} \rvert \leq \phi_{\mathrm{st}} 
%\end{aligned} \\
\\
0 \ &\quad \mathrm{otherwise}
\end{cases} 
\end{align*}

Substituting $D_{\mathrm{s}i}$ in \eqref{eq:MAP}, we will get the desired result.
\end{IEEEproof} 

Here, a protection zone exists in the intersection region of the main lobes of the primary receiver's and secondary transmitter's antenna, but there is no protection zone required in the side lobes' direction of either side's antenna (due to the zero side lobe gain). The use of antennas with narrower beamwidths restricts the angular spread of the primary receiver's protection zone, resulting more secondary transmitters to have unit MAP. However, at the same time, the MAP $p_{\mathrm{m}i, \mathrm{ideal}}$ decreases inside the protection zone. To quantitatively understand this effect, let us see the following example.
%------------------------------
\begin{example}
Let us consider the example shown in Fig. 
%\ref{fig:SectionE--pi},
3(b), 
%%%%%%%
with ULA-type antennas to focus on the impact of the primary receiver's antenna beamwidth. Let us first consider a constant fading $G_i = h$. For the protection zone $\mathsf{S}_1$, the maximum radius increases and angular spread reduces with $M_\mathrm{pr}$. Its area is given as $c \apr^{2/\alpha} /M_\mathrm{pr}$ where $c$ is constant equal to $(\kappa/2) (p_\seco h \ast /\rho)^{2/\alpha}$. Assuming $\apr = M_\mathrm{pr}$, the area is $c M_\mathrm{pr}^{2/\alpha-1}$. 
Thus, the area of the primary receiver's protection zone decreases with $M_\mathrm{pr}$ (assuming $\alpha > 2$). Since it is true for each value of fading, we can say directional CCS increases the secondary activity.
\end{example}
%------------------------------
To quantify this effect for the general scenario, we define a new metric {\em activity factor} next.

%--------------------------------------------------------------------------------------
\subsection{Activity factor (AF) of secondary network} 
%--------------------------------------------------------------------------------------

For a given region, the AF of a network represents the fraction of transmitters that are allowed to transmit at any instant of the time. In the context of cognitive networks, it indicates the availability of secondary transmission opportunities within a considered region. Let us consider a region of interest comprising of a ball of radius $R$ {\em i.e.} $\Ball(\mathbf{o}, R)$. The mean number of secondary transmitters inside this ball is $\pi \lambda_{\mathrm{s}} R^2$. Since $U_{i}$ represents the transmission activity of the $i^{\mathrm{th}}$ secondary transmitter, the AF can be given as
\begin{align}
 	\eta_{\mathrm{s}} &= \frac{ \mathbb{E} \left[ \sum_{\X_{\mathrm{s}i} \in \Phi_\seco \cap \Ball(0, R)} U_i \right] }{\pi \lambda_{\mathrm{s}} R^2}. \label{eq:AF}
\end{align}

Here, $\eta_{\mathrm{s}}$ denotes the MAP of secondary transmitters averaged over their locations. Interestingly, despite the presence of $\lambda_{\mathrm{s}}$ in \eqref{eq:AF}, the final expression of AF is independent of $\lambda_{\mathrm{s}}$ as given in the following theorem.

%---------------------------------- THEOREM 2 -----------------------------------
\begin{theorem}\label{theorem:AF}
The AF of the secondary network is given as
\begin{align}
 \eta_{\mathrm{s}} &= 1 \! - \! \frac1{\pi R^2}\frac1\alpha\int_0^{2\pi} \! \expects{\omega}{\!
	{\left(\!{ \frac{\rho}{p_\seco
			\gpr{\theta}
			\gst{\theta \! - \! \pi  \! - \! \omega }}}\!\right)\!\!}^{-\frac{2}{\alpha}} \!
	\Gamma \!
	\left(\!\frac{2}{\alpha},\frac{\rho R^\alpha}{p_\seco
			\gpr{\theta}
			\gst{\theta \! - \! \pi \! - \! \omega }}\!\right)
\!}
\dd\theta. \label{eq:AFsolution}
\end{align}
\end{theorem}

%--------------------------- PROOF --------------------------
\begin{IEEEproof}
See Appendix \ref{thrm:proof:AF}.
\end{IEEEproof}

From \eqref{eq:AFsolution}, we can observe that the AF of the secondary devices in the network can be increased by carefully choosing the ratio $\rho/p_{\seco}$ and antenna beam patterns of the primary and secondary devices. We can simplify Theorem \ref{theorem:AF} for the sectorized beam pattern to get the following corollary.

%---------------------------------- COROLLARY 2 -----------------------------------
\begin{corr}\label{corollary:AFsectorBeam} 
Under sectorized beam pattern approximation, AF of the secondary network is given as
\vspace*{-0.5cm}
\begin{align}
\eta_{\mathrm{s},\mathrm{sec}} &= 1 - \frac1{ R^2} \frac2\alpha \left[ \qpr\qst\psi {\left(\frac{ \rho/p_\seco}{\apr\ast}\right)} + \qpr(1-\qst)\psi {\left(\frac{ \rho/p_\seco}{\apr \bst}\right)} 
 		\right.\nonumber\\
 		&\qquad  +\left.
		(1-\qpr) \qst \psi {\left(\frac{ \rho/p_\seco}{\bpr \ast}\right)} + (1-\qpr)(1-\qst)\psi {\left(\frac{ \rho/p_\seco}{\bpr \bst}\right)}\right], %\nonumber
		\label{eq:AFsector}
\end{align}
where $\psi(u)= {u}^{-\frac2\alpha} \Gamma \left( 2/\alpha, {uR^\alpha} \right)$.
\end{corr}

%--------------------------- PROOF --------------------------
\begin{IEEEproof}
See Appendix \ref{corl:proof:AFsectorBeam}.
\end{IEEEproof}

Note that the probability $q_{jk}$ is directly proportional to the main-lobe beamwidth $\phi_{jk}$ of the antenna and thus the trade-off between $\phi_{jk}$ and $a_{jk}$ also exists for improving the AF of the secondary network. To better understand the inherent characteristics of this trade-off, we further simplify \eqref{eq:AFsector} with the help of ideal beam pattern approximation. Substituting $ b_{\mathrm{pr}} = b_{\mathrm{st}} = 0 $ in Corollary \ref{corollary:AFsectorBeam}, we get following result.
%---------------------------------- COROLLARY 3 -----------------------------------
\begin{corr}
Under ideal beam pattern, AF of the secondary network is given as
\begin{align}
\eta_{\mathrm{s},\mathrm{ideal}} &= 1-\frac1{ R^2}\frac2\alpha
 	\qpr\qst\psi {\left(\frac{ \rho/p_\seco}{\apr\ast} \right)}. \label{eq:AFideal}
\end{align}
\end{corr}

\begin{remark}
Note that AF of the secondary network under omnidirectional CCS is $\eta_\mathrm{omni} =  1 - (2/\alpha R^2) \psi \left(\rho/p_\seco\right)$. 
\end{remark} 

%------------------------------
\label{subsect:AFBeamwidthGainTrade-offAnalysis}
%------------------------------
Now, let us define the following auxiliary and intermediate terms for the cases with ideal and omni beam patterns.
\begin{align*}
\Gamma_{\mathrm{s},\mathrm{ideal}} = \Gamma \left(2/\alpha, (\rho/p_\seco)/(\apr\ast)\right), \  \ \psi_{\mathrm{s},\mathrm{ideal}} &= \psi((\rho/p_\seco)/(\apr\ast)), \  \ \bar{\eta}_{\mathrm{s},\mathrm{ideal}} = 1 - \eta_{\mathrm{s},\mathrm{ideal}}, \\ 
\Gamma_{\mathrm{omni}} = \Gamma \left(2/\alpha, \rho/p_\seco \right), \  \ \psi_{\mathrm{omni}} &= \psi(\rho/p_\seco), \  \ \bar{\eta}_{\mathrm{omni}} = 1 - \eta_{\mathrm{omni}}. 
\end{align*}

Here, function $\psi(u)$ is led by the term $u^{-2/\alpha}$ which contributes $(\apr\ast)^{2/\alpha}$ term to $\psi_{\mathrm{s},\mathrm{ideal}}$ for directional CCS. Note that $\apr\ast$ is reciprocal to $\qpr\qst$ which always decreases with antenna directionality. Now, the resultant term $\bar{\eta}_\seco$  is equal to the product of $\psi$ and $\qpr\qst$. Owing to this term being proportional to $(\qpr\qst)^{1-2/\alpha}$, $\bar{\eta}_{\seco, \mathrm{ideal}}/\bar{\eta}_\mathrm{omni}$ decreases with directionality. Thus an increment in AF is observed under directional CCS due to dominating effect of antenna-beamwidth ($\qpr\qst$) on $\bar{\eta}_{\mathrm{s},\mathrm{ideal}}  = \qpr\qst \psi_{\mathrm{s},\mathrm{ideal}}$ in comparison to the antenna-gain ($\apr\ast$) for the considered mmWave cognitive system. In the section \ref{section:EffectsGainBeamwidthAF} also, we show that for $\alpha > 2$, ratios $\bar{\eta}_{\mathrm{s}, \mathrm{ideal}}/\bar{\eta}_{\mathrm{omni}}$ and $\Gamma_{\mathrm{s},\mathrm{ideal}}/\Gamma_{\mathrm{omni}}$ are less than $1$ while $\psi_{\mathrm{s},\mathrm{ideal}}/\psi_{\mathrm{omni}} > 1$. 

%-------------------------------------------------------------
\subsection{Performance of the primary link} \label{section:PrimaryRemarks}
%-------------------------------------------------------------

The instantaneous SINR at the primary receiver is given as
\begin{align}
	\SINR_\prim &= \frac{ H_{\mathrm{p}} p_\prim \gpt{0}\gpr{0} r_\prim^{- \alpha} }{ \sigma^2 \ + \ I_{ \mathrm{s} }  \left( \Phi_{\mathrm{s}}  \right) }. \label{eq:PrimarySINR}
\end{align}

Here, $\noise$ is the thermal noise, and $I_{ \mathrm{s} }  \left( \Phi_{\mathrm{s}} \right)$ is the total interference due to all active secondary transmitters at the primary receiver, given by
\begin{align}
I_{ \mathrm{s} }  \left( \Phi_{\mathrm{s}} \right) &= \sum_{  \X_{\mathrm{s}i} \in \Phi_\seco} p_\seco U_{i} G_{i}  g_{\mathrm{pr}} \left( \theta_{\mathrm{s}i} \right) g_{\mathrm{st}} \left( \theta_{\mathrm{s}i} - \pi - \omega_{\mathrm{s}i} \right) \lvert \lvert \X_{\mathrm{s}i} \rvert \rvert^{- \alpha}.
\label{eq:PrimaryInterference}
\end{align}

Note that transmission-protection threshold $\rho$ affects the number of active secondary transmitters which in turn affect $I_{ \mathrm{s} }  \left( \Phi_{\mathrm{s}} \right)$. To evaluate the performance of the primary link, we now compute its coverage probability which is defined as the complementary cumulative density function (CCDF) \textit{i.e.} $p_\mathrm{cp}(\SThres, \, \rho) = \mathbb{P} \left[ \SINR_\prim > \tau \lvert \, \rho \right] $. It is given in the following theorem.

%-------------------------- THEOREM 3 -----------------------------------

\begin{theorem} \label{theorem:PrimaryCoverage}
The SINR coverage  of the primary link in the presence of secondary links with transmit-restriction threshold $\rho$ is given as
\begin{align}
p_\mathrm{cp}(\SThres,\, \rho) &= \expU{-\frac{\SThres \noise r_\prim^{\alpha}}{p_\prim \gpr{0}\gpt{0} }} \expS{-\lambda_\seco\int_0^{2\pi}\int_0^\infty \left(1- \expects{\omega_\seco}{e^{-\frac{\rho x_\seco^{\alpha}}{p_\seco  \gpr{\theta_\seco}\gst{\theta_\seco-\omega_\seco-\pi}} }
\right.\right.\right.+ \nonumber \\
&\qquad \qquad \left.\left.\left.
\frac{1 - e^{-\frac{\rho  r_\prim^{\alpha}}{p_\seco \gpr{0}\gpt{0}} \left(\SThres \frac{p_\seco }{p_\prim} + {\left(\frac{x_\seco }{r_\prim}\right)^{\alpha}} \frac{\gpr{0}\gpt{0}}{\gpr{\theta_\seco}\gst{\theta_\seco-\omega_\seco - \pi} }\right)}}{\SThres \frac{p_\seco}{p_\prim} \frac{1}{\left(x_\seco/r_\prim\right)^{\alpha}} \frac{\gpr{\theta_\seco} \gst{\theta_\seco - \omega_\seco - \pi} }{\gpr{0}\gpt{0}} + 1} \!\!}\! \right) \!\! x_\seco\dd x_\seco \dd \theta_\seco }.  
\label{eq:PrimaryCov}
\end{align}

It can be further simplified to
%--------------- Simple expression -------------
\begin{align}
p_\mathrm{cp}(\SThres,\, \rho)&= \exp \left( - \frac{\sigma^2 \kappa_\prim \tau}{\rho} - \frac{\lambda_\seco}{\alpha} \frac{p_\seco^{{2}/{\alpha}}}{\rho^{{2}/{\alpha}}} \left( \kappa_\prim^{{2}/{\alpha}} \, \tau^{{2}/{\alpha}} \, \mathsf{n}_{1} (\alpha) - \Gamma \left( {2}/{\alpha} + \mathsf{n}_{2} (\alpha, \kappa_\prim \tau) \right) \right) \mathsf{n}_{3} \right),
\label{eq:PrimaryCov:Simplified}
\end{align}
where
\vspace*{-0.5cm}
\begin{align*}
\mathsf{n}_{3} = \mathbb{E}_{\omega_\seco}  \left[ \int_0^{2\pi} \left[ \gpr{ \theta_\seco} \gst{ \theta_\seco - \pi - \omega_\seco} \right]^{{2}/{\alpha}} \mathrm{d}\theta_\seco \right],
\end{align*}
denoting the average secondary directivity, $\kappa_\prim = \rho r_\prim^{\alpha}/ p_\prim \gpt{0}\gpr{0}$ denoting the noise-to-signal-ratio (NSR) at the primary receiver, $\mathsf{n}_{1} (\alpha) = \pi \csc (2 \pi/\alpha)$ and $\mathsf{n}_{2} (\alpha, \nu) = \int_{\nu}^\infty \frac{e^{- u}}{u (u - \nu)^{- \frac{2}{\alpha}} } \dd u$. 
\end{theorem}

%--------------------------- PROOF --------------------------
\begin{IEEEproof}
	See Appendix \ref{thrm:proof:PrimaryCoverage}.
\end{IEEEproof}

From \eqref{eq:PrimaryCov}, we can observed that the dependence of $p_\mathrm{cp}$ on $\sigma^2$, $p_\prim$, $p_\seco$ and $\tau$ is via the ratios $\tau \sigma^2/p_\prim$, $\rho/p_\seco$ and $\tau p_\seco/p_\prim$ only. Therefore, we can get the same value of the primary coverage $p_\mathrm{cp}$ for the same values of these ratios. For example, in the considered cognitive system if we increase the values of $\rho$, $\sigma^2$ and $p_\seco$ by $10$ times and $p_\prim$ by $100$ times, the value of $\tau$ needs to be increased by $10$ times. In other words, a positive shift of $10$ dB is observed in the $p_\mathrm{cp}(\SThres,\, \rho)$ {\em vs.} $\tau$ curve. From \eqref{eq:PrimaryCov:Simplified}, we can also observe that secondary directionality affects $p_\mathrm{cp}(\SThres,\, \rho)$ {\em via} $\mathsf{n}_{3}$ term only. To investigate the impact of directionality, we can simplify Theorem \ref{theorem:PrimaryCoverage} for the sectorized beam to get the following result.

%---------------------------------- COROLLARY 4 -----------------------------------
\begin{corr} \label{corollary:PrimaryCoverageSectorBeam}
Under sectorized beam approximation, coverage probability $p_\mathrm{cp, sec}(\SThres, \rho)$ of the primary link is given by \eqref{eq:PrimaryCov:Simplified} with 
%--------------- Simple expression -------------
\begin{align}
\mathsf{n}_{3} &= 2 \pi \left( \qpr \apr^{{2}/{\alpha}}  + (1 - \qpr) \bpr^{{2}/{\alpha}} \right) \left( \qst \ast^{{2}/{\alpha}} + (1 - \qst) \bst^{{2}/{\alpha}} \right),
\label{eq:PrimaryCov:Simplified:Sector}
\end{align} 
and $\kappa_\prim = \rho r_\prim^{\alpha}/(p_\prim \apt \apr)$. 
\end{corr}

%--------------------------- PROOF --------------------------
\begin{IEEEproof}
See Appendix \ref{corl:proof:PrimaryCoverageSectorBeam}.
\end{IEEEproof}

To better understand the dependence of $p_{\mathrm{cp}} (\tau,\, \rho)$ on the probability $q_{jk}$, we further simplify \eqref{eq:PrimaryCov:Simplified:Sector} for the ideal beam pattern. Substituting $ b_{\mathrm{pr}} = b_{\mathrm{st}} = 0 $ in Corollary \ref{corollary:PrimaryCoverageSectorBeam}, we get following result.

%---------------------------------- COROLLARY 5 -----------------------------------
\begin{corr}
Under the ideal beam pattern, the coverage probability $p_\mathrm{cp, ideal}(\SThres, \rho)$ of the primary link is given by \eqref{eq:PrimaryCov:Simplified} with  $\mathsf{n}_{3} = 2 \pi \qpr \qst (\apr\ast)^{{2}/{\alpha}}$ and $\kappa_\prim = \rho r_\prim^{\alpha}/(p_\prim \apt \apr)$. 
\end{corr}

%---------------------------------------

\begin{remark}
Note that coverage probability $p_\mathrm{cp, omni}(\SThres, \rho)$ of the primary link under omni-directional CCS is given by \eqref{eq:PrimaryCov:Simplified} with $\mathsf{n}_{3} = 2 \pi $ and $\kappa_\prim$ replaced by $\kappa_\mathrm{o} = {\rho r_\prim^{\alpha}}/{p_\prim}$. 
\end{remark}

Comparing $p_\mathrm{cp, ideal}(\SThres, \rho)$ and $p_\mathrm{cp, omni}(\SThres, \rho)$, we observe that primary coverage under directional CCS is affected by parameters $\qpr$, $\qst$, $\apr$, $\apt$ and $\ast$. Here, with directionality, $\qpr$ and $\qst$ decrease, while $\apt$, $\ast$, and $\apr$ increase. Let us understand the impact of each of these variables. 
\begin{enumerate}
\item Due to a decrease in $\qpr$ and $\qst$, the number of interfering secondary transmitters reduces which should improve the primary coverage (see the pre-factor). 
\item Due to an increase in $\apt$ and $\apr$, the serving signal power for the primary link improves, improving the coverage (see the term $\kappa_\prim$ which decreases with directionality) {\em s.t.} $\kappa_\prim < \kappa_\mathrm{o}$.
\item An increase in $\ast$ and $\apr$ increases the strength of the cross-link between the secondary transmitter and the primary receiver which has two types of effects. First, the interference increases due to an increase in the signal power from the secondary transmitters with main lobes aligned towards the primary receiver, reducing the primary coverage. Second, the activity of such transmitters also decreases improving coverage. It can observed from the expression of $p_\mathrm{cp, ideal}(\SThres, \rho)$ that increase in $\ast$ decreases primary coverage as combined effect of terms $\mathsf{n}_\mathrm{2}$ and $\mathsf{n}_\mathrm{3}$. The impact of $\apr$ is similar and it decreases the primary coverage. However, $\apr$ also has the reverse impact (see point 2), leading to a more complex effect. 
\end{enumerate}

Due to the competing effects of the above variables, the exact behaviour of $p_{\mathrm{cp}} (\tau, \rho)$ with respect to directionality is non-trivial and depends on system parameters including antenna patterns. 

%----------------------------------------------------------------------------
\begin{lemma} \label{example:PrimaryCov:Simplified:Example1a}
Let us assume ULA as antennas. In this case, $\mathsf{n}_\mathrm{3}$ in \eqref{eq:PrimaryCov:Simplified:Sector} is simplified as
\begin{align}
\mathsf{n}_\mathrm{3} &= 2 \pi  (\kappa')^2 (M_\prim M_\seco)^{\frac{2}{\alpha} - 1} \left[ 1  + \frac{1 - \kappa'}{\kappa'}  \left(\frac{1 - \kappa'}{M_\prim - \kappa' }\right)^{\frac{2}{\alpha} - 1} \right] \left[ 1 + \frac{1 - \kappa'}{\kappa'}  \left(\frac{1 - \kappa' }{M_\seco - \kappa' }\right)^{\frac{2}{\alpha} - 1} \right].
\label{eq:PrimaryCov:Simplified:Example1a}
\end{align}
\end{lemma}

Hence, $\mathsf{n}_\mathrm{3}$ decreases with $M$ (both $M_\prim$ and $M_\seco$). However, this decrease in $\mathsf{n}_\mathrm{3}$ with respect to $M$ is very slow and it becomes constant for large $M$. Thus, the impact of $\mathsf{n}_\mathrm{3}$ and hence secondary directionality ($\ast$) is not significant. We will observe this behaviour in numerical results also. On the other hand, it can be shown that $(\kappa_\prim \tau)^{2/\alpha} \mathsf{n}_\mathrm{1} + \mathsf{n}_\mathrm{2} (\alpha, \kappa_\prim \tau)$ decreases (rapidly) with decrease in $\kappa_\prim \tau$ \cite[pp. 18-20]{TripGupTheoremFile2023} where $\kappa_\prim = \rho r_\prim^{\alpha}/(p_\prim \apt \apr)$. Hence, primary directionality ($\apt$ and $\apr$) plays a more significant role in improving the primary coverage probability. After characterizing the primary link's performance, we now turn our focus to the performance of the typical secondary link.

%----------------------------------------------------------------------------------------------------- 
\subsection{Performance of the secondary link}
%-----------------------------------------------------------------------------------------------------

\begin{figure}[ht!]
\vspace*{-0.5cm}
\centering
{\includegraphics[scale=0.15]{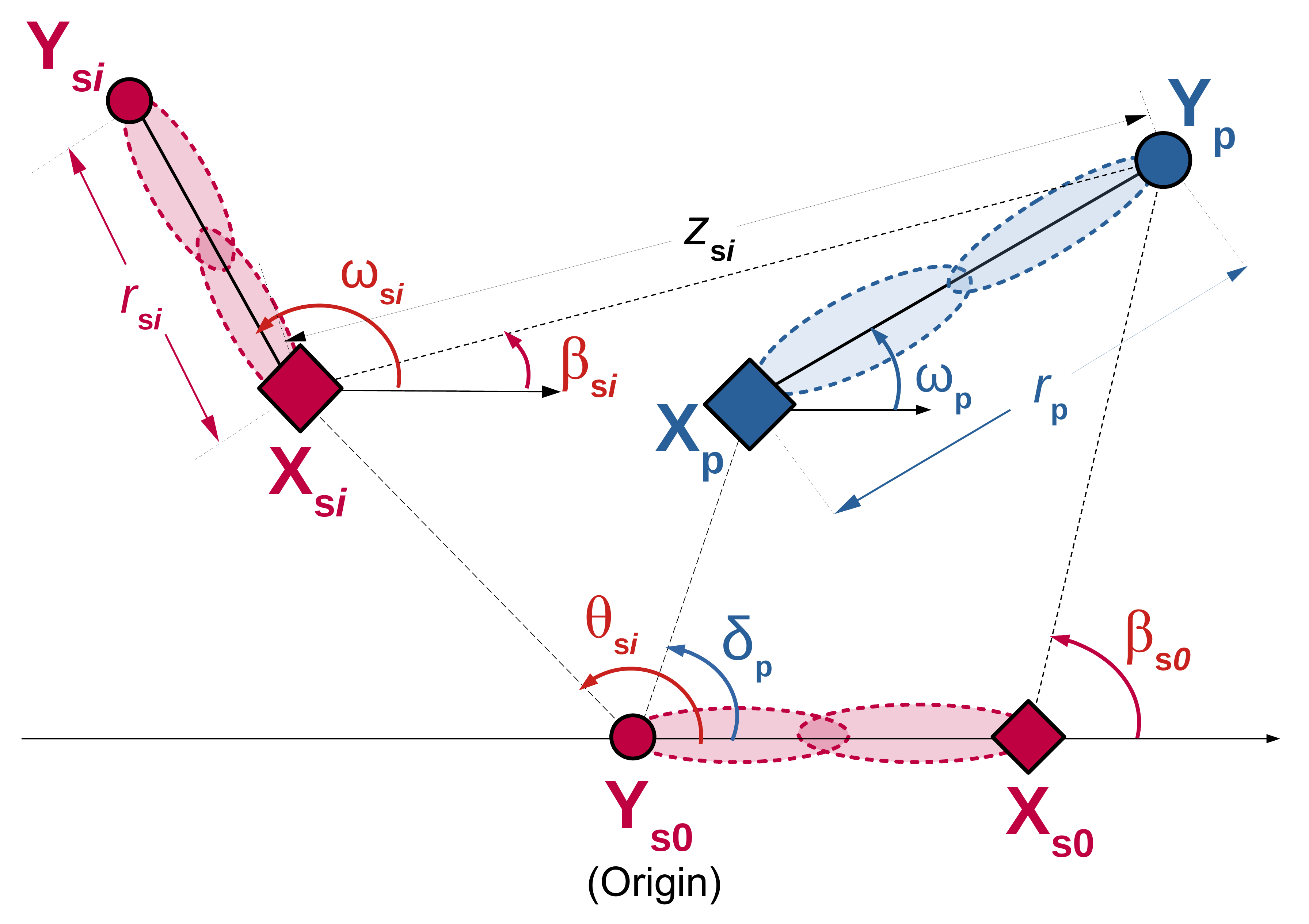}}%\label{fig:SecondarySystemModel}}
%{\includegraphics[scale=0.15]{figure/SimpleCase2a-Rotated.pdf}\label{fig:SecondarySystemModel-Rotated}}\\
%\vspace*{-0.3cm}
%{\small (a) \hspace{5cm} (b)}
\vspace*{-0.3cm}
\caption{An illustration of the new co-ordinate reference demonstrating (a) the locations of the primary link and $i^{\mathrm{th}}$ secondary link from the perspective of the location of the typical secondary link.}%, and (b) the computation of angle $ \beta_{\mathrm{s}i}$ with $\delta_\prim = \omega_\prim = \pi$.}
\label{fig:SecondaryModel}	
\vspace*{-0.5cm}
\end{figure}

To evaluate the performance of the secondary links, we consider a typical secondary transmitter and receiver pair. Let $ \{\X_{ \mathrm{s}0},\, \Y_{ \mathrm{s}0} \}$ represents the locations of the typical secondary transmitter-receiver pair. For ease of computation, we transform the axis to make the typical secondary receiver at the origin {\em i.e.} $\Y_{ \mathrm{s}0} = \mathbf{o}$ and the typical secondary link aligned with the x-axis (see the 
%--------------------------------------------------------
Fig. \ref{fig:SecondaryModel}).
%--------------------------------------------------------
Note that this does not affect the performance. In the new coordinate reference, $ \X_{ \mathrm{p}} = x_\prim \angle \delta_\prim $ represents the location of the primary transmitter {\em i.e.} primary link is oriented at the $\delta_\prim$ angle. The angle $ \beta_{\mathrm{s}i} $ represents the orientation of the primary receiver with respect to the $ i^{\mathrm{th}} $ secondary transmitter and is given as
\vspace*{-0.2cm}
\begin{align*}
\beta_{\mathrm{s}i} &= \angle (\X_{\mathrm{s}i} - \Y_{\mathrm{p}}) = \angle (x_{\seco i}\angle \theta_{\seco i} - \mathbf{Y}_\prim) \\
&= \theta_{\mathrm{s}i} - \sin^{-1} \left(({y_\prim}/{x_{\seco i}}) \sin \left(\delta_\prim - \omega_\prim + \sin^{-1} \left( ({x_\prim}/{y_\prim}) \sin \left(\delta_\prim - \omega_\prim \right)\right)\right)\right) \\
&= \theta_{\mathrm{s}i} - \sin^{-1} \left(({y_\prim}/{z_{\seco i}}) \sin \left(\theta_{\mathrm{s}i} - \delta_\prim + \sin^{-1} \left( ({r_\prim}/{y_\prim}) \sin \left(\delta_\prim - \omega_\prim\right)\right)\right)\right),
%\tan^{-1} \left[{(\sin \theta_{\seco i})}/{ (({y_\prim}/{x_{\seco i}}) + \cos \theta_{\seco i}})\right].
\end{align*}
\vspace*{-0.2cm}
where
%Thus, the angle of the primary receiver with respect to the typical secondary transmitter is denoted by $\beta_{\mathrm{s}0}$. Further, 
$z_{\seco i}$ denotes the distance between $ i^{\mathrm{th}} $ secondary transmitter and the primary receiver and is given as
\vspace*{-0.2cm}
\begin{align*}
z_{\seco i} &= \lvert \lvert \X_{\mathrm{s}i} - \Y_{\mathrm{p}} \rvert \rvert = \lvert \lvert x_{\seco i} \angle \theta_{\seco i} - \Y_{\mathrm{p}} \rvert \rvert \\
& = \big(x_{\seco i}^2 + r_\prim^2 + x_\prim^2 - 2 r_\prim x_{\seco i} \cos (\theta_{\seco i} - \omega_\prim) - 2 x_\prim x_{\seco i} \cos (\theta_{\seco i} - \delta_\prim) + 2 x_\prim r_\prim \cos (\delta_\prim - \omega_\prim) \big)^{1/2}.
\end{align*}

The received power at the primary receiver due to $ i^{\mathrm{th}} $ secondary transmitter is
\begin{align*}
P^{'}_{\mathrm{s}i \to \prim} = p_\seco G_{i} g_{\mathrm{st}} \left( \beta_{\mathrm{s}i} - \omega_{\mathrm{s}i} \right) g_{\mathrm{pr}}  \left(\omega_\prim - \beta_{\mathrm{s}i} \right) z_{\seco i}^{- \alpha},
\end{align*} 

The transmission indicator $U^{'}_{i}$ represents the transmission activity of the $ i^{\mathrm{th}} $ secondary transmitter, given as
\vspace*{-0.5cm}
\begin{align*}
U^{'}_{i} &= \mathbbm{1} \left ( p_\seco G_{i} g_{\mathrm{st}} \left( \beta_{\mathrm{s}i} - \omega_{\mathrm{s}i} \right) g_{\mathrm{pr}}  \left(\omega_\prim - \beta_{\mathrm{s}i} \right) z_{\seco i}^{- \alpha} < \rho \right ). 
\end{align*}

Due to the change in the coordinate system, the expression of MAP in \eqref{eq:MAP} is slightly modified for typical and $i^{\mathrm{th}}$ secondary transmitters (see the following lemmas). However, their proof is similar to the Lemma \ref{lemma:MAP}.

%-------------------- Lemma 3 -------------------
\begin{lemma} \label{lemma:MAPTypicalSecondary}
The MAP of the typical secondary link in the presence of primary link $ \X_\prim -  \Y_\prim$ is  $p^{'}_{\mathrm{m0}} = 1 - \exp \left( - { \rho z_{\mathrm{s0}}^{\alpha}}/{ p_\seco g_{\mathrm{st}} \left( \beta_{\mathrm{s0}} - \pi \right) g_{\mathrm{pr}}  \left(\omega_\prim - \beta_{\mathrm{s0}} \right) } \right)$, where $ z_{\mathrm{s0}} = \| r_\seco \angle 0 - \Y_\prim \| = \big(r_\prim^2 + r_\seco^2 + x_\prim^2 - 2 r_\prim r_\seco \cos \omega_\prim - 2 x_\prim r_\seco \cos \delta_\prim + 2 x_\prim r_\prim \cos (\delta_\prim - \omega_\prim)\big)^{1/2}$.
\end{lemma}

%-------------------- Lemma 4 ----------------
\begin{lemma} \label{lemma:MAPforShiftedPPP}
The MAP of the $i^{\mathrm{th}}$ secondary transmitter at location $\X_{\mathrm{s}i} = x_{\mathrm{s}i} \angle \theta_{\seco i}$ in the presence of primary link $ \X_\prim -  \Y_\prim$ is $p^{'}_{\mathrm{m}i} = 1 - \exp \left( - { \rho z_{\seco i}^{\alpha}}/{ p_\seco g_{\mathrm{st}} \left( \beta_{\mathrm{s}i} - \omega_{\mathrm{s}i} \right) g_{\mathrm{pr}}  \left(\omega_\prim - \beta_{\mathrm{s}i} \right) } \right). $
\end{lemma}

\begin{remark}
Note that the indicator function under omnidirectional CCS is $U^{'}_{i, \mathrm{omni}} = \mathbbm{1} \left ( p_\seco G_{i} z_{\seco i}^{- \alpha} < \rho \right )$ and corresponding MAP is $p^{'}_{\mathrm{m}i, \mathrm{omni}} = 1 - \exp \left(- \rho z_{\seco i}^{ \alpha } / p_\seco \right)$.
\end{remark} 

From lemmas \ref{lemma:MAPTypicalSecondary} and \ref{lemma:MAPforShiftedPPP}, we can observe that the MAP of a secondary link depends not only on its distance from the primary transmitter but also on the orientation of the primary link. The instantaneous SINR at the secondary receiver at the origin $\mathrm{o}$ is given as
\begin{align}
\SINR_{\seco0} &= \frac{ F_{\seco0} U^{'}_0 p_\seco \gst{0}\gsr{0} r_{\seco}^{- \alpha} }{ \sigma^2 + I_\prim +  I_\seco \left(\Phi_\seco \right)}. \label{eq:secondarySINR}
\end{align}

Here, $I_\prim$ is the interference at the secondary receiver due to the primary link while $I_\seco$ is the total interference at the secondary receiver due to the rest of the active secondary transmitters. $I_\prim$ and $I_\seco$ are respectively given as
\begin{align*}
I_\prim &= G^{'}_0 p_\prim \gsr{\delta_\prim}  \gpt{\delta_\prim - \pi - \omega_\prim} \| \X_\prim \|^{- \alpha}, \\
I_\seco  &= \!\!\!\!  \sum_{ \X_{\mathrm{s}i} \in \Phi_\seco / \{\X_{\mathrm{s}0}\} } \!\! \!\! \!\!  U^{'}_{i} G^{''}_{i}  p_\seco \gsr{ \theta_{\seco i} } \gst{ \theta_{\seco i} - \pi - \omega_{\seco i} } \| \X_{\mathrm{s}i} \|^{- \alpha}.
\end{align*}
where $G^{''}_{i}$ represent the fading coefficient for the $i^{\mathrm{th}}$-secondary-transmitter-to-typical-secondary-receiver. We now provide the coverage probability of the typical secondary link in the following theorem. 

%----------------------------------- THEOREM 5 ---------------------------------------------
\begin{theorem}\label{theorem:SecondaryCoverage}
The coverage probability of the typical secondary link in the presence of other secondary links while ensuring interference at the primary link to be below $\rho$ is given as
\begin{align}
p_\mathrm{cs} \left(\tau,\, \rho\right) 
&= \exp \left(-\frac{\tau \sigma^2}{A_\mathrm{0}}\right) \frac{ 1 - e^{ - {1}/{A_\mathrm{1} } }}{ 1 + \tau \frac{A_\mathrm{2}}{A_\mathrm{0}}} \exp \left( - \lambda_\seco 
\!\!
\int_0^{2\pi} 
\!\!
\int_0^\infty \mathbb{E}_{\omega_\seco} \left[ \frac{1 - e^{- A_\mathrm{s} z_\seco^{\alpha}} }{1 + C_\mathrm{s} x_\seco^\alpha} \right] x_\seco \dd x_\seco\dd\theta_\seco \right),
\label{eq:SecondaryCov:Simplified}
\end{align}
where $A_\mathrm{0} = p_\seco \gsr{0}\gst{0} r_\seco^{-\alpha}$ denotes the secondary link serving power, $A_\mathrm{1} = p_\seco g_\mathrm{st} (\beta_{\seco0} - \pi) g_\mathrm{pr} (\omega_\prim-\beta_{\seco0}) z_{\seco 0}^{-\alpha}/\rho $ denotes the typical secondary-to-primary co-link interference relative to the primary-transmit-protection threshold, $A_\mathrm{2} = p_\prim g_\mathrm{pt} (\delta_{\prim}-\pi-\omega_{\prim}) g_\mathrm{sr} (\delta_{\prim}) x_{\prim}^{-\alpha}$ denotes the primary-to-typical secondary co-link interference, $A_\mathrm{s} (\omega_\seco, \beta_\seco (x_\seco,\theta_\seco)) = {\rho}/(p_\seco g_\mathrm{st} (\beta_\seco(x_\seco,\theta_\seco) - \omega_\seco) g_\mathrm{pr} (\omega_\prim - \beta_\seco (x_\seco,\theta_\seco)))$ denotes the ratio between interference threshold and the directional gain from the interfering secondary transmitter to the primary receiver and $C_\mathrm{s} (\omega_\seco,\theta_\seco) = {A_\mathrm{0}}/\SThres p_\seco g_\mathrm{st} (\theta_\seco - \omega_\seco - \pi) g_\mathrm{sr} (\theta_\seco)$. Here, $x_{\prim}$ and $\delta_{\prim}$ is the distance and the angle of the primary transmitter with respect to the typical secondary receiver, $z_{\mathrm{s0}} = \| r_\seco \angle 0 - \Y_\prim \|$, $\beta_{\seco0} = \angle (r_{\seco}\angle 0-\mathbf{Y}_\prim)$,  $z_{\seco} = \lvert \lvert x_\seco \angle \theta_\seco - \Y_{\mathrm{p}} \rvert \rvert$, and $\beta_\seco = \angle (x_\seco\angle \theta_\seco-\mathbf{Y}_\prim)$. 
\end{theorem}

%--------------------------- PROOF --------------------------
\begin{IEEEproof}
See Appendix \ref{thrm:proof:SecondaryCoverage}.
\end{IEEEproof}

To derive the insights from secondary coverage expression, we can divide \eqref{eq:SecondaryCov:Simplified} into four terms as listed below:
\begin{enumerate}
\item[(a)] $\mathsf{Term-1}: \exp ( -{\tau \sigma^2}/{A_\mathrm{0}} )$ arises due to system-noise $\sigma^2$ and depends on the level of SINR threshold $\tau$ in comparison to SNR of the typical secondary link ({\em i.e.} $A_\mathrm{0}/\sigma^2$). Here, $A_\mathrm{0}$ represents the signal strength of the typical secondary link.
		
\item[(b)] $\mathsf{Term-2}: (1 - e^{- 1/{A_\mathrm{1} } })$ represents the MAP of the typical secondary pair and depends on the received signal strength from the typical secondary transmitter to the primary receiver in terms of $A_\mathrm{1}$.
		
\item[(c)] $\mathsf{Term-3}: {1}/{(1 + \tau \frac{A_\mathrm{2}}{A_\mathrm{0}})}$ depends on the level of primary interference relative to the typical secondary link's serving power. Here, $A_\mathrm{2}$ represents the received signal strength from the primary transmitter to the typical secondary receiver.
		
\item[(d)] $\mathsf{Term-4}:$ The integral term $I = \int_0^{2\pi} \mathbb{E}_{\omega_\seco} \left[ \int_0^\infty  (1 - e^{- A_\mathrm{s} z_\seco^{\alpha}})/(1 + C_\mathrm{s} x_\seco^\alpha) x_\seco \dd x_\seco \right] \dd  \theta_\seco$ in which numerator and denominator represent the MAP and the interference power of the secondary transmitter located at distance $x_\seco$.
\end{enumerate}

A mismatch between signals corresponding to links of $\mathsf{Term-2}$ and $\mathsf{Term-3}$ occurs because the orientation of primary-to-typical-secondary and typical-secondary-to-primary cross-links are not symmetric. Hence, a low secondary interference at the primary receiver does not guarantee a low primary interference $I_\prim$ at the typical secondary receiver. If the two links are symmetric, both links will see the same gain {\em s.t.} $\rho A_\mathrm{1} = A_\mathrm{2}$ and low value of $A_\mathrm{1}$ will ensure high activity and low $I_\prim$, leading to high values for terms $\mathsf{2}$ and $\mathsf{3}$. A similar mismatch occurs in $\mathsf{Term-4}$ due to difference between the variables $ C_\mathrm{s}$ and $ A_\mathrm{s}$, and their respective multipliers $x_\seco$ and $z_\seco$. Further, this makes the simplification of the integral in \eqref{eq:SecondaryCov:Simplified} difficult. Following two examples approximate the $\mathsf{Term-4}$ in \eqref{eq:SecondaryCov:Simplified} for special cases with symmetry (see \cite[pp. 27-29]{TripGupTheoremFile2023} for full proof).  

%---------------------------------------------

\begin{example}	
Consider the case when the typical secondary transmitter is very close to the primary receiver compared to the mean contact distance ($= 1/(2\sqrt{\lambda_\seco})$) of secondary transmitters. Here, almost all of the secondary transmitters are far away relative to the primary link, hence, $z_\seco \approx x_\seco$. Also, all secondary transmitter gain to the primary receiver is approximately the same as their gain to the typical secondary receiver {\em i.e.} $g_\mathrm{st} (\beta_\seco(x_\seco,\theta_\seco) - \omega_\seco) \approx g_\mathrm{st} (\theta_\seco - \omega_\seco - \pi)$. Then, $A_\seco/C_\seco$ is not a function of $\omega_\seco$ and $\mathsf{Term-4}$ can be simplified as
\begin{align}
I = \frac{1}{\alpha} \ \Gamma \left( \frac{2}{\alpha} \right) 
\int_0^{2\pi} 
%\!
\left[ \Gamma 
%\!
\left(
%\!
1 - \frac{2}{\alpha}
%\!
\right) 
\! - \!
e^{- \frac{A_\mathrm{s}}{C_\mathrm{s}}} \Gamma 
%\!
\left(
%\!
1 - \frac2\alpha, \frac{A_\mathrm{s}}{C_\mathrm{s}}
%\!
\right) 
%\!
\right] 
%\!
\mathbb{E}_{\omega_\seco} 
%\!
\left[ \frac{1}{ C_\seco^{2/\alpha}} \right] 
%\!
\dd\theta_\seco.
\label{eq:SecondaryCovSimplified3}
\end{align}
\end{example} 

%---------------------------------------------

\begin{example}	
When the typical secondary link is taken at a very far away distance from the primary receiver, dominant interference comes from nearby secondary transmitters. For these transmitters, we have $z_{\seco} = \lvert \lvert x_\seco \angle \theta_\seco - \Y_{\mathrm{p}} \rvert \rvert \implies z_{\seco} \in (y_\prim - x_\seco, y_\prim + x_\seco)$ where $y_\prim = \lvert\lvert \Y_\prim\rvert\rvert$. For $y_\prim \gg x_\seco$, we get $z_{\seco} \approx y_\prim$. For such cases, $\mathsf{Term-4}$ can be simplified as 
\begin{align}
I &= \frac{1}{\alpha} \ \Gamma \left(\frac2\alpha\right) \Gamma \left(1 - \frac2\alpha\right) \int_0^{2\pi} \mathbb{E}_{\omega_\seco} \left[ \frac{\left(1 - e^{- A_\mathrm{s} y_\prim^{\alpha}}\right)}{C_\seco^{2/\alpha}} \right] \dd\theta_\seco.
\label{eq:SecondaryCovSimplified4}	
\end{align}	
\end{example}

%---------------------------------------------

The secondary coverage given in \eqref{eq:SecondaryCov:Simplified} can be further simplified for the sectorized beam approximation as given in the following corollary (see \cite[pp. 29-34]{TripGupTheoremFile2023} for full proof).  

\begin{corr}
Under sectorized beam approximation, $\mathsf{Term-4}$ in the expression of the secondary coverage probability $p_\mathrm{cs} (\tau, \rho)$ in \eqref{eq:SecondaryCov:Simplified} can be simplified as
\begin{align}
I &= \int_0^\infty \sum_{k = 1}^{4} \int_{\mathcal{F}_k} \sum_{i = 1}^{4} q_i (\delta_\seco (x_\seco,\theta_\seco)) \left[ \left( \frac{ 1 - \exp \left( - \frac{ \rho z_\seco^{\alpha}}{ p_\seco \mathcal{A}_i \mathcal{C}_k} \right) }{ 1 + \frac{1}{\tau} \! \left(\frac{ x_\seco}{r_{\seco}}\right)^{\alpha} \!\! \frac{\ast \asr}{ \mathcal{B}_i \mathcal{D}_k } } \right) \right] x_\seco \dd x_\seco \dd \theta_\seco.
%\label{eq:SimplifyIntegral3}
\end{align}
where $\delta_\seco (x_\seco,\theta_\seco) = \theta_\seco - \beta_\seco(x_\seco,\theta_\seco)$. Here, $\mathcal{A}_i$ and $\mathcal{B}_i$ represent the transmit gains from the interfering-secondary-transmitter-to-primary-receiver and the interfering-secondary-transmitter-to-typical-secondary-receiver with probability $q_i (\delta_\seco (x_\seco,\theta_\seco))$ in Table \ref{table:SecondaryTable}. Similarly, $\mathcal{C}_k$ and $\mathcal{D}_k$ represent the gains from primary-receiver-to-the-interfering-secondary-transmitter and typical-secondary-receiver-to-the-interfering-secondary-transmitter corresponding to events $\mathcal{F}_k$ as given in Table \ref{table:SecondaryTable}.
\vspace*{-0.5cm}
\begin{table}[!htb]
\centering
\begin{minipage}{.5\linewidth}
\centering
\begin{tabular}{|l|l|l|}
\hline
Probability		&  $\mathcal{A}_i$ &  $\mathcal{B}_i$ \\ \hline
$q_\mathrm{1} (\delta_\seco(x_\seco,\theta_\seco))$	&  $\mathcal{A}_\mathrm{1} = \ast$ &  $\mathcal{B}_\mathrm{1} = \ast$ \\ \hline
$q_\mathrm{2} (\delta_\seco(x_\seco,\theta_\seco))$	&  $\mathcal{A}_\mathrm{2} = \ast$ &  $\mathcal{B}_\mathrm{2} = \bst$ \\ \hline
$q_\mathrm{3} (\delta_\seco(x_\seco,\theta_\seco))$	&  $\mathcal{A}_\mathrm{3} = \bst$ &  $\mathcal{B}_\mathrm{3} = \ast$ \\ \hline
$q_\mathrm{4} (\delta_\seco(x_\seco,\theta_\seco))$ 	&  $\mathcal{A}_\mathrm{4} = \bst$ &  $\mathcal{B}_\mathrm{4} = \bst$ \\ \hline
\end{tabular}
\end{minipage} 
\hspace*{-3cm}
\begin{minipage}{.5\linewidth}
\centering
\begin{tabular}{|l|l|l|}
\hline
$\mathcal{C}_k$	&  $\mathcal{D}_k$	& $\mathcal{F}_k$	 										\\ \hline
$\apr$			&  $\asr$ 			&  $E_\mathrm{1} \cap E_\mathrm{2}$							\\ \hline
$\apr$			&  $\bsr$			&  $E_\mathrm{1} \cap E_\mathrm{2}^\mathrm{c}$ 				\\ \hline
$\bpr$			&  $\asr$			&  $E_\mathrm{1}^\mathrm{c} \cap E_\mathrm{2}$				\\ \hline
$\bpr$			&  $\bsr$			&  $E_\mathrm{1}^\mathrm{c} \cap E_\mathrm{2}^\mathrm{c}$	\\ \hline
\end{tabular}
\end{minipage}
\caption{$E_\mathrm{1} = \left\{ \theta_\seco \, : \,  \theta_\seco - \omega_{\seco0} \in \left( -\frac{\phi_{\mathrm{sr}}}{2}, \frac{\phi_{\mathrm{sr}}}{2}\right) \right\}$ and $E_\mathrm{2} = \left\{ \theta_\seco \, : \,  \beta_\seco(x_\seco,\theta_\seco) - \omega_{\prim} \in \left( -\frac{\phi_{\mathrm{pr}}}{2}, \frac{\phi_{\mathrm{pr}}}{2}\right) \right\}$.}
\label{table:SecondaryTable}
\end{table}
\vspace*{-1cm}
\end{corr}

%---------------------------------------------

\begin{remark}
The expression \eqref{eq:SecondaryCov:Simplified} simplifies significantly under omnidirectional CCS as the terms under the integral are no longer a function of $\omega_\seco$ or $\theta_\seco$. Hence, for this case, the coverage probability of the typical secondary link is given as
\begin{align*}
p_\mathrm{cs, omni}(\SThres,\, \rho) &=  \frac{1 - \exp \left(- \rho z_{\seco0}^{ \alpha } / p_\seco \right)}{1 + (\tau p_\prim/p_\seco) (r_\seco/x_\prim)^\alpha} \exp\left(-\frac{\tau \sigma^2 r_\seco^\alpha }{p_\seco} - 2 \pi \lambda_\seco \int_0^\infty \frac{1 - \exp \left(- \rho z_{\seco}^{ \alpha } / p_\seco \right)}{1 + {1}/{\tau} ({x_\seco}/{r_\seco})^{\alpha}} x_\seco \dd x_\seco\right).
\end{align*}
\end{remark} 

%--------------------------------------------------------------------------
\subsection{Network performance} 
%--------------------------------------------------------------------------
To characterize the simultaneous performance of the primary and secondary links, we define a new metric the \textit{cumulative} performance $p_{\mathrm{c}} (\tau, \rho)$ as the sum of primary and secondary coverage probabilities {\em i.e.} $p_{\mathrm{c}} (\tau, \rho) = p_{\mathrm{cp}} (\tau, \rho) + p_{\mathrm{cs}} (\tau, \rho)$.

%------------------------------------------------------------------------------------------------------
\section{Numerical results}
%----------------------------------------------------------------------------------------------------
In this section, we provide some numerical results to derive insights. All mmWave (primary as well as secondary) devices are operating at carrier frequency $f = 60$ GHz with bandwidth $\mathrm{BW} = 200$ MHz and path-loss exponent $\alpha = 3.3$. We have taken a simulation region with radius $R = 4000$ m, centred at the origin $\mathbf{o}$. The density of $\Phi_\seco$ is $\lambda_\seco = 8 \times 10^{-5}$ /$\text{m}^2$, resulting on average $4000$ secondary devices pair. The primary and secondary transmit powers are $p_\prim  = 27$ dBm and $p_\seco = 17$ dBm with link distances $r_\prim = 50$ m and $r_\seco = 20$ m, respectively. The Johnson-Nyquist noise power is taken as $N_0 = -174 + 10\log_\mathrm{10}(\mathrm{BW})= 7.962 \times 10^{-13}$ W. The normalised noise power is $\sigma^2 = N_0/C_\mathrm{L} = 7.962 \times 10^{-07}$ where $C_\mathrm{L} = 10^{-6}$ is the near field gain. We assume that devices associated with each link have the same number of antenna elements {\em i.e.} $M_\mathrm{pt} = M_\mathrm{pr} = M_\prim$ in primary-link and $M_\mathrm{st} = M_\mathrm{sr} = M_\seco$ in secondary-link. For each $k^\mathrm{th}$ device ($k \in \{\mathrm{t}, \mathrm{r}\}$) with type $j \in \{\mathrm{p}, \mathrm{s}\}$, we have assumed a sectorized beam pattern \cite{AndrewsmmWaveTut2016,li2017design} for each device with beamwidth $\phi_{jk} = {\kappa / M_{jk}}; M_{jk} > 1$ where $\kappa = 121^{\circ}$. The main lobe gain is $M_{jk}$ and the total emitted power is normalized to 1. This also means that when the device antenna switches from omnidirectional to directional CCS, most of the power is concentrated in the main lobe of antenna radiation. 
\begin{figure}%[ht!]
\vspace*{-0.5cm}
\centering
\includegraphics[scale=0.35]{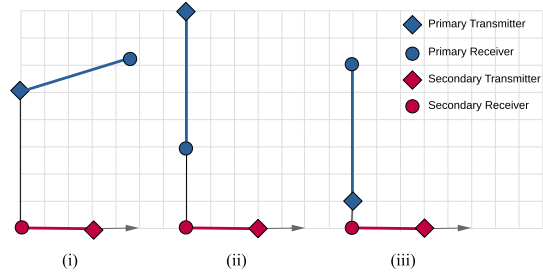}
\vspace*{-0.3cm}
\caption{An illustration of the three set-ups for the primary link's location with parameters $\{\angle \delta_\prim, x_\prim, \angle \omega_\prim\}$ taken as (i) $\{\pi/2, 50 \,\text{m}, \pi/12\}$, (ii) $\{\pi/2, 80 \, \text{m}, -\pi/2\}$ and (iii) $\{\pi/2, 10 \,\text{m}, \pi/2\}$.}	
\label{fig:SetUps}
\vspace*{-0.7cm}
\end{figure}
When analyzing the performance of the typical secondary link, we take the following three specific types of secondary links based on their environment and location/orientation relative to the primary link (See Fig. \ref{fig:SetUps}). 
\begin{enumerate}
\item[\textbf{(i)}]  \textbf{Type 1 - Minimal mutual effect of the links:} Here, the distance of cross-links is large and cross-links are not aligned ({\em i.e.} receivers of both cross-links doesn't fall in the main lobe of the link's transmitters under directional CCS).
\item[\textbf{(ii)}] \textbf{Type 2 - Moderate mutual effect:} Here, the primary receiver is outside the main lobe of the secondary transmitter under directional CCS but may get affected by secondary transmission under omni or antenna with wider beamwidths. Also, the secondary receiver falls in the main lobe of the primary transmitter and hence gets affected by the primary transmitter.
\item[\textbf{(iii)}] \textbf{Type 3 - High mutual effect:} Here, cross links are at short distances. The primary transmitter may fall in the main lobe of the secondary receiver and its associated transmitter under both omni and directional CCS. Thus primary transmission is affected significantly by both secondary devices and vice-versa.
\end{enumerate}

We also observe the performance of an average typical secondary link, termed as \textbf{Type 4}. Here, for the primary-link, we take $x_\prim = \sqrt{u_\prim}$ where $u_\prim = \mathcal{U}(0, R^2)$ with $\omega_\prim$ and $\delta_\prim$ to be distributed uniformly between $0$ and $2\pi$.

%----------------------------------------
\subsection{MAP of secondary links}
%----------------------------------------

\begin{figure}%[ht!]
\vspace*{-0.5cm}
\centering
\hspace{-0.8cm}
%{\small \bf (a)}
{\includegraphics[scale=0.3]{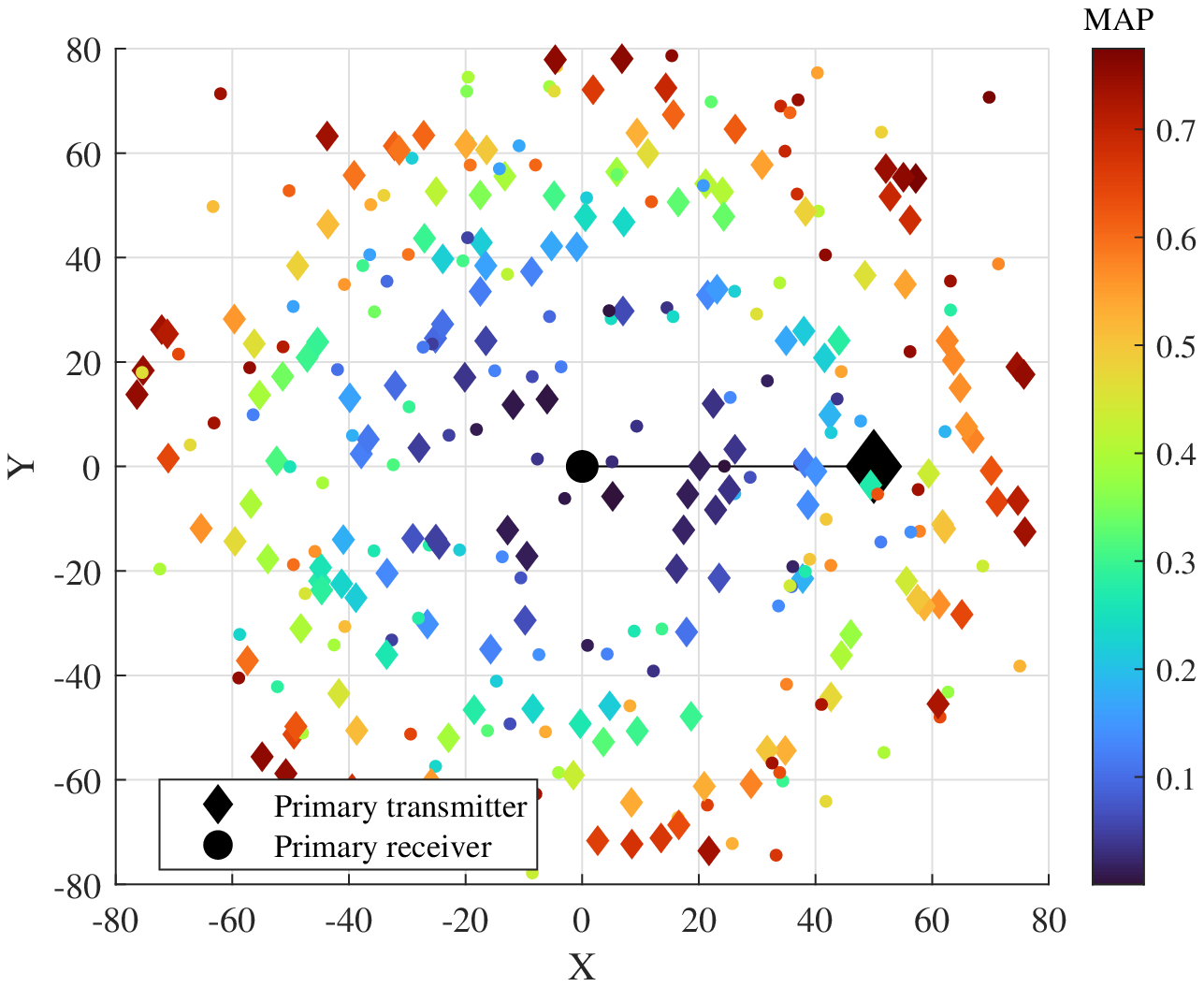} \label{numres:MAP11} }
\hspace{-0.7cm}
%{\small \bf (b)}
{\includegraphics[scale=0.3]{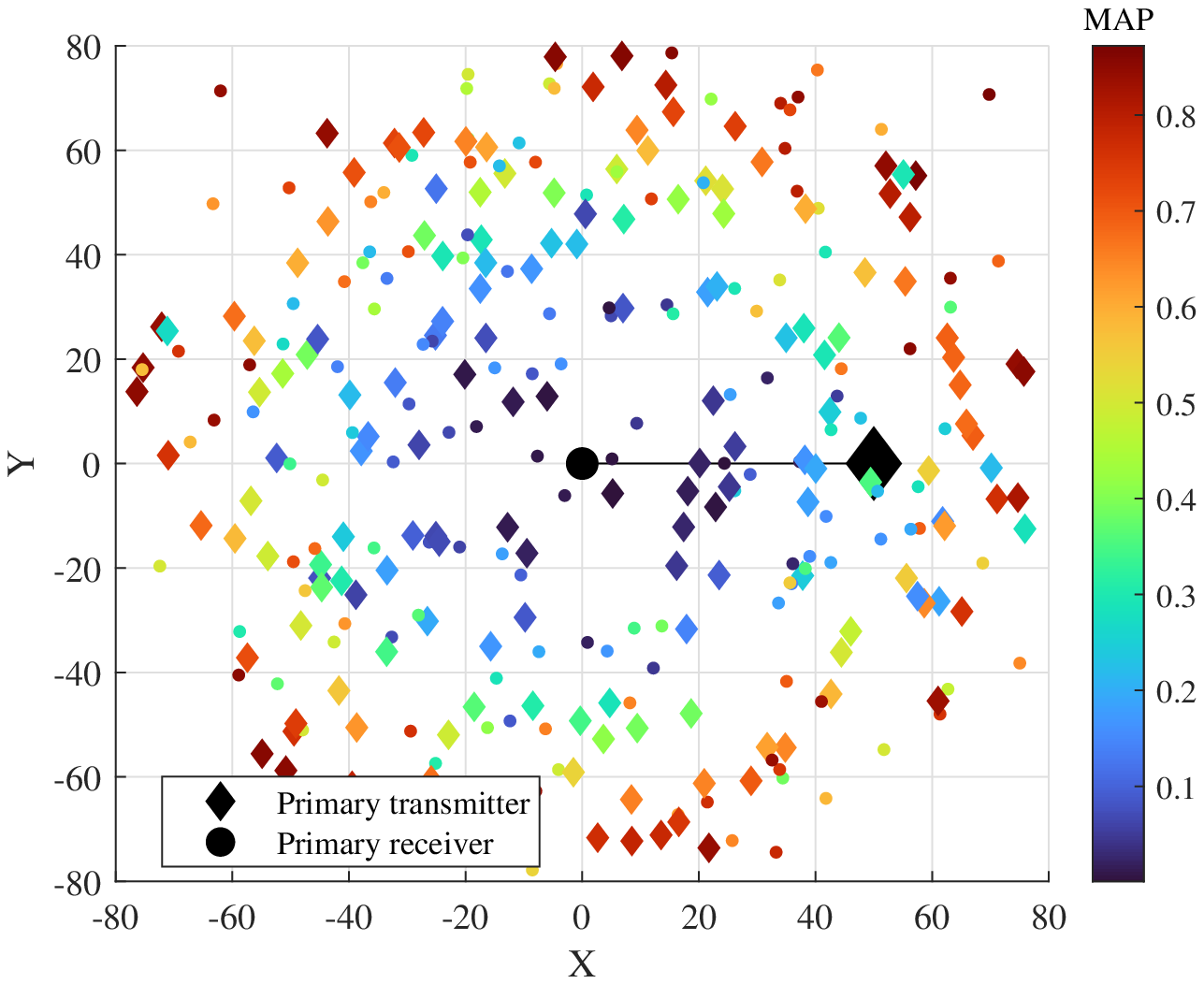} \label{numres:MAP14} }
\hspace{-0.7cm}
%{\small \bf (c)}
{\includegraphics[scale=0.3]{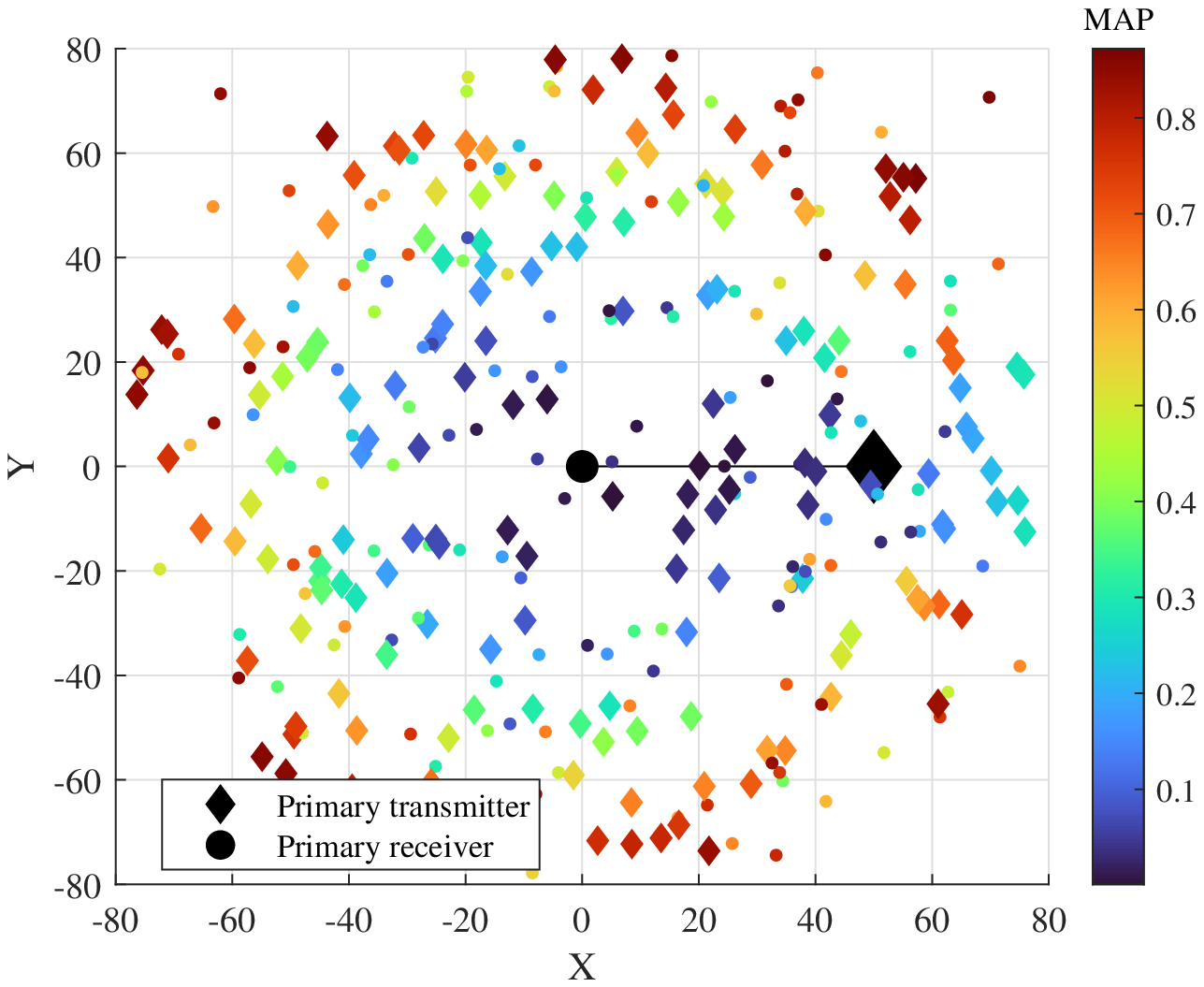} \label{numres:MAP41} }
\hspace{-0.7cm}
%{\small \bf (d)}
{\includegraphics[scale=0.3]{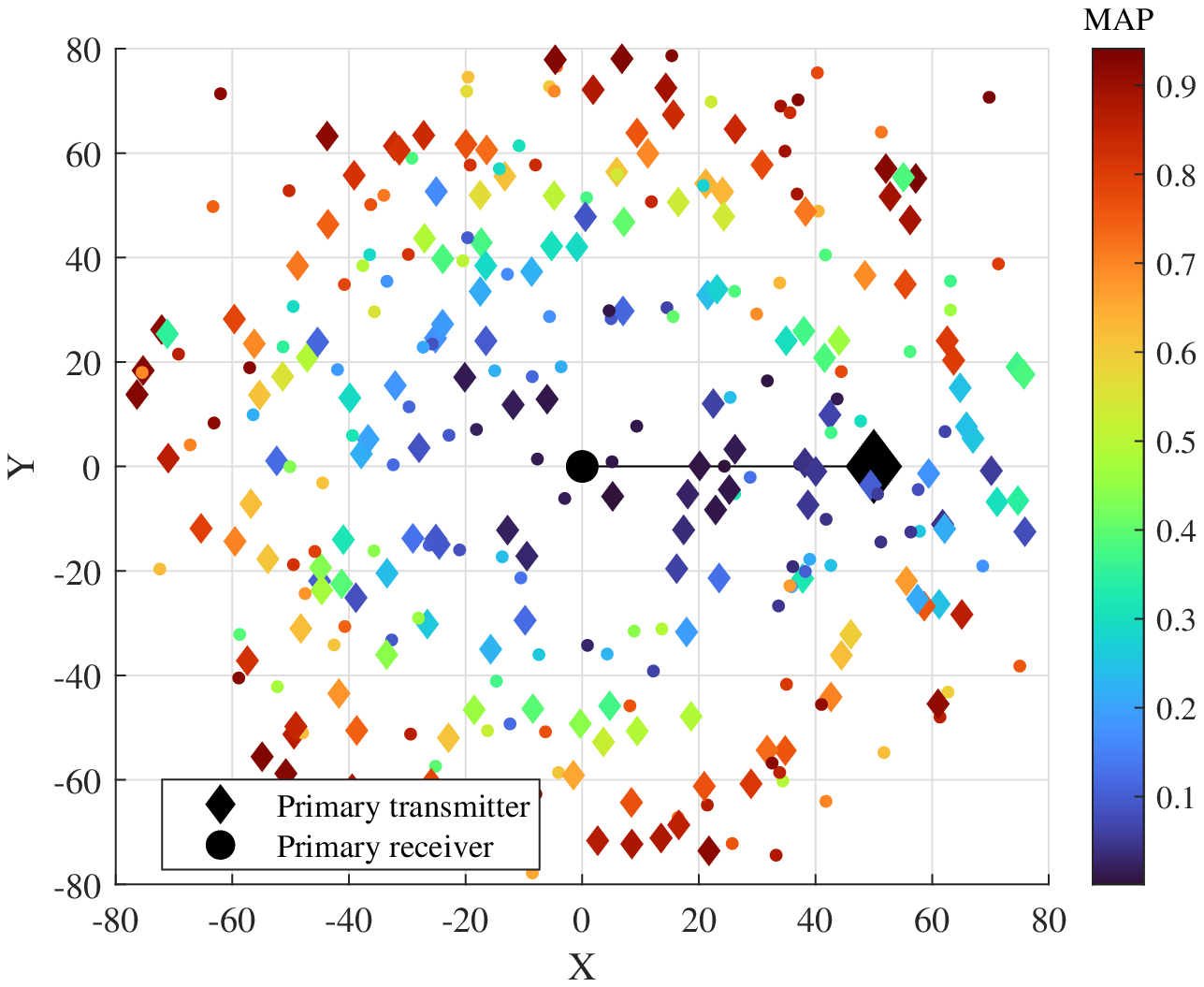} \label{numres:MAP44} } \\
\vspace*{-0.3cm}
{\small (a) $M_\prim = 1, \, M_\seco = 1$ \hspace{0.7cm} (b) $M_\prim = 1, \, M_\seco = 4$ \hspace{0.7cm} (c) $M_\prim = 4, \, M_\seco = 1$ \hspace{0.7cm} (d) $M_\prim = 4, \, M_\seco = 4$}
\vspace*{-0.3cm}
\caption{The spatial variation of $p_{\mathrm{m}i}$ for omni and directional CCS with $\rho = 40 $ nW, $\alpha = 3.3$ and $\lambda_\seco = 8 \times 10^{-3}$ /$\text{m}^2$. Both ends of the primary and secondary links have $M_\prim$ and $M_\seco$ antennas.}
%-------> Previous result is for $\rho = 0.01 \, \mu$W and $p_\seco = 10$ dBm.
\label{numres:MAP}
\vspace*{-0.8cm}
\end{figure}

%-------------------------------------------------
Fig. \ref{numres:MAP} 
%-------------------------------------------------
shows the impact of the location of a secondary transmitter on its MAP. We can observe that the secondary transmitters, located at the same distance from the primary receiver have the same MAP under omnidirectional CCS. However, when any one or both of the primary and secondary links employ directional CCS, MAP depends on the orientation of the secondary transmitter and link ($\theta_\seco$ and $\omega_\seco$) as per \eqref{eq:MAP}. We also observe that the impact of primary directionality ($M_\prim$) on MAP is higher in comparison to that of secondary directionality ($M_\seco$). 

%--------------------------------------------------------------------------------------
\subsection{AF of secondary network} 
%--------------------------------------------------------------------------------------
Now, we investigate the AF of the secondary network and the impact of various factors on it.

%------------------
\subsubsection{Trade-off between beamwidth and gain of directional antenna}
\label{section:EffectsGainBeamwidthAF}
\begin{figure}[ht!]
\vspace*{-0.5cm}
\centering
\hspace{-0.7cm}
%{\small \bf (a)}
{\includegraphics[trim = 0 0 0 0, clip, scale=0.4]{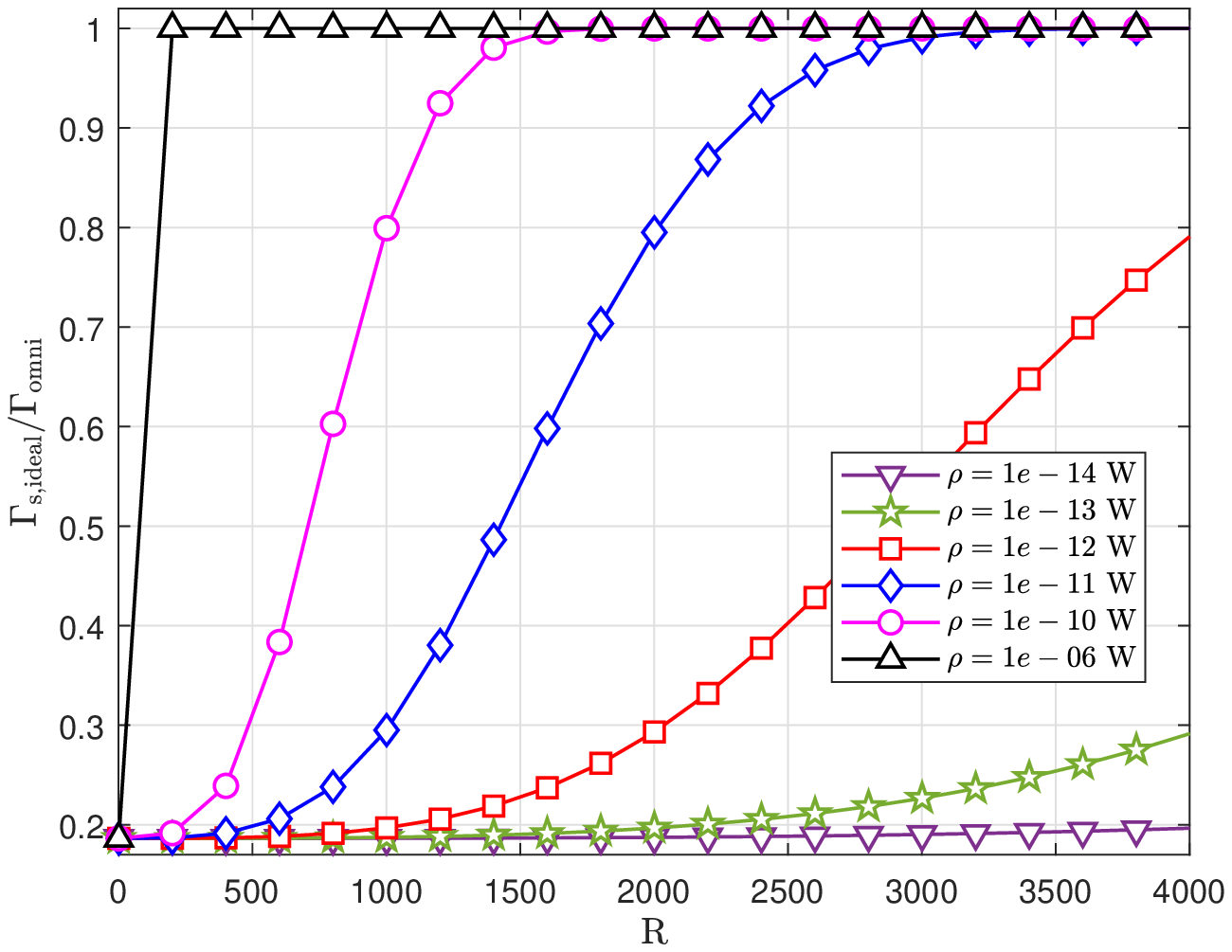} \label{numres:Gamma44} }
\hspace{-0.8cm}
%{\small \bf (b)}
{\includegraphics[scale=0.4]{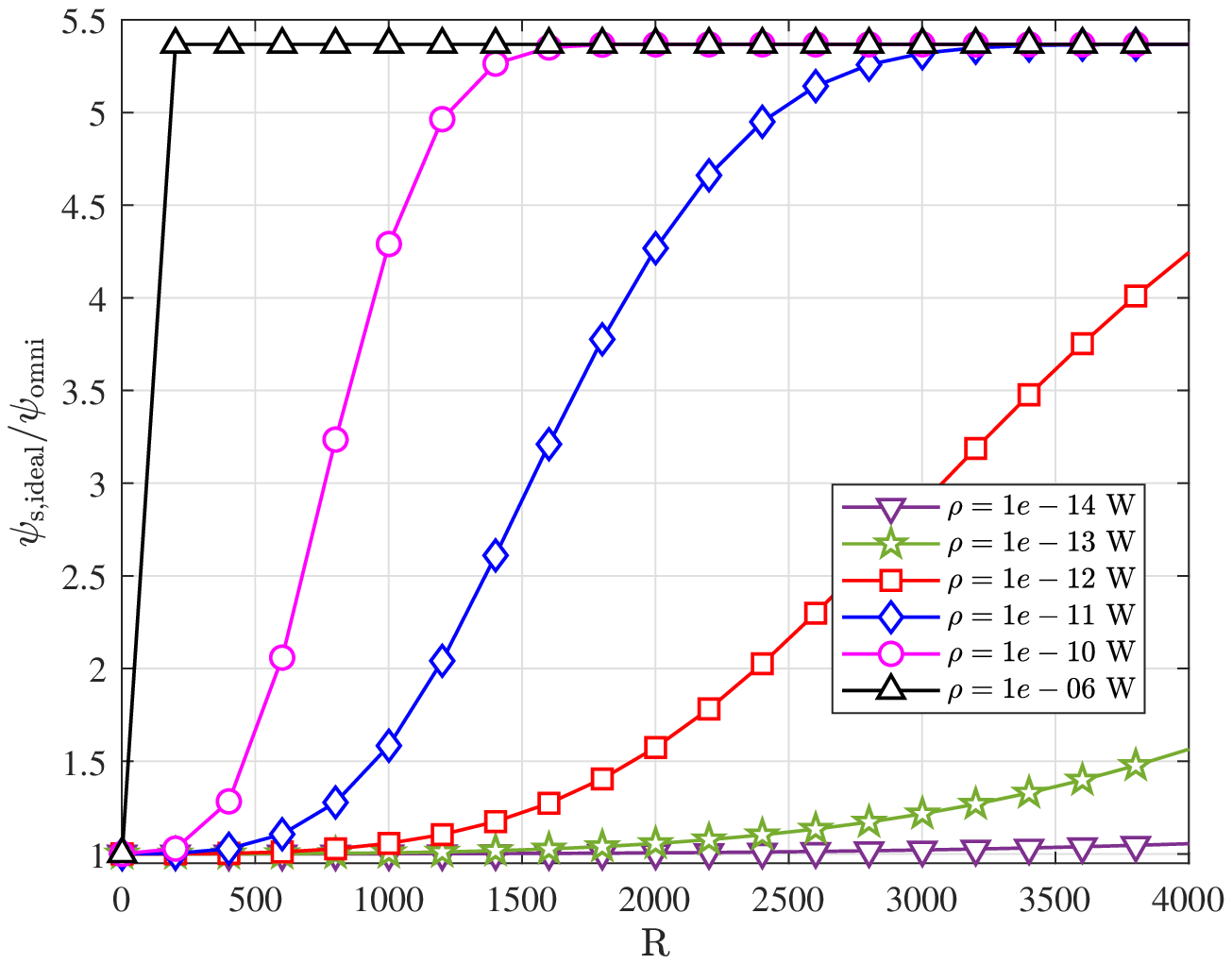} \label{numres:EtaRatioWithoutProb44} }
\hspace{-0.8cm}
%{\small \bf (c)}
{\includegraphics[scale=0.4]{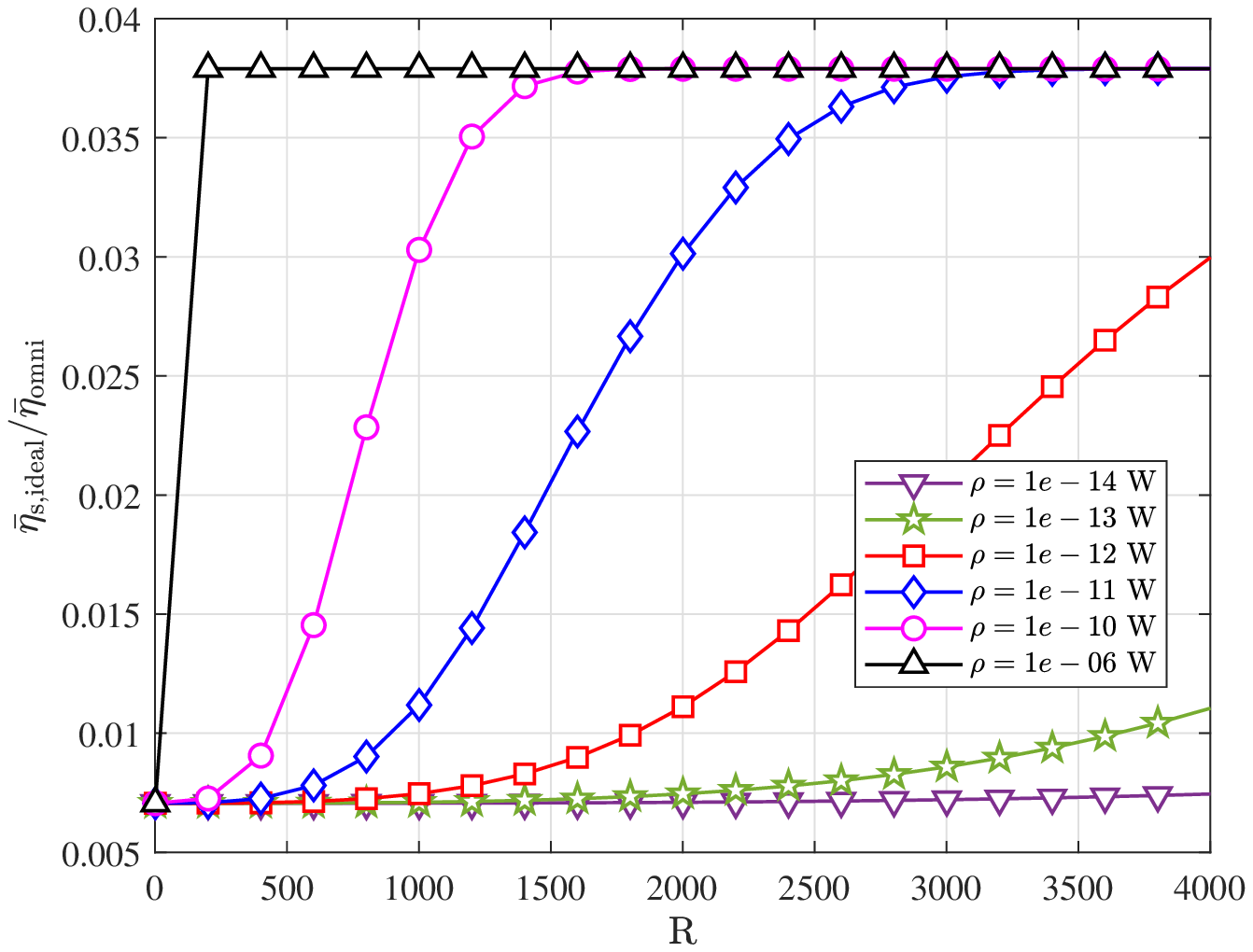} \label{numres:EtaRatio44} } \\ 
\vspace*{-1cm}
{\small \bf (a) \hspace{4.8cm} (b) \hspace{4.8cm} (c)}
\vspace*{-0.3cm}	
\caption{The variation of the relative value of (a) $\bar{\eta}$, (b) $\psi$ and (c) $\Gamma$ under directional sensing with ideal beam pattern compared to omni sensing with respect to deployment radius $R$ for $M_\prim = M_\seco = 4$ and $\alpha = 3.3$.}	
\label{numres:EtaRatio}
\vspace*{-0.5cm}
\end{figure}

%------------------
%-------------------------------------------------
Fig. \ref{numres:EtaRatio} 
%-------------------------------------------------
shows the variation of the relative value of $\Gamma$, $\psi$ and $\bar{\eta}$ under directional CCS with ideal beam pattern compared to omni sensing with respect to deployment radius $R$. Note that these variables determine the behaviour of AF as per \eqref{eq:AFideal} and observation of Fig. \ref{numres:EtaRatio} is true for all $\alpha > 2$. As discussed in section \ref{subsect:AFBeamwidthGainTrade-offAnalysis}, the contribution of $(\apr\ast)^{2/\alpha}$ term in $\psi(u)$ is overshadowed by the contribution of term $\qpr\qst$ in $\bar{\eta}_{\seco, \mathrm{ideal}}$ under directional CCS. This results in $\psi_{\mathrm{s},\mathrm{ideal}}/\psi_{\mathrm{s},\mathrm{omni}} > 1$ and  $\bar{\eta}_{\seco, \mathrm{ideal}}/\bar{\eta}_\mathrm{omni} < 1$ as shown in
%-------------------------------------------------
Fig. 7(b)
%\ref{numres:EtaRatioWithoutProb44} 
%-------------------------------------------------
and
%-------------------------------------------------
Fig. 7(c).
%\ref{numres:EtaRatio44}. 
%-------------------------------------------------
Hence, the increment in AF under directional CCS is observed due to the far prominent effect of antenna beamwidth ($\qpr\qst$) in comparison to the antenna gain ($\apr\ast$). 

%------------------
\subsubsection{Impact of directional CCS}  
%------------------
%-------------------------------------------------
Fig. 8(a)
%\ref{numres:AFRatio} 
%-------------------------------------------------
shows the variation of secondary AF with interference threshold $\rho$ for different values of $M_\prim$ and $M_\seco$ for a region of interest with radius $R = 4000$ m. We can observe that AF improves with directionality. This happens because omni CCS imposes distance-based transmit restrictions on all neighbouring secondary devices around the primary receiver, making them to remain inactive. On the other hand, directional CCS allows orientation-based transmit-restriction on all neighbouring secondary devices, providing them with increased primary channel access opportunities that saturate with higher values of $M_\prim$ and $M_\seco$. This verifies our observation in section \ref{section:EffectsGainBeamwidthAF}, that the effect of antenna beamwidth dominates over antenna gain resulting in higher AF.
\begin{figure}[ht!]
\vspace*{-0.5cm}
\centering
%{\small \bf (a)}
{\includegraphics[trim = 0 0 0 0, clip, scale=0.4]{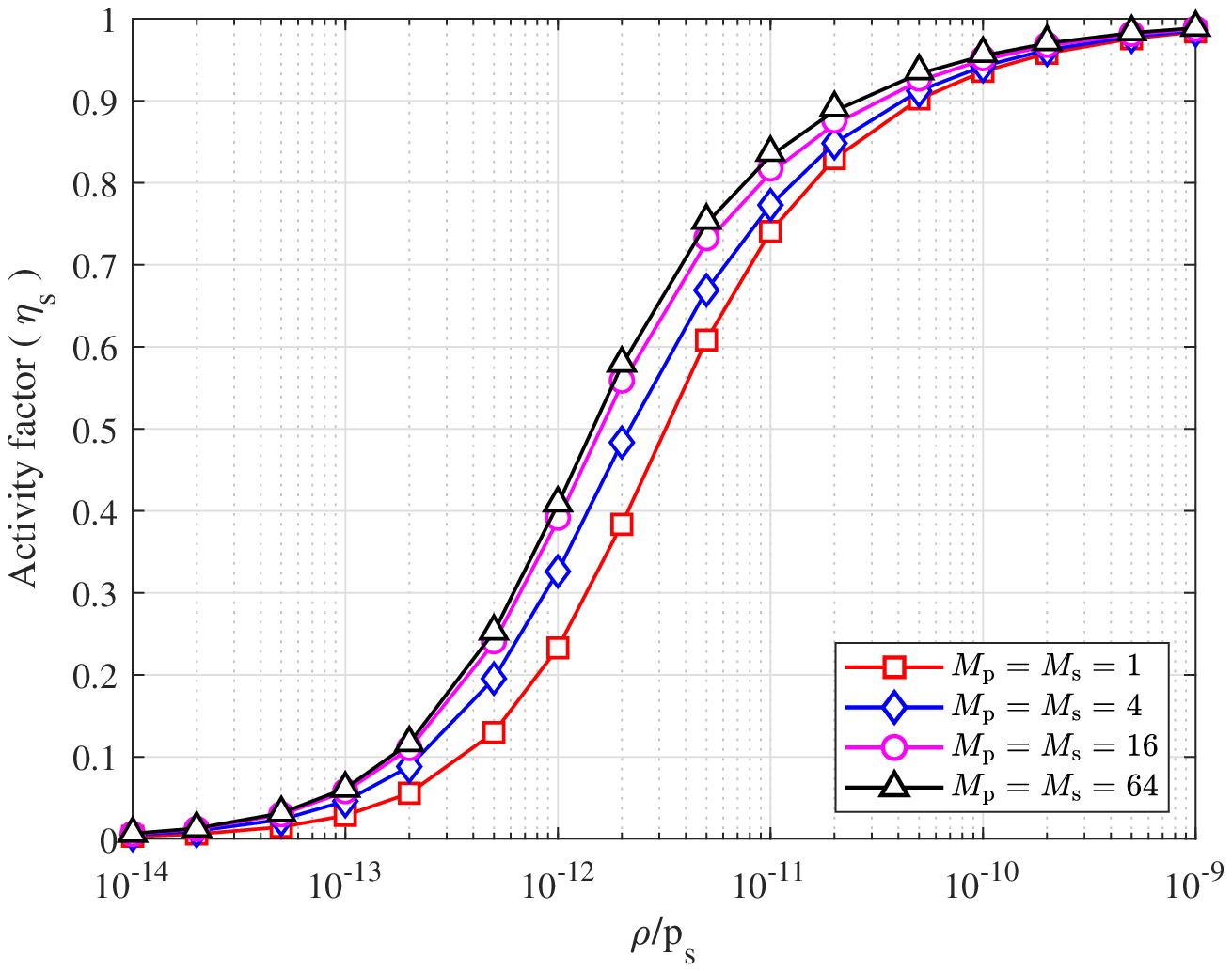} \label{numres:AFRatioN} }
%{\small \bf (b)}
{\includegraphics[scale=0.4]{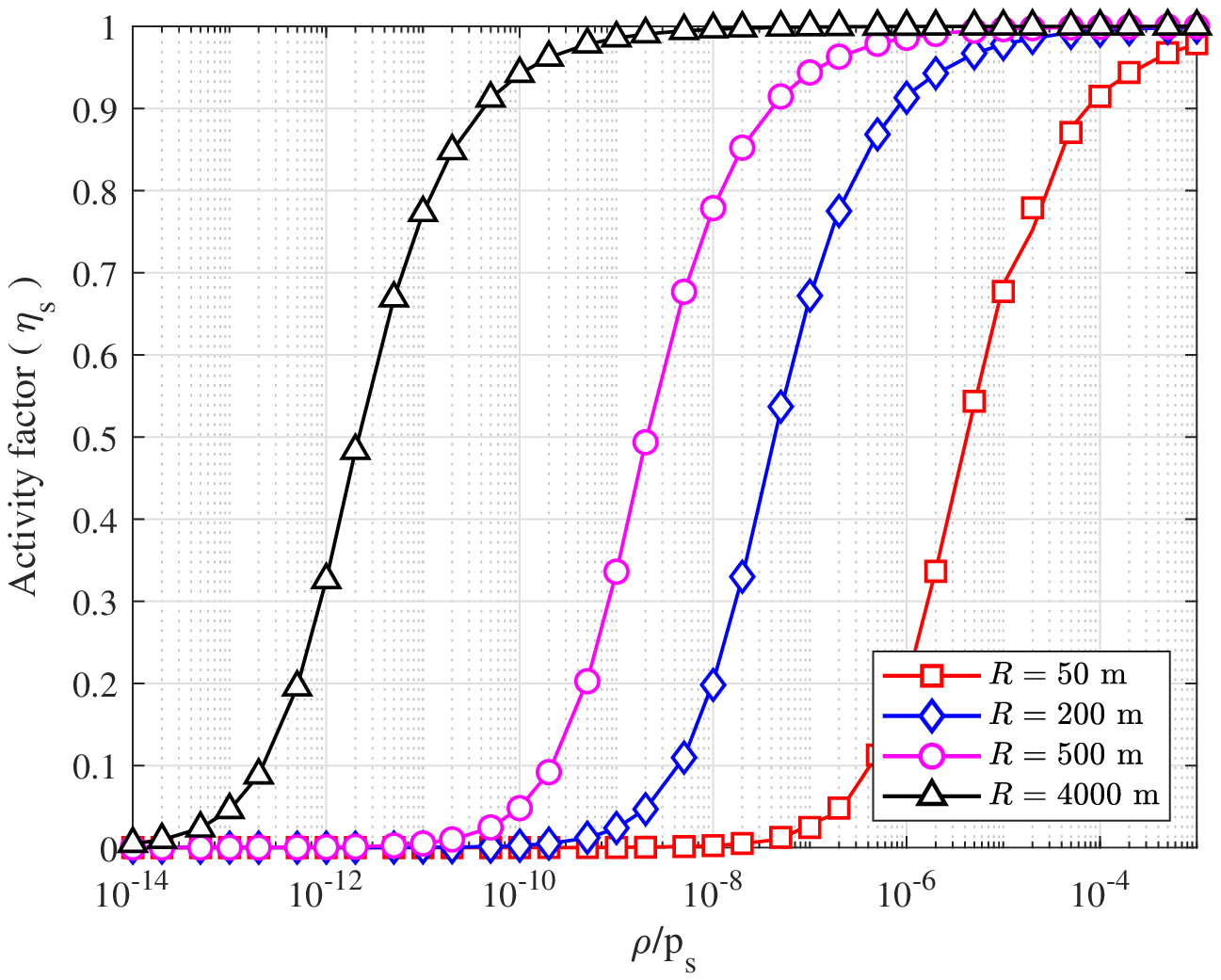} \label{numres:AFRatioR} } \\
\vspace*{-0.3cm}
{\small {\bf (a)} \hspace{6cm} {\bf (b)} }
\vspace*{-0.3cm}
\caption{The variation of secondary AF $\eta_\seco$ with the interference threshold $\rho$ for (a) different values of the number of antennas with $R = 4000$ m and (b) different values of region-of-interest radius $R$ with $M_\prim = M_\seco = 4$.}
\label{numres:AFRatio}
\vspace*{-0.5cm}
\end{figure}

%------------------
\subsubsection{Impact of secondary link proximity from the primary receiver}  
%------------------
%-------------------------------------------------
Fig. 8(b)
%\ref{numres:AFRatioR} 
%-------------------------------------------------
shows the variation of secondary AF with $\rho/p_\seco$ for different values of $R$. A large value of $R$ can capture a long-range effect. We observe that a small restriction (high $\rho$) affects nearby devices and its impact vanishes over large $R$. A higher restriction (smaller $\rho$) has a more distant effect as seen for larger values of $R$. For example, for $R = 50$ m, and $M_\prim = M_\seco = 4$ antennas, the mean $\rcvdp{\seco i}{\prim}/{p_\seco}$ at the edge of the region of interest is $3.958 \times 10^{-5}$ while for $R = 4000$ m, it is $2.076 \times 10^{-11}$.  As seen from  
%-------------------------------------------------
Fig. 8(b),
%\ref{numres:AFRatioR}, 
%-------------------------------------------------
a threshold of this order will change AF significantly.
We also observe from 
%-------------------------------------------------
Figs. 8(a) and (b)
%Figs. \ref{numres:AFRatioN} and \ref{numres:AFRatioR},
%------------------------------------------------- 
that AF saturates {\em i.e.} $\eta_\seco \to 1$ for higher values of $\rho$, irrespective of the values of $M_\prim$, $M_\seco$ and $R$. This saturation represents the removal of any restriction on secondary transmission. 

%--------------------------------------------------------------------------------------
\subsection{Impact of transmit-restrict threshold $\rho$}
%--------------------------------------------------------------------------------------
As seen from the analysis, $\rho$ has a reciprocal impact on the primary and secondary link performances. A lower value of $\rho$ restricts more secondary devices to remain idle improving primary coverage at the expense of secondary coverage. On the other hand, higher $\rho$ increases the number of active secondary transmitters, thus increasing the interference at the primary receiver and causing a degradation in primary QoS.
\begin{figure}%[ht!]
\vspace*{-0.5cm}
\centering
%{\small \bf (a)}
{\includegraphics[trim = 0 0 0 0, clip, scale=0.4]{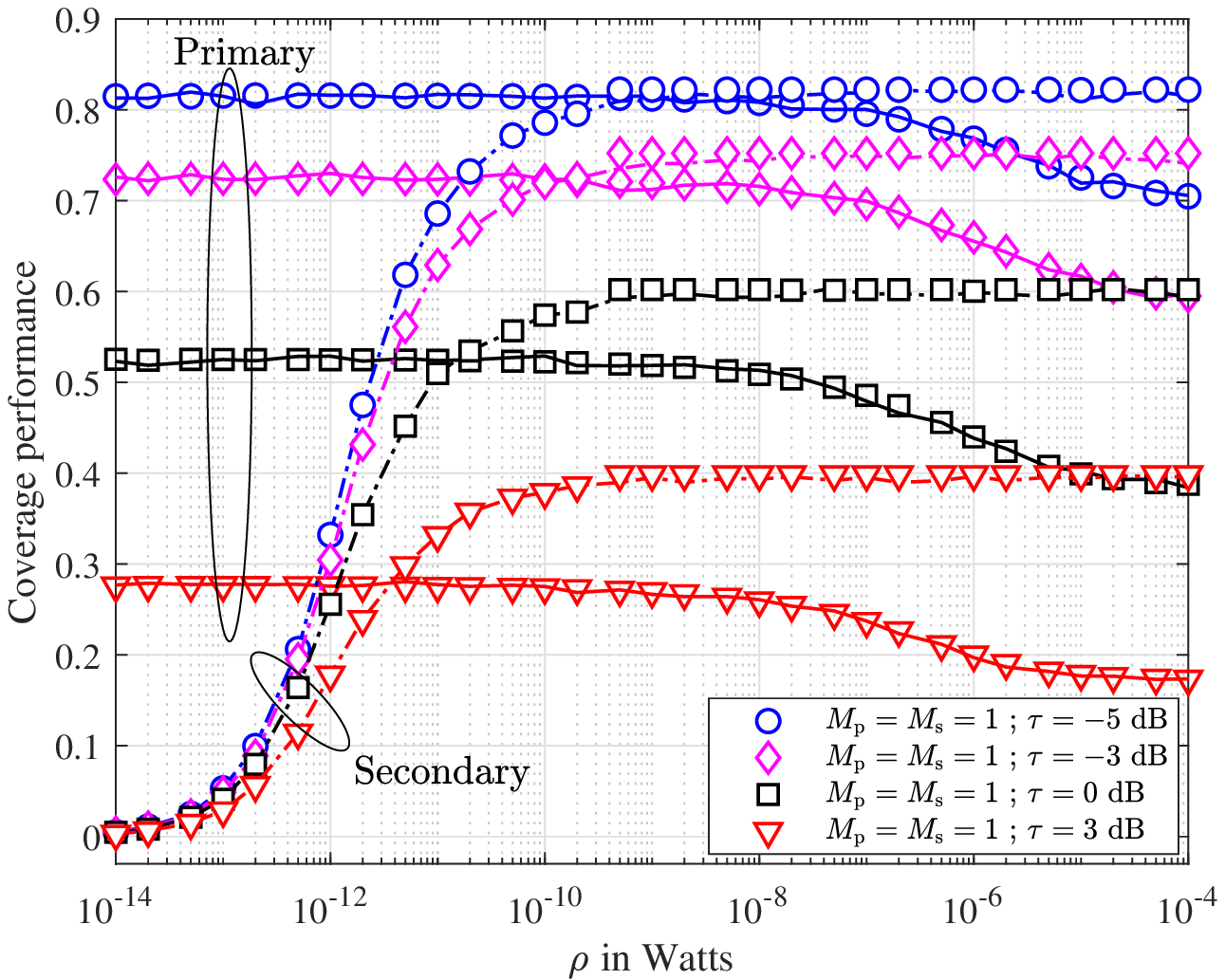} \label{numres:TauN1-Randomized} }
%{\small \bf (b)}
{\includegraphics[scale=0.4]{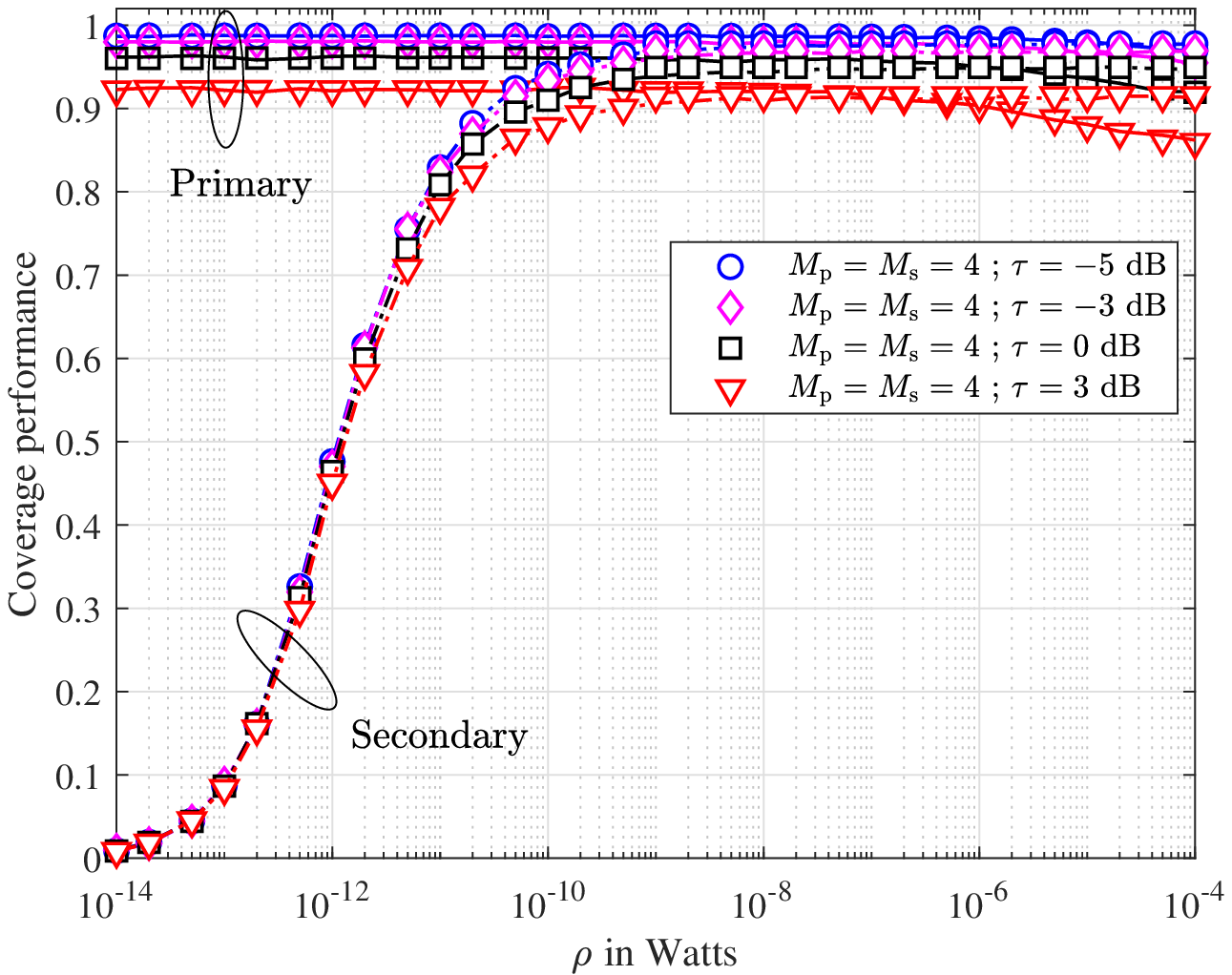} \label{numres:TauN4-Randomized} } \\
\vspace*{-0.3cm}
{\small {\bf (a)} \hspace{4cm} {\bf (b)}}
\vspace*{-0.3cm}
\caption{The variation of primary and secondary coverage probabilities with transmit-restriction-threshold $\rho$ for (a) omnidirectional CCS and (b) directional CCS for Type 4 (the average secondary user).}
\label{numres:TauN-Randomized}
\vspace*{-0.5cm}
\end{figure}

%------------------------------------------------- 
Figs. 9(a) and (b)
%\ref{numres:TauN1-Randomized} and \ref{numres:TauN4-Randomized}
%------------------------------------------------- 
show the variation of primary and secondary coverage with threshold $\rho$ for omni and directional CCS. We can see that directionality can help to improve the secondary coverage without affecting the primary coverage significantly for appropriate thresholds, {\em e.g.} $\rho = 10$ pW. This shows that if given enough directionality, a suitable value $\rho^\dagger$ of threshold $\rho$ can be obtained which not only preserve the primary-link's performance but also guarantee a reasonable secondary coverage. We now discuss how to select $\rho^\dagger$.

One way to find the suitable/appropriate value of the primary-transmit-protection threshold is to observe the first saturation points of $p_{\mathrm{cs}}$. This gives us different suitable values of $\rho^\dagger$ based on SINR threshold $\tau$ and number of antenna elements $M_{jk}$, as shown in 
%------------------------------------------------- 
Fig. \ref{numres:TauN-Randomized}.
%-------------------------------------------------  
Another way is to utilize the sum of primary and secondary coverage performance to determine the value of threshold $\rho^\dagger$. Note that this method may not be desirable because it does not provide any guarantee for QoS to primary and secondary links. To overcome this problem, an extra condition can be added which sets a lower limit on the primary and secondary coverage probabilities. For example, consider the constraints $\{p_{\mathrm{cp}} > 70 \%, p_{\mathrm{cs}} > 50 \%\}$. In this case, from 
%------------------------------------------------- 
Fig. 9(a)
%\ref{numres:TauN1-Randomized} 
%------------------------------------------------- 
$\rho^\dagger = 8.89$ pW given that $\tau \leq 0$ dB and from 
%------------------------------------------------- 
Fig. 9(b)
%\ref{numres:TauN4-Randomized} 
%------------------------------------------------- 
$\rho^\dagger = 1.3$ pW for $\tau \leq 3$ dB. Therefore, guidelines for the selection of suitable $\rho^\dagger$ can be written as: For some $\prim^\star, \, \seco^\star$ such that $0 \leq \{\prim^\star, \seco^\star\} \leq 1$ and $\tau^\star$, find
\vspace*{-0.7cm}
\begin{align}
&\rho^\dagger = \min \rho \nonumber \\
s.t. \qquad & p_{\mathrm{cp}} (\tau^\star,\, \rho) \geq \prim^\star \ ; \ p_{\mathrm{cs}} (\tau^\star,\, \rho) \geq  \seco^\star.
\label{eq:OptCondition}
\end{align}

Note that if $\tau^\star$ exists then the condition is also true for all $\tau \leq \tau^\star$. For the exposition of the system's behaviour, Table  \ref{table:SuitableRho} lists the solution for \eqref{eq:OptCondition} for some sets of parameter values. 

\begin{table}[ht!]
\begin{tabular}{|c|c|c|c|c|cccc|}
\hline
\multirow{3}{*}{Set-up} 	& \multirow{3}{*}{$\angle \delta_\prim$} 	& \multirow{3}{*}{$x_\prim$} 	& \multirow{3}{*}{$\omega_\prim$ } 	& 		\multirow{3}{*}{$\tau^\star$} 	&  \multicolumn{1}{c|}{$M_{jk} = 1$} 	& \multicolumn{1}{c|}{$M_{jk} = 2$}  & \multicolumn{1}{c|}{$M_{jk} = 4$} & $M_{jk} = 8$  		\\ \cline{6-9} 
%----------------------------------------------------------------------------------------------------------------------------------------------
& & & & &  \multicolumn{1}{c|}{$\phi_{jk} = 360^{\circ}$} &  \multicolumn{1}{c|}{$\phi_{jk} = 60.5^{\circ}$} &  \multicolumn{1}{c|}{$\phi_{jk} = 30.25^{\circ}$} &  $\phi_{jk} = 15.125^{\circ}$       \\ \cline{6-9} 
%----------------------------------------------------------------------------------------------------------------------------------------------
& & & & &  \multicolumn{4}{c|}{$\rho^\dagger$}  \\ \hline
%----------------------------------------------------------------------------------------------------------------------------------------------
$1$	& $\pi/2$ & $50$ m & $\pi/12$  & $\leq -3$ dB  & \multicolumn{1}{c|}{$75$ nW}  & \multicolumn{1}{c|}{$23$ nW} & \multicolumn{1}{c|}{$17$ nW} & $14$ nW \\ \hline
$2$	& $\pi/2$ & $80$ m & $-\pi/2$  & $\leq 0$ dB  & \multicolumn{1}{c|}{$0.85\, \mu$W} &  \multicolumn{1}{c|}{$0.21\, \mu$W} &  \multicolumn{1}{c|}{$0.15\, \mu$W} & $0.13\, \mu$W  \\ \hline
$3$	& $\pi/2$ & $10$ m & $\pi/2$   & $\leq - 13$ dB & \multicolumn{1}{c|}{$40.4$ nW} & \multicolumn{1}{c|}{$0.216\, \mu$W} & \multicolumn{1}{c|}{$26.1$ nW} & $19.9$ nW  \\ \hline
$4$	& $\mathcal{U} (0,\, \pi)$ & $\frac{R}{2}\sqrt{\mathcal{U} (0, 1)}$ & $\mathcal{U}(0, \pi)$ & $\leq 0$ dB  & \multicolumn{1}{c|}{$8.89$ pW} & \multicolumn{1}{c|}{$1.98$ pW} & \multicolumn{1}{c|}{$1.21$ pW} & $0.95$ pW \\ \hline
%----------------------------------------------------------------------------------------------------------------------------------------------
\end{tabular}
\caption{The appropriate value of $\rho^\dagger$ satisfying $\{p_{\mathrm{cp}} > 70 \%, p_{\mathrm{cs}} > 50 \%\}$ criteria for the different set-ups of the typical secondary location with $\alpha = 3.3$.}
\label{table:SuitableRho}
\vspace*{-0.5cm}
\end{table}

%--------------------------------------------------------------------------------------
\subsection{Impact of antenna directionality $M_{jk}$} 
%--------------------------------------------------------------------------------------
From 
%------------------------------------------------- 
Figs. 9(a) and (b),
%\ref{numres:TauN1-Randomized} and \ref{numres:TauN4-Randomized}
%------------------------------------------------- 
we can observe that directionality can improve feasibility of the cognitive communication. For example, for $\tau = 3$ dB, no feasible value of $\rho^{\dagger}$ exists under omnidirectional CCS. However, under $M_{jk} = 8$, it is feasible to satisfy the constraints for $\tau = 3$ dB. From Table \ref{table:SuitableRho}, we observe that on average, directionality allows a higher restriction to be put on secondary links while providing the same performance guarantee for primary and secondary links. We now investigate the impact of directionality more carefully. For the rest results, we fix $\rho = 40$ nW unless stated otherwise.
\begin{figure}[ht!]
\vspace*{-0.5cm}
\centering
%{\small \bf (a)}
{\includegraphics[trim = 0 0 0 0, clip, scale=0.4]{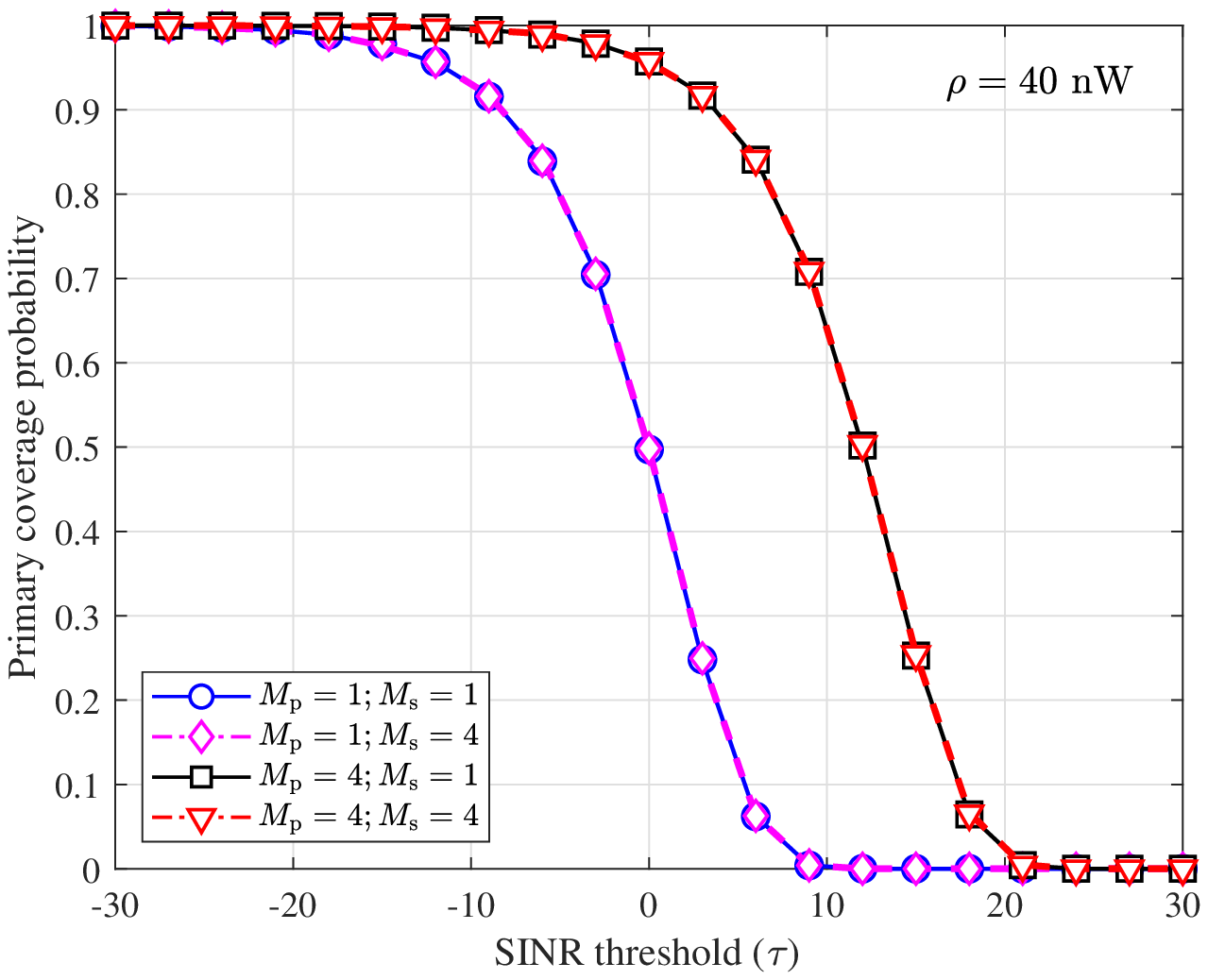} \label{numres:PP:SameDensity} }
%{\small \bf (b)}
{\includegraphics[scale=0.4]{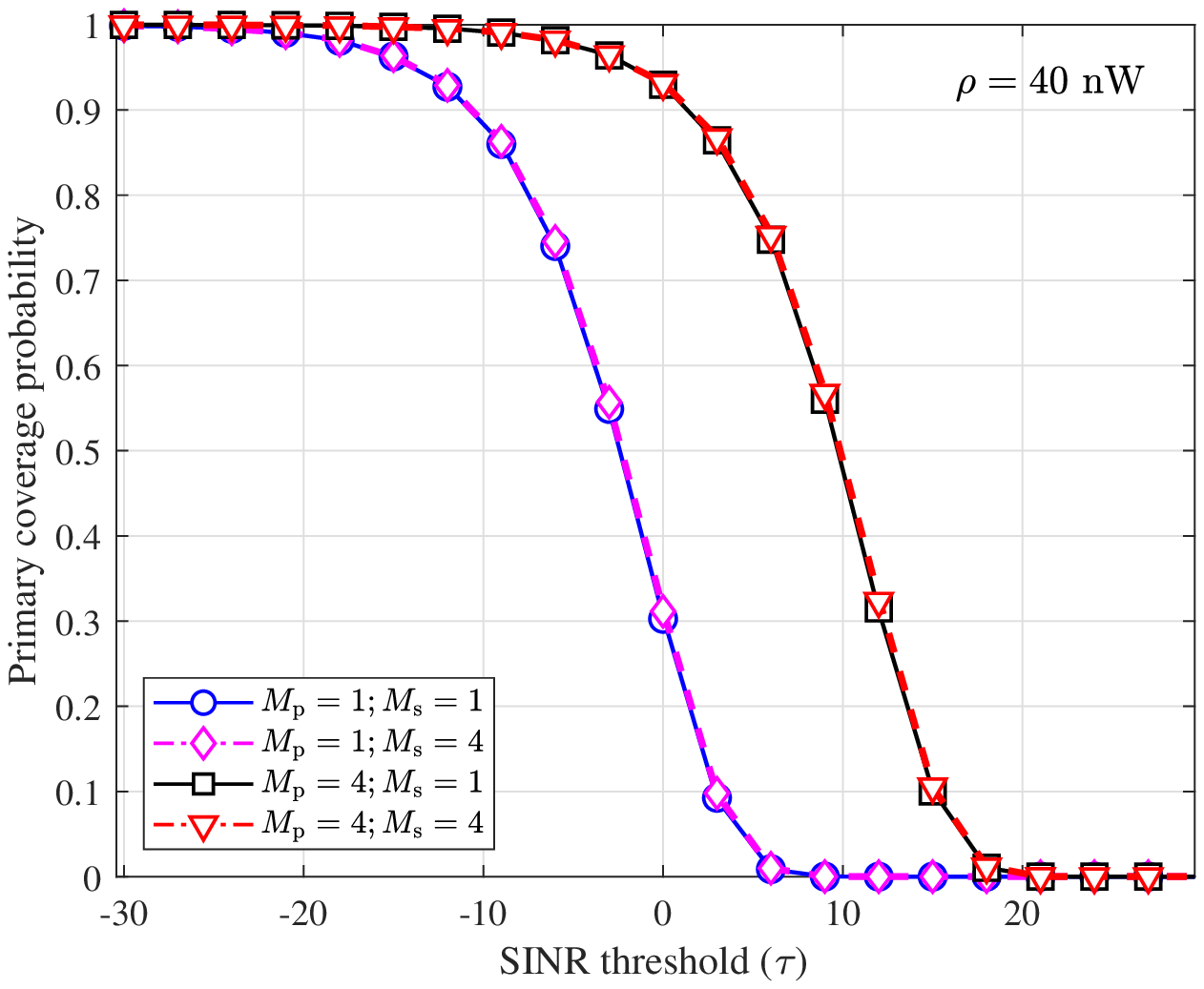} \label{numres:PP:10TimesDensity} } \\
\vspace*{-0.3cm} 
{\small {\bf (a)}  \hspace{6cm} {\bf (b)}  }
\vspace*{-0.3cm} 
\caption{Impact of the primary and secondary antenna directionality on the primary-link performance at different values of secondary density for $R = 4000$ m, $\alpha = 3.3$ with (a) $\lambda_\seco = 8 \times 10^{-5}$ /$\text{m}^2$ and (b) $\lambda_\seco = 8 \times 10^{-4}$ /$\text{m}^2$.}
\label{numres:PP}
\vspace*{-0.5cm}
\end{figure}

%--------------------------------------------------------------------------------------
\subsubsection{On the primary coverage performance}
%------------------------------------------------- 
Fig. \ref{numres:PP} 
%------------------------------------------------- 
shows the variation of primary coverage with different values of $M_\prim$ and $M_\seco$. We observe that primary-link's directionality has prominent effects on its coverage performance $p_\mathrm{cp}(\tau, \rho)$ irrespective of the secondary antenna directionality. A positive shift of $+12$ dB is observed in $p_\mathrm{cp}$ when primary antenna is changed from $\phi = 360^{\circ} (M_\prim = 1)$ to $\phi = 30.25^{\circ} (M_\prim = 4)$. On the other hand, it is not affected much by secondary directionality which is consistent with our analytical result given in lemma \ref{example:PrimaryCov:Simplified:Example1a}. We can observe a very small positive shift between $M_\seco = 1$ and $4$. A small change in the primary coverage implies that directional CCS with threshold $\rho$ is able to limit the secondary interference on the primary.

%--------------------------------------------------------------------------------------
\subsubsection{On the secondary coverage performance} 
\begin{figure}[ht!]
\vspace*{-0.5cm}
\centering
\hspace{-0.6cm}
%{\small \bf (a)}
{\includegraphics[trim = 0 0 0 0, clip, scale=0.3]{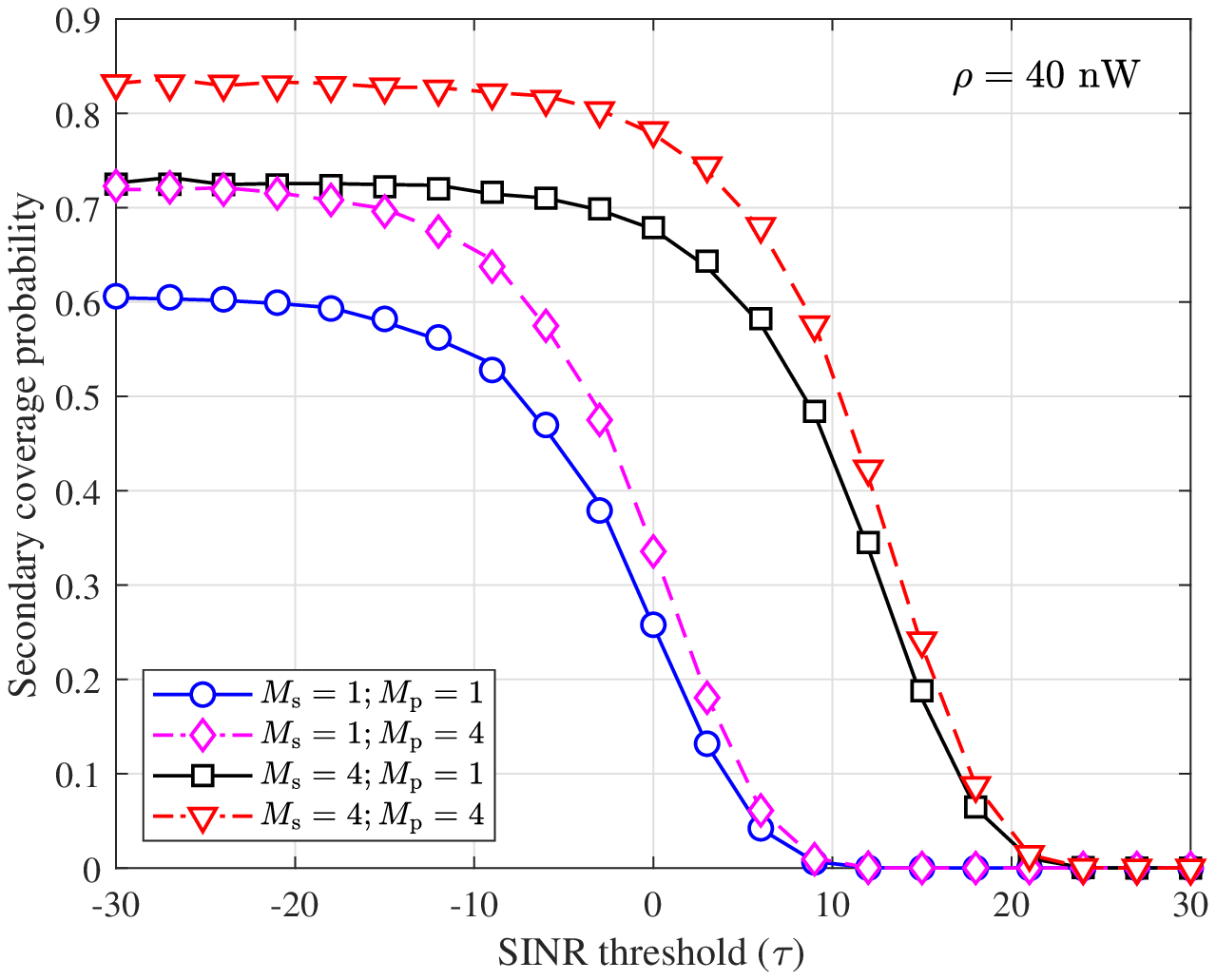} \label{numres:SP:Set1} }
\hspace{-0.7cm}
%{\small \bf (b)}
{\includegraphics[scale=0.3]{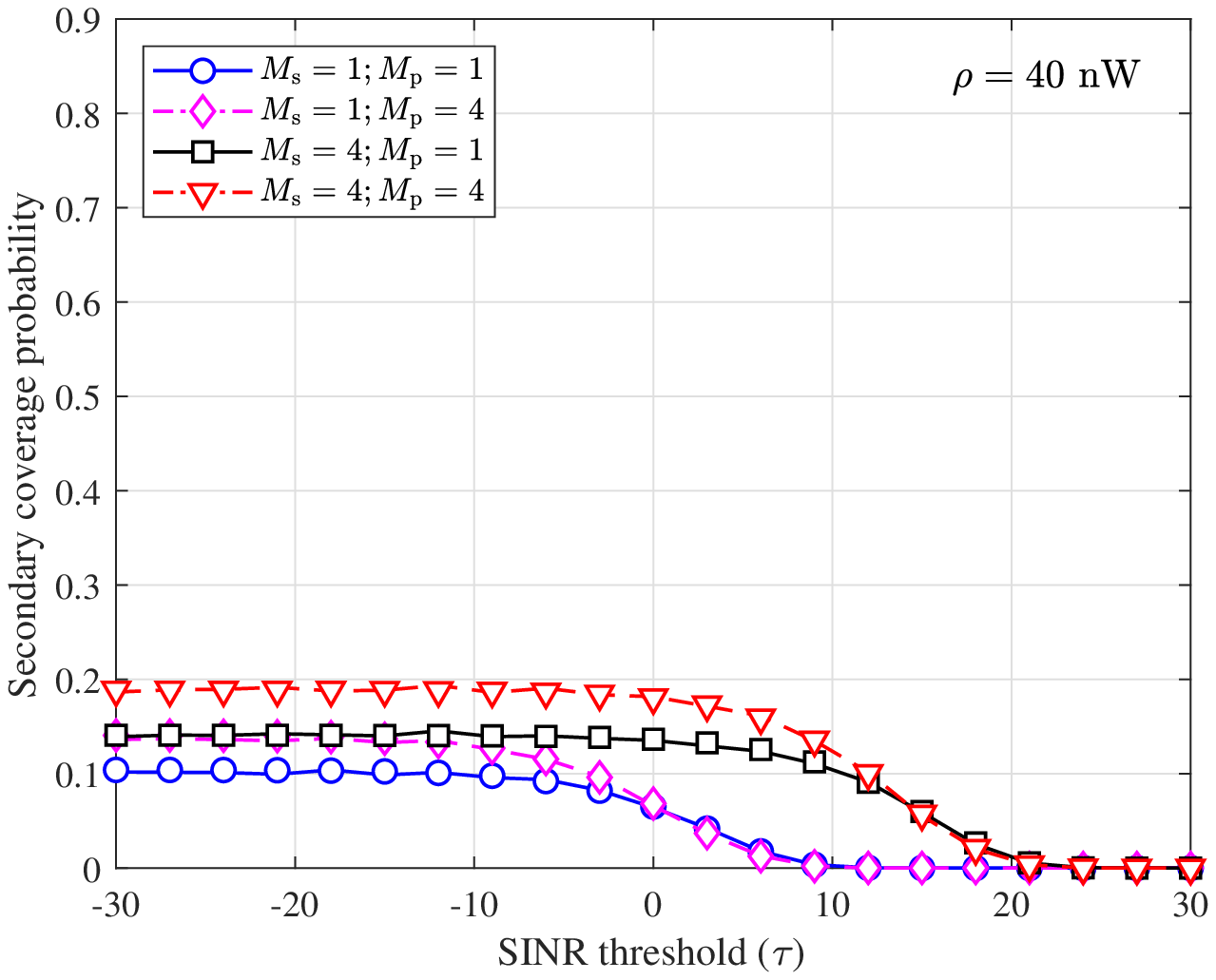} \label{numres:SP:Set2} }
\hspace{-0.7cm}
%{\small \bf (c)}
{\includegraphics[scale=0.3]{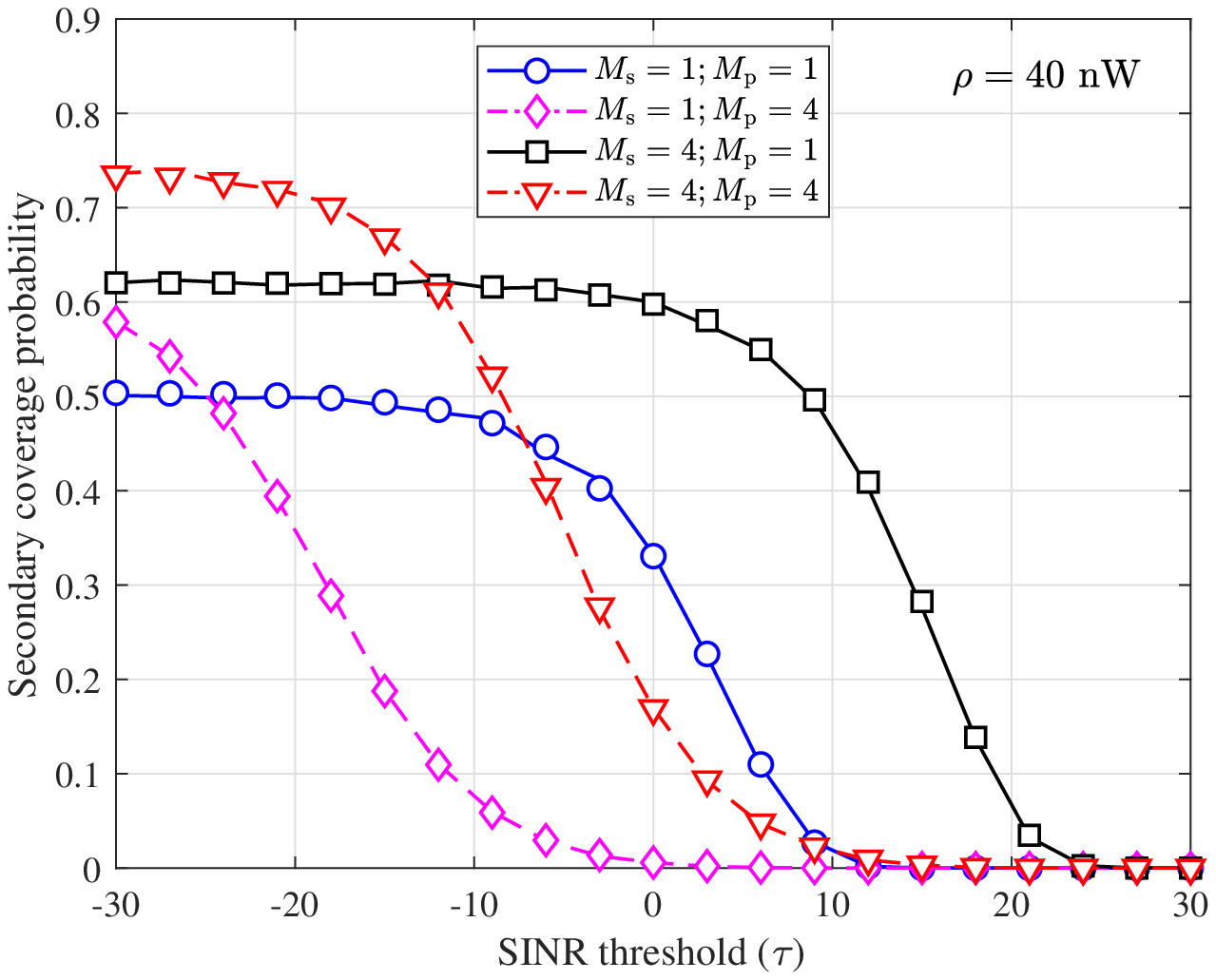} \label{numres:SP:Set3} }
\hspace{-0.7cm}
%{\small \bf (d)}
{\includegraphics[scale=0.3]{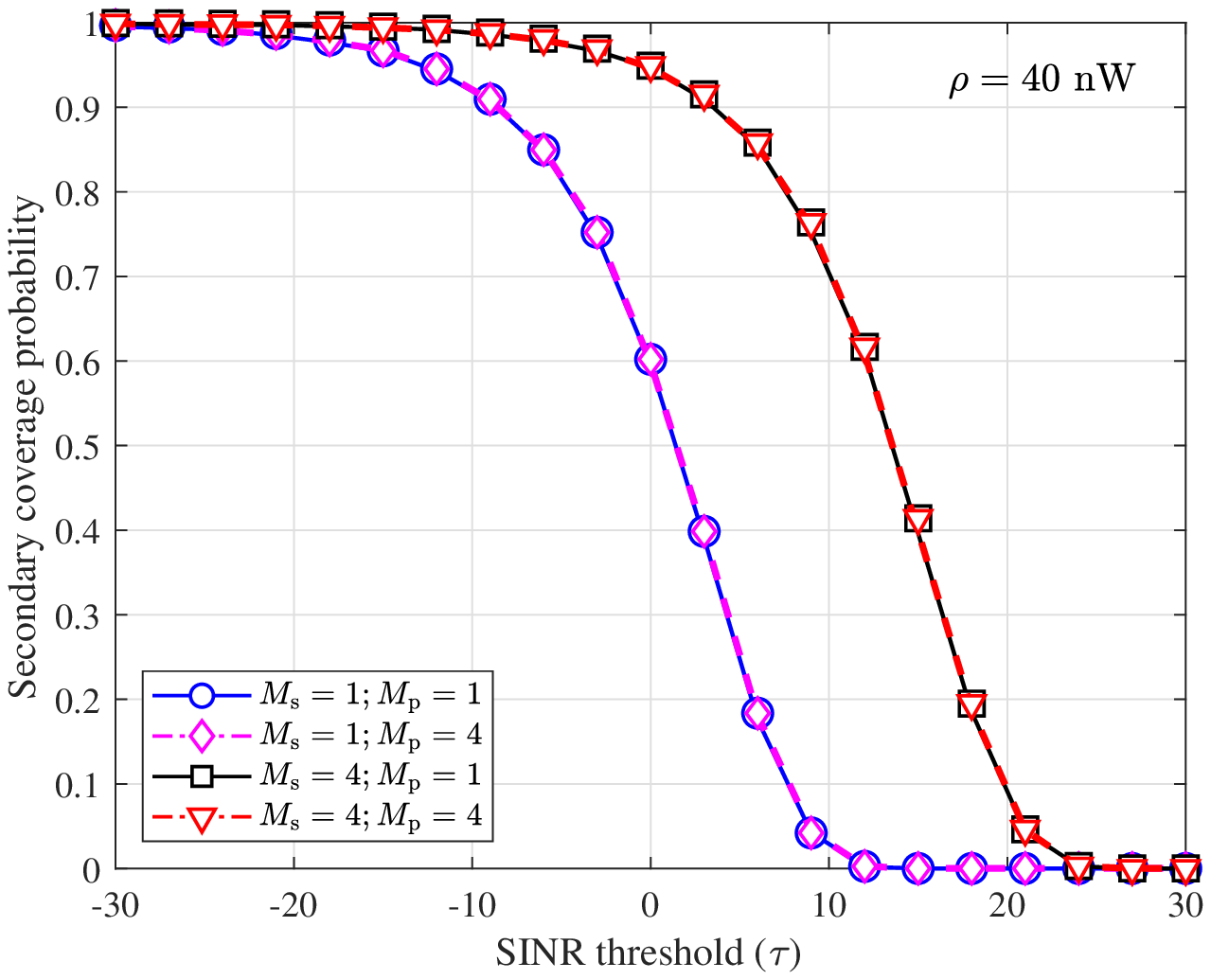} \label{numres:SP:Set4} } \\
\vspace*{-0.3cm}
{\small \hspace{0.2cm} {\bf(a)} Type 1 \hspace{2.5cm} {\bf(b)} Type 2 \hspace{2.4cm} {\bf(c)} Type 3 \hspace{1.5cm} {\bf(d)} Average user }
\vspace*{-0.3cm}
\caption{Impact of the primary and secondary antenna on the coverage of the typical secondary link for different types of secondary users with $R = 4000$ m and $\alpha = 3.3$.}
\label{numres:SP}
\vspace*{-0.5cm}
\end{figure}

%------------------------------------------------- 
Fig. \ref{numres:SP} 
%------------------------------------------------- 
shows the variations of secondary coverage with different combinations of $M_\prim$ and $M_\seco$ for all four types of secondary users. We observe that the effect of both primary and secondary antenna directionality are significant. This observation is in line with our theoretical analysis which showed that if the primary link uses directional antennas, the chances of link-access increase for neighbouring secondary devices due to the reduction of the primary receiver's protection zone. 

From
%-------------------------------------------------  
Fig. \ref{numres:SP}, 
%------------------------------------------------- 
we observe that primary antenna directionality ($M_\prim$) \textit{does not always aid to} secondary coverage probability and the exact effect depends on the location and orientation of the secondary link. Both positive and negative shifts are observed in $p_\mathrm{cs}$ when the primary antenna characteristic is changed from $M_{p} = 1$ to $M_p = 4$. Specifically, a positive shift is observed for the secondary users of Type 1. For the rest of the two Types (2 and 3), a positive shift was observed at low thresholds while a negative shift was observed at high thresholds. As per our understanding, this happens because (a) primary link has a more favourable location for Type-1 secondary users, as the transmitter and receiver of cross-links do not fall in the main lobes of each other under directional CCS, (b) the primary transmission direction is towards the typical secondary receiver of Type 2, therefore, $M_\prim$ increases the primary interference at the secondary receiver, and (c) cross-links are very near to each other, primary transmission directed outwards from the typical secondary receiver, but the secondary transmitter's signal hits the primary receiver under directional CCS of Type 3, reducing secondary activity significantly. Further, $M_\prim$ also increases AF resulting in a higher secondary interference. On the other hand, secondary antenna directionality ($M_\seco$) \textit{always aids in the improvement} of secondary coverage probability. For example, a positive shift of $+12$ dB is observed in $p_\mathrm{cs}$ when secondary antenna characteristics are changed from $M_\seco = 1$ to $M_\seco = 4$ for all three types of secondary users. For the average effect, we see that primary directionality has only a small positive effect on the secondary performance. However, the secondary directionality improves the coverage significantly.

%--------------------------------------------------------------------------------------
\subsection{Adaptive Directional Sensing} \label{section:CompilePerformance}
%--------------------------------------------------------------------------------------
The three setups described earlier represent three different types of secondary pairs. 
%------------------------------------------------- 
Fig. 12(a)
%\ref{numres:Sets123combined} 
%-------------------------------------------------
shows the variation of cumulative performance $p_\mathrm{c} (\tau, \rho)$ (which is the sum of primary and secondary coverage) for $\tau^\star = - 5$ dB for three types. We observe that the performance of Type 1 and 2 scenarios is decent with $M = 4$. However, even after adjusting $\rho$, the performance of the Type 3 scenario is not satisfactory due to high mutual interference between primary and secondary devices. But allowing the secondary user of Type 3 to use $8$ or $16$ antennas ({\em i.e.} $M_\seco = 8$ or $16$) can improve the performance significantly. On similar lines, 
%------------------------------------------------- 
Fig. 12(b) and (c)
%\ref{numres:CoveragePerformanceCombined} and \ref{numres:SumPerformanceCombined}
%-------------------------------------------------
show the cumulative and individual performance of the primary and secondary links with respect to threshold $\tau$ for three types of secondary users. We observe that the coverage of Type 3 secondary users improves significantly by increasing $M_\seco$ to $8$ and $16$ without changing the primary coverage.
\begin{figure}[ht!]
\vspace*{-0.5cm}
\centering
\hspace{-0.9cm}
%{\small \bf (a)}
{\includegraphics[trim = 0 0 0 0, clip, scale=0.4]{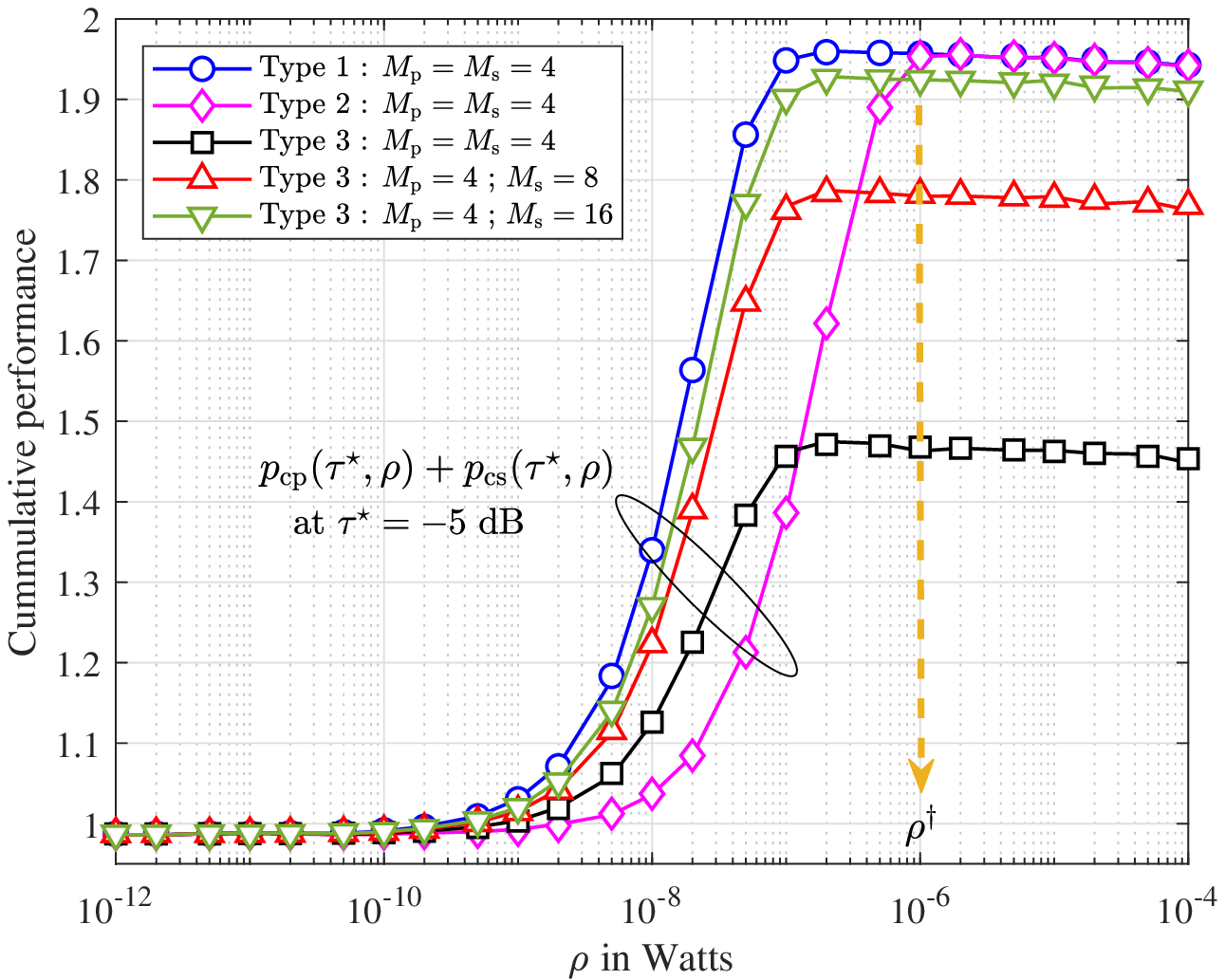}  \label{numres:Sets123combined} } \hspace{-0.9cm}
%{\small \bf (b)}
{\includegraphics[scale=0.4]{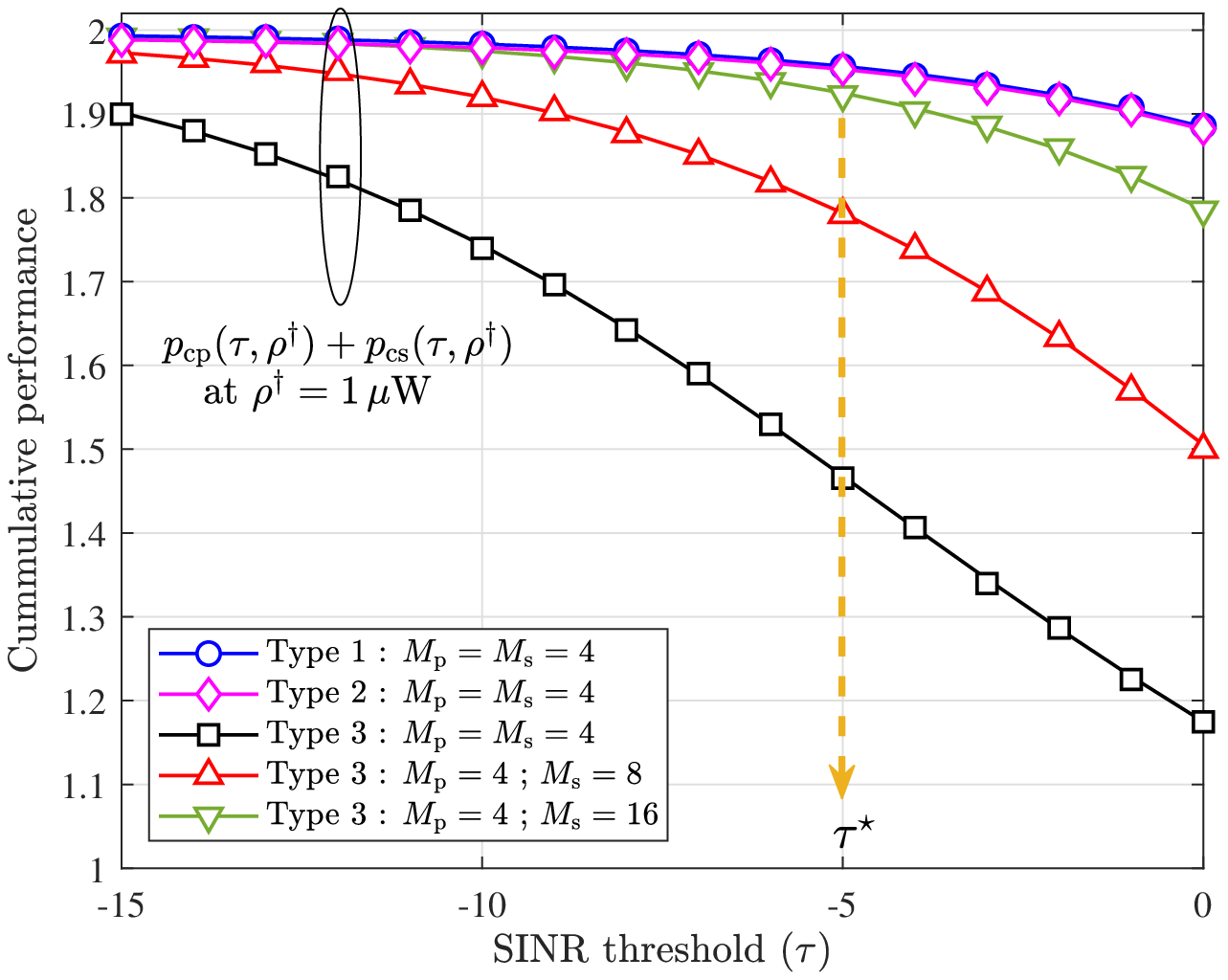} \label{numres:SumPerformanceCombined} }  \hspace{-0.9cm}
%{\small \bf (c)}
{\includegraphics[scale=0.4]{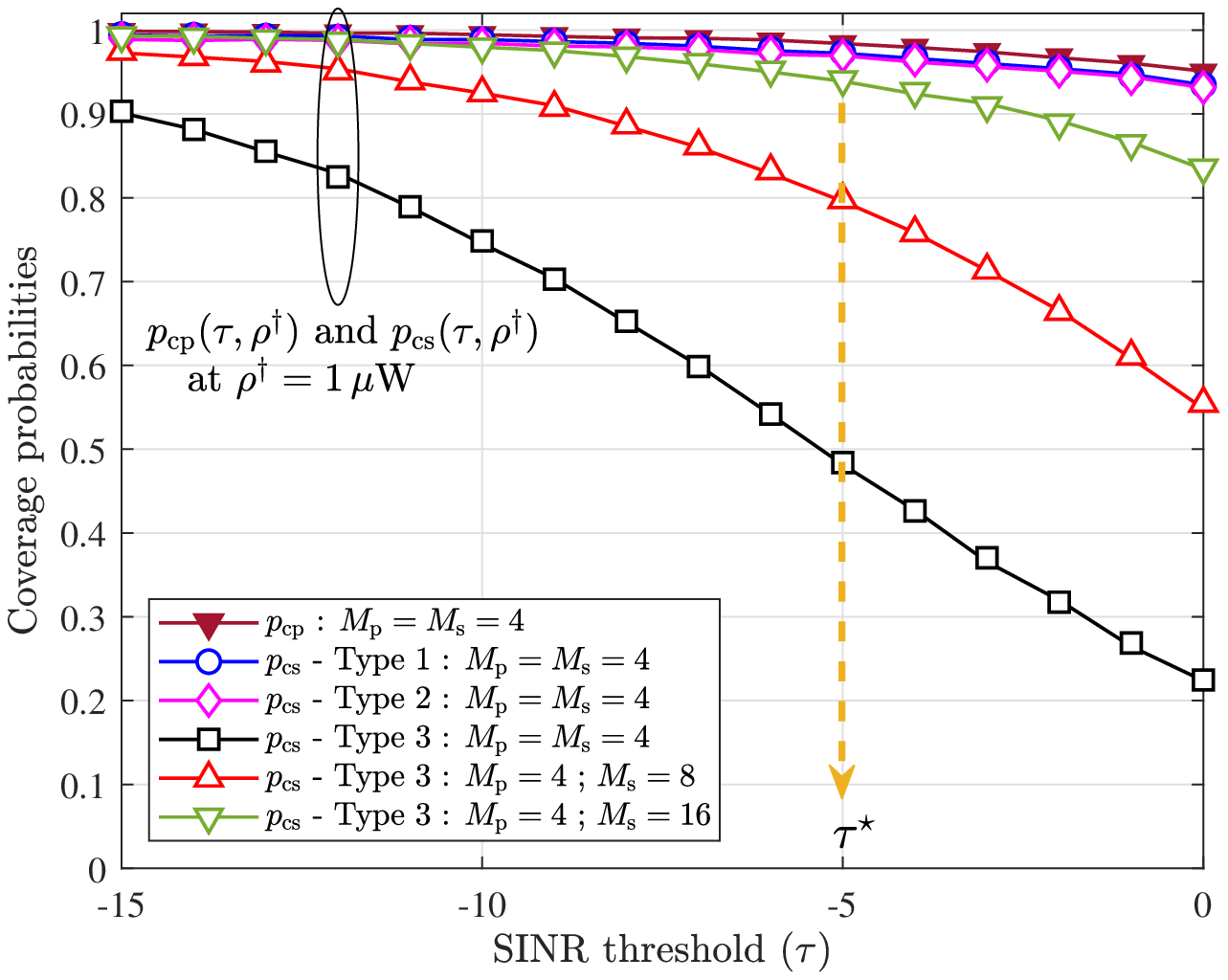} \label{numres:CoveragePerformanceCombined} } \\
\vspace*{-0.3cm}
{\small {\bf (a)} \hspace{5cm} {\bf (b)} \hspace{5cm} {\bf (c)} }
\vspace*{-0.3cm}
\caption{The figures demonstrate roles of $M_\prim$ and $M_\seco$ to achieve similar coverage performances for $\tau^\star = - 5$ dB and $\rho^\dagger = 1 \ \mu$W.}
\label{numres:Set123}
\vspace*{-0.5cm}
\end{figure}

For the particular configurations ({\em e.g.} $\tau^\star \leq - 5$ dB and $\rho^\dagger = 1 \ \mu$W), the cumulative performance and the secondary probability, both, increased by $14.32\%$ when $M_\seco$ is changed from $8$ to $16$ (see the 
%------------------------------------------------- 
Fig. 12(b) and (c)).
%\ref{numres:CoveragePerformanceCombined} and \ref{numres:SumPerformanceCombined}
%-------------------------------------------------
This shows that the Type 3 secondary users can use $M_\seco = 8$ or $16$ while the other two types of secondary users can still use $M_\seco = 4$ to achieve pre-defined levels of cumulative and individual performances. Thus, a dynamic directionality for sensing where secondary links can choose higher or lower directionality based on their orientation and location can result in a higher level of equality and fairness among devices over the network.

%-------------------------------------------------------------------------------------------------------- 
\section{Conclusions}
%-------------------------------------------------------------------------------------------------------- 
With the use of stochastic geometry, we established an analytical framework to derive the performance of a cognitive mmWave network, which consists of a single primary link and multiple secondary links. In contrast to omnidirectional CCS, which forced secondary devices to broadcast to be outside a certain distance in all directions, we have found that the use of directional CCS in mmWave networks produces better spatial reuse for the secondary transmitters. Additionally, directionality can enhance the performance of the primary and secondary networks. The primary-transmit-protection threshold $\rho$ plays a significant role in achieving a trade-off between primary and secondary performance. If $\rho$ is very small, more secondary devices become idle. On the other hand, the priority for primary transmission is compromised when  $\rho$ is very large. Further, directionality in sensing and communication can improve this trade-off by improving the coverage of both types of links. We also found that primary-link directionality does not always help from the perspective of secondary coverage performance, especially for secondary devices close to the primary. In such scenarios, one can improve the coverage performance of the network by making these secondary devices switch on a narrow beam-width regime.

%-------------------------------------------------------------------------------------------------------- 
 \appendices
 
%-------------------------------------------------------------------------------------------------------- 
\section{} \label{thrm:proof:AF}
Apply Campbell's rule \cite{AndGupDhi2016} on \eqref{eq:AF} and using the value of $p_{\mathrm{m}}$ from \eqref{eq:MAP}, we get
\begin{align}
\eta_\seco &= 1 -  \frac{1}{\pi R^2}\int_0^{2\pi} 
		\mathbb{E}_{\omega_\seco} 
%\!\! 
		\left[ 
%\! 
		\int_0^R 
%\! 
		\exp 
%\! 
		\left(
%\! 
		- \frac{ \rho x_\seco^{\alpha}}{p_\seco \gpr{\theta_\seco} \gst{\theta_\seco 
%\! 
		- 
%\! 
		\pi 
%\! 
		- 
%\! 
		\omega_\seco }} 
%\! 
		\right) 
%\! 
		x_\seco
\, 
		\dd x_\seco 
%\! 
		\right] \dd \theta_\seco.  \nonumber
\end{align}
Substituting $t = { \rho x_\seco^{\alpha} / p_\seco \gpr{\theta_\seco} \gst{\theta_\seco - \pi - \omega_\seco}} $ and using the definition of incomplete Gamma function will give the desired result.

%-------------------------------------------------------------------------------------------------------- 
\section{}\label{corl:proof:AFsectorBeam}

%For the ease of notation, we will drop the argument of the function $ \gst{\theta_\seco - \pi - \omega_\seco} $ in the below proof. 
Using \eqref{eq:gainApproximation} in \eqref{eq:AFsolution}, we get
\begin{align}
 \eta_\seco &= 1 - \frac{1}{\pi R^2} \frac{2}{\alpha} 
		\int_{0}^{\phi_{\mathrm{pr}}/2} 
		\mathbb{E}_{\omega_\seco} \left[ \left(
		\frac{\rho}{p_\seco \apr \gst{\theta_\seco - \pi - \omega_\seco} }
		\right)^{-2/\alpha}
		\Gamma \left( 
		\frac{2}{\alpha}, {\frac{ \rho R^\alpha}{p_\seco \apr \gst{\theta_\seco - \pi - \omega_\seco} }} 
		\right) \right] \dd \theta_\seco 
\nonumber \\
&\, 
		- \frac{1}{\pi R^2} \frac{2}{\alpha} 
		\int_{\phi_{\mathrm{pr}}/2}^{\pi} 
		\mathbb{E}_{\omega_\seco} \left[ \left(
		\frac{\rho}{p_\seco \bpr \gst{\theta_\seco - \pi - \omega_\seco} }
		\right)^{-2/\alpha}
		\Gamma \left(
		\frac{2}{\alpha}, {\frac{ \rho R^\alpha}{p_\seco \bpr \gst{\theta_\seco - \pi - \omega_\seco} }} 
		\right) \right] \dd \theta_\seco \nonumber \\
		& \overset{(a)}{=} 1 
%\! 
		- 
%\! 
		\frac{1}{R^2} \frac{2}{\alpha}\qpr 
%\! 
		\left[ 
%\! 
		\qst 
%\! 
		\left( 
%\! 
		\frac{\rho}{p_\seco \apr \ast }
%\! 
		\right)^{
\!\!\!
		- 2/\alpha} 
\!\!\!\!
		\Gamma 
%\! 
		\left( 
%\! 
		\frac{2}{\alpha}, \frac{ \rho R^\alpha}{p_\seco \apr \ast}
%\! 
		\right) 
\!
%\right. \nonumber \\
%&\quad 
 		+ 
%\left.
		(1 - \qst) \left( 
%\! 
		\frac{\rho}{p_\seco \apr \bst } 
%\!
		\right)^{
\!\!\!
		-2/\alpha} 
\!\!\!\! 
		\Gamma 
%\! 
		\left( 
%\! 
		\frac{2}{\alpha}, \frac{ \rho R^\alpha}{p_\seco \apr \bst} 
%\! 
		\right) 
%\! 
		\right] \nonumber \\
		&
\, 
%\! 
		- 
%\! 
		\frac{1}{ R^2} \frac{2}{\alpha} ( 1 - \qpr ) 
%\! 
		\left[ 
%\! 
		\qst 
%\! 
		\left( 
%\! 
		\frac{\rho}{p_\seco \bpr \ast} 
%\!
		\right)^{
\!\!\!
		-2/\alpha} 
\!\!\!\! 
		\Gamma 
%\! 
		\left(
%\! 
		\frac{2}{\alpha}, \frac{ \rho R^\alpha}{p_\seco \bpr \ast} 
%\! 
		\right) 
\!
%\right. \nonumber \\
%&\quad 
 		+ 
%\left.
\!
		(1 - \qst) \left( 
%\! 
		\frac{\rho}{p_\seco \bpr \bst } 
%\!
		\right)^{
\!\!\!	
		- 2/\alpha} 
\!\!\!\! 
		\Gamma 
%\! 
		\left( 
%\! 
		\frac{2}{\alpha}, \frac{ \rho R^\alpha}{p_\seco \bpr \bst} 
%\! 
		\right) 
%\! 
		\right] \nonumber 
\end{align}
where $(a)$ is due to the fact that $\omega_{\mathrm{s}} \sim \mathcal(0, 2\pi)$ resulting $g_\mathrm{st}$ to be equal to $\ast$ and $\bst$ with probabilities $\qst$ and $1 - \qst$, respectively. By taking $\psi(u)= {u}^{-\frac2\alpha} \Gamma \left( 2/\alpha, {u R^\alpha} \right)$ we get the desired result.

%-------------------------------------------------------------------------------------------------------- 
\section{} \label{thrm:proof:PrimaryCoverage}

From \eqref{eq:PrimarySINR}, the SINR coverage $p_\mathrm{cp}(\SThres. \, \rho)$ of the primary link is
\begin{align}
p_\mathrm{cp}(\SThres, \, \rho) %&= \mathbb{P} 
%		\left[ H_{\mathrm{p}}  >  \tau   \left(  \sigma^2 + I_{ \mathrm{s} }  \left( \Phi_{\mathrm{s}}  \right) \right) /  \left( p_\prim \gpt{0}\gpr{0} r_\prim^{- \alpha} \right) \right]  
%\nonumber \\
&\overset{(a)}{=} 
		\mathbb{E} \left[ \exp \left( -  \tau \left( \sigma^2 + I_{ \mathrm{s} }  \left( \Phi_{\mathrm{s}}\right) \right) /  \left( p_\prim \gpt{0}\gpr{0} r_\prim^{- \alpha} \right) \right) \right]  
%\nonumber \\
%&
		= e^{- \frac{ \kappa_\prim \tau \sigma^2 }{\rho } } \mathcal{L}_{ I_{ \mathrm{s} } } \left( { \kappa_\prim \tau / \rho} \right). \label{eq:PrimaryCov1}
\end{align}
where  $\kappa_\prim = {\rho r_\prim^{\alpha}}/{p_\prim \gpt{0}\gpr{0}}$ and $\mathcal{L}_{ I_\seco}$ is the Laplace transform (LT) of $I_\seco$. Here, (a) is due to $H_\prim \sim \exp(1)$. If we let $\chi_{\mathrm{s}i} = p_\seco g_{\mathrm{pr}} \left( \theta_{\mathrm{s}i} \right) g_{\mathrm{st}} \left( \theta_{\mathrm{s}i} - \pi - \omega_{\mathrm{s}i} \right) x_{\seco i}^{- \alpha} $ in \eqref{eq:PrimaryInterference}, we get 
\begin{align}
\mathcal{L}_{ I_\seco } \left( s \right) =& \mathbb{E} 
		\left[ e^{-s \sum_{ \X_{\mathrm{s}i} \in \Phi_\seco} 
%\!\! 
		\chi_{\mathrm{s}i} 
\, 
		G_i 
\, 
		\mathbbm{1} \left ( G_i < \frac{\rho}{\chi_{\mathrm{s}i} }  \right ) } \right] 
%\nonumber \\
\overset{(a)}{=} 
		\exp 
%\! 
		\left( 
%\! 
		- 
%\! 
		\lambda_\seco 
\!\!
		\int_0^{2\pi} 
\!\!\! 
		\int_0^\infty 
\!\!
		\left( 1 - \mathbb{E} 
%\! 
		\left[ e^{- s  \chi_\seco G
\, 
		\mathbbm{1} 
		\left ( G < \frac{\rho}{\chi_{\mathrm{s}i} } \right )  }
		\right] \right) 
%\!
		x_\seco  \mathrm{d}x_\seco \mathrm{d}\theta_\seco \right) 
\nonumber \\	
\overset{(b)}{=}& 
		\exp 
%\! 
		\left( 
%\! 
		- 
%\! 
		\lambda_\seco
%\!\!
		\int_0^{2\pi} 
%\!\!\!\!
		\int_0^\infty 
%\!\!\!
		\left( 
%\! 
		1 
%\! 
		- 
%\! 
		\mathbb{E}_{\omega_\seco} 
%\! 
		\left[ 
%\!
		e^{ - { \frac\rho{\chi_\seco}} }
%\!
		+
%\! 
		\frac{ 1 
%\! 
		- 
%\! 
		e^{ -  \rho \left(s 
%\! 
		+ 
%\! 
		{1}/{\chi_\seco} \right)} }
		{ 1 + s \chi_\seco }
%\! 
		\right] 
%\! 
		\right) 
%\! 
		x_\seco \mathrm{d}x_\seco \mathrm{d}\theta_\seco 
%\! 
		\right),
\label{eq:PrimLaplace1}
\end{align}
with $\chi_{\mathrm{s}} = p_\seco g_{\mathrm{pr}} \left( \theta_{\mathrm{s}} \right) g_{\mathrm{st}} \left( \theta_{\mathrm{s}} - \pi - \omega_{\mathrm{s}} \right) x_{\seco}^{- \alpha} $. Here, $(a)$ is due to application of PGFL on $\Phi_\seco$ \cite{AndGupDhi2016} and $(b)$ is due to $G \sim \exp(1)$. For $ s = {\kappa_\prim \tau }/{\rho}$ along with $A = \kappa_\mathrm{p} \tau$ and $C = \rho x_\seco^{-\alpha}/\chi_\seco$ in \eqref{eq:PrimLaplace1}, we can write
\vspace*{-0.5cm}
\begin{align}
\mathcal{L}_{ I_\seco } \left( {\kappa_\prim \tau }/{\rho} \right) =& \exp \left( -\lambda_\seco 
\,
\mathbb{E}_{\omega_\seco} \left[ \int_0^{2\pi} \mathsf{n}_\mathrm{p} (\theta_\seco, \omega_\seco) \mathrm{d}\theta_\seco \right] \right)
\label{eq:PrimLaplace1new}
\end{align}
where
\vspace*{-0.5cm}
\begin{align}
\mathsf{n}_\mathrm{p} (\theta_\seco, \omega_\seco) &= \int_0^\infty \left( 1 - e^{ - C x_\seco^{ \alpha} } - \frac{\left( 1 - e^{ -  A - C x_\seco^{\alpha}} \right)}{ 1 + {A}/{C x_\seco^{\alpha}}} 
\right) x_\seco \mathrm{d}x_\seco \nonumber \\
\overset{(c)}{=}& \frac{1}{\alpha} \left({p_\seco}/{\rho} \right)^{{2}/{\alpha}} \left[ (\kappa_\prim \tau)^{{2}/{\alpha}} \mathsf{n}_\mathrm{1} (\alpha) - \Gamma \left( {2}/{\alpha} \right) + \mathsf{n}_\mathrm{2} (\alpha, \kappa_\prim \tau) \right] \mathsf{n}_{3},
\label{eq:S-final}
\end{align}	 
where $\mathsf{n}_\mathrm{1} (\alpha)$, $\mathsf{n}_\mathrm{2} (\alpha, \nu)$ and $\mathsf{n}_\mathrm{3}$ are defined in Theorem \ref{theorem:PrimaryCoverage}. We use the change of variables $A + C x_\seco^\alpha = A y$ to obtain $(c)$ (see \cite[pp. 9-12]{TripGupTheoremFile2023} for full proof). Finally, substituting \eqref{eq:S-final} in \eqref{eq:PrimLaplace1} along with \eqref{eq:PrimaryCov1} will give the desired result.

%-------------------------------------------------------------------------------------------------------- 
\section{}\label{corl:proof:PrimaryCoverageSectorBeam}

Since $\mathsf{n}_\mathrm{3} = \mathbb{E}_{\omega_\seco}  \left[ \int_{-\pi}^{\pi}  \left[g_{\mathrm{pr}} \left( \theta_\seco \right) g_{\mathrm{st}} \left( \theta_\seco - \pi - \omega_\seco \right) \right]^{{2}/{\alpha}}  \mathrm{d}\theta_\seco \right]$, using \eqref{eq:gainApproximation} and \eqref{eq:WsiPrabability}, we get
\begin{align}
\mathsf{n}_\mathrm{3} &= 2 \left[ \apr^{{2}/{\alpha}} \int_0^{\phi_{\mathrm{pr}}/2}  \mathbb{E}_{\omega_\seco} \left[ \gst{ \theta_\seco - \pi - \omega_\seco}^{{2}/{\alpha}} \right]  \mathrm{d} \theta_\seco  + \bpr^{{2}/{\alpha}} \int_{\phi_{\mathrm{pr}}/2}^{\pi}  \mathbb{E}_{\omega_\seco} \left[\gst{ \theta_\seco - \pi - \omega_\seco}^{{2}/{\alpha}} \right]  \mathrm{d}\theta_\seco  \right] \nonumber \\
&= 2 \left[ \apr^{{2}/{\alpha}} \int_0^{\phi_{\mathrm{pr}}/2} \left[\qst \ast^{{2}/{\alpha}} + (1 - \qst) \bst^{{2}/{\alpha}} \right] \mathrm{d} \theta_\seco  + \bpr^{{2}/{\alpha}} \int_{\phi_{\mathrm{pr}}/2}^{\pi}  \left[\qst \ast^{{2}/{\alpha}} + (1 - \qst) \bst^{{2}/{\alpha}} \right]  \mathrm{d}\theta_\seco  \right] \nonumber \\
&= 2 \pi \left( \qpr \apr^{{2}/{\alpha}}  + (1 - \qpr) \bpr^{{2}/{\alpha}} \right) \left( \qst \ast^{{2}/{\alpha}} + (1 - \qst) \bst^{{2}/{\alpha}} \right).
\label{eq:PrimaryCov:Simplified:Sector:2}
\end{align} 

%-------------------------------------------------------------------------------------------------------- 
\section{} \label{thrm:proof:SecondaryCoverage}
From \eqref{eq:secondarySINR}, the SINR coverage of the typical secondary link is %$p_\mathrm{cs}(\SThres, \, \rho) =$ 
\begin{align}
p_\mathrm{cs}(\SThres, \, \rho) &= \mathbb{P} 
		\left[ 	F_{\seco0} \cdot 1 >  \frac{ \tau r_{\seco}^{ \alpha} \left(\sigma^2 + I_\prim +  I_\seco \left(\Phi_\seco \right)\right) }{ p_\seco \gst{0} \gsr{0} } 
\, 
		\Big \lvert U^{'}_0 = 1 \right]  \mathbb{P} \left[ U^{'}_0 = 1 \right] 
\nonumber \\
&\qquad 
		+ \mathbb{P} \left[ F_{\seco0} \cdot 0 > \frac{ \tau r_{\seco}^{ \alpha} \left(\sigma^2 + I_\prim +  I_\seco \left(\Phi_\seco \right)\right) }{ p_\seco \gst{0} \gsr{0} }
\,
		\Big \lvert U^{'}_0 = 0 \right] \mathbb{P} \left[ U^{'}_0 = 0 \right] \nonumber \\
		&\overset{(a)}{=} \mathbb{P} \left[ F_{\seco0} >  \frac{ \tau r_{\seco}^{ \alpha} \left(\sigma^2 + I_\prim +  I_\seco \left(\Phi_\seco \right)\right) }{ p_\seco \gst{0} \gsr{0} } \right]  p^{'}_{\mathrm{m0}} + 0 \cdot (1 - p^{'}_{\mathrm{m0}}) \nonumber \\
		&\overset{(b)}{=} p^{'}_{\mathrm{m0}} \cdot e^{- { \tau \sigma^2 r_{\seco}^{ \alpha} / p_\seco \gst{0} \gsr{0}}} 
		\mathcal{L}_{I_\prim} 
%\! 
		\left( \frac{ \tau r_{\seco}^{ \alpha} }{ p_\seco \gst{0} \gsr{0} } 
%\! 
		\right) 
%\! 
		\mathcal{L}_{I_\seco} 
%\! 
		\left( 
%\! 
		\frac{ \tau r_{\seco}^{ \alpha} }{ p_\seco \gst{0} \gsr{0} } 
%\! 
		\right). \! \label{eq:SecondaroCov}
\end{align}
where $\mathcal{L}_{ I_\prim}$ and $\mathcal{L}_{ I_\seco}$ are the LTs of $I_\prim$ and $I_\seco$, respectively. Here, (a) is due to $U_0 \sim \mathrm{Bernoulli}(p_{\mathrm{m}0}).$ and (b) is due to $F_{\seco 0} \sim \exp(1)$. The LT of primary interference ${ I_\prim}$ is %$\mathcal{L}_{ I_\prim}(s) =$
\begin{align}
\mathcal{L}_{ I_\prim}(s) &= \mathbb{E}_{G^{'}_0} 
		\left[ \exp \left( - s G^{'}_0 p_\prim \gsr{\delta_\prim}  \gpt{\delta_\prim - \pi - \omega_\prim} x_\prim^{- \alpha} \right) \right] 
\nonumber \\
		&=  { 1 / \left(1 + sp_\prim \gsr{\delta_\prim}  \gpt{\delta_\prim - \pi - \omega_\prim} x_\prim^{- \alpha}\right) }.
		\label{eq: Prim2SecoInterference}
\end{align} 
where the last step is due to $G^{'}_{0} \sim \exp(1)$. If $\chi_{\mathrm{s}i} = p_\seco g_{\mathrm{sr}} \left( \theta_{\mathrm{s}i} \right) g_{\mathrm{st}} \left( \theta_{\mathrm{s}i} - \pi - \omega_{\mathrm{s}i} \right) x_{\seco i}^{- \alpha} $, we can write
\vspace*{-0.5cm}
\begin{align}
\mathcal{L}_{ I_\seco} (s) &= \mathbb{E}_{\Phi_\seco } 
%\! 
		\left[ \exp \left(- 
\,
		s 
\hspace{-0.5cm}
		\sum_{ \X_{\mathrm{s}i} \in \Phi_\seco \cap \mathcal{B}(0,R) / \{\X_{\mathrm{s}0}\} } 
\hspace{-0.5cm}
		\chi_{\mathrm{s}i} 
\, 
		G^{''}_{i} 
\, 
		U^{'}_{i} \right) \right] 
%\nonumber \\
%&
		\overset{(a)}{=}  \mathbb{E}_{\Phi_\seco} \left[ \exp \left(- 
\hspace{-0.4cm}		
		\sum\limits_{ \X_{\mathrm{s}i} \in \Phi_{\seco} \cap \mathcal{B}(0,R) } 
\hspace{-0.5cm}		
		s
\, 		
		\chi_{\mathrm{s}i}
\, 
		G^{''}_{i}
\, 
		U^{'}_{i}  \right) \right] 
		\nonumber \\
    	&
		\overset{(b)}{=}  \exp 
%\! 
		\left( 
%\! 
		- \lambda_\seco 
%\!\! 
		\int_0^{2\pi} 
%\!\!\! 
		\int_0^{R} 
%\!\! 
		\left(1 - \mathbb{E}_{G^{''}}  \left[ \mathbb{E}_{U^{'}} 
		\left[ e^{ - s\, \chi_{\mathrm{s}}
\, 
		G^{''}
\, 
		U^{'} } \right] 
		\right] \right)  x_\seco
\, 
		\dd x_\seco
\, 
		\dd \theta_\seco 
%\! 
		\right) 
\nonumber \\
&
		\overset{(c)}{=}  \exp 
%\! 
		\left( 
%\! 
		- \lambda_\seco 
%\!\!
		\int_0^{2\pi} 
%\!\!\! 
		\int_0^{R} 
%\!\!
		\mathbb{E} 
%\! 
		\left[ 
		\frac{ p^{'}_{\mathrm{m}} }{ 1 + {1 /s \chi_\seco}} 
		\right] x_\seco
\, 
		\dd x_\seco
\, 
		\dd \theta_\seco 
%\! 
		\right). 	\label{eq: SecoLaplace}
\end{align}
where $(a)$ is due to Silvnyak theorem \cite{AndGupDhi2016}, (b) is due to application of PGFL on $\Phi_\seco$ and (c) is due to $U^{'} \sim \mathrm{Bernoulli}(p^{'}_\mathrm{m})$ and $G^{''} \sim \exp(1)$. Substituting \eqref{eq: Prim2SecoInterference} and \eqref{eq: SecoLaplace} in \eqref{eq:SecondaroCov} along with values of $s$, $\chi_\seco$, $p^{'}_{\mathrm{m}}$ and $p^{'}_{\mathrm{m0}}$ and using change of variables
$A_\mathrm{0}$, $A_\mathrm{1}$, $A_\mathrm{2}$, $A_\mathrm{s}$ and $C_\mathrm{s}$ will give the desired result.

%-------------------------------------- Bibliography -------------------------------------------- 
\nocite{*}
\bibliographystyle{IEEEtran}
\bibliography{IEEEabrv,PapersName}

%----------------------------------------------------------------------------------------------------- 
\end{document}